\documentclass[fleqn,usenatbib]{mnras}

\usepackage{times}
\usepackage[T1]{fontenc}

\DeclareRobustCommand{\VAN}[3]{#2}
\let\VANthebibliography\thebibliography
\def\thebibliography{\DeclareRobustCommand{\VAN}[3]{##3}\VANthebibliography}

\usepackage{graphicx}	
\usepackage{amsmath}	
\usepackage{amssymb}	

\usepackage{rotating}
\usepackage{pdflscape}
\usepackage{colortbl}
\usepackage[normalem]{ulem}
\usepackage{color}
\usepackage{bm} 
\definecolor{darkblue}{rgb}{0.0,0.0,0.8}
\definecolor{darkred}{rgb}{0.75,0.,0.25}
\definecolor{darkorange}{rgb}{1,0.3,0.0}
\definecolor{darkgreen}{rgb}{0.0,0.6,0.0}
\definecolor{darkpurple}{rgb}{0.8,0.,0.9}
\definecolor{brown}{rgb}{0.65,.16,0.16}
\definecolor{grey}{rgb}{0.4,0.5,0.6}
\definecolor{white}{rgb}{1,1,1}
\definecolor{trolleygrey}{rgb}{0.5, 0.5, 0.5}
\definecolor{lavender}{rgb}{0.835,0.812,0.969}
\definecolor{pastelorange}{rgb}{0.99,0.92,0.82}
\definecolor{pastelblue}{rgb}{0.85,0.93,0.99}

\newcommand{\white}[1]{\textcolor{white}{#1}}
\newcommand{\darkg}[1]{\textcolor{trolleygrey}{#1}}

\newcommand{\darkr}[1]{\textcolor{darkred}{#1}}

\newcommand{\darkor}[1]{\textcolor{darkorange}{#1}}

\renewcommand{\d}{{\rm d}}
\newcommand{\mpo}{{\sc MAMPOSSt-PM}}
\newcommand{\balrogo}{{\sc BALRoGO}}
\newcommand{\agama}{{\sc Agama}}
\newcommand{\gaia}{{\sl Gaia}}
\newcommand{\hst}{{\sl HST}}
\newcommand{\cmc}{{\tt CMC}}
\newcommand{\parsec}{{\sc Parsec}}

\newcommand{\msun}{\rm M_\odot}

\newcommand{\masyr}{\rm mas\,yr^{-1}}

\newcommand{\posr}{{\rm R}}
\newcommand{\post}{{\rm T}}

\title[Compact objects in globular clusters]{Stellar graveyards: Clustering of compact objects in globular clusters NGC~3201 and NGC~6397}

\author[Vitral, Kremer, Libralato, Mamon \& Bellini]{
Eduardo Vitral,$^{1}$\thanks{E-mail: vitral@iap.fr}
Kyle Kremer,$^{2,3}$\thanks{E-mail: kkremer@caltech.edu}
Mattia Libralato,$^{4}$
Gary A. Mamon$^{1}$
and Andrea Bellini$^{5}$
\\
$^{1}$Sorbonne Universit\'e, CNRS, UMR 7095, Institut d’Astrophysique de Paris, 98 bis bd Arago, 75014 Paris, France \\
$^{2}$TAPIR, California Institute of Technology, Pasadena, CA 91125, USA\\
$^{3}$The Observatories of the Carnegie Institution for Science, Pasadena, CA 91101, USA\\
$^{4}$AURA for the European Space Agency (ESA), ESA Office, Space Telescope Science Institute, 3700 San Martin Drive, Baltimore, MD 21218, USA \\
$^{5}$Space Telescope Science Institute, 3700 San Martin Drive, Baltimore, MD 21218, USA
}

\date{Accepted 2022 May 4.}

\pubyear{2015}

\begin{document}
\label{firstpage}
\pagerange{\pageref{firstpage}--\pageref{lastpage}}
\maketitle

\begin{abstract}

We analyse \gaia\ EDR3 and re-calibrated \hst\ proper motion data from the core-collapsed and non core-collapsed globular clusters NGC~6397 and NGC~3201, respectively, with the Bayesian mass-orbit modelling code \mpo. We use Bayesian evidence and realistic mock data sets constructed with \agama\ to select between different mass models. In both clusters, the velocities are consistent with isotropy within the extent of our data. We robustly detect a dark central mass (DCM) of roughly $1000\,\msun$ in both clusters. Our \mpo\ fits strongly prefer an extended DCM in NGC~6397, while only presenting a mild preference for it in NGC~3201, with respective sizes of a roughly one and a few per cent of the cluster effective radius. We explore the astrophysics behind our results with the \cmc\ Monte Carlo $N$-body code, whose snapshots best matching the phase space observations lead to similar values for the mass and size of the DCM. The internal kinematics are thus consistent with a population of hundreds of massive white dwarfs in NGC~6397, and roughly 100 segregated stellar-mass black holes in NGC~3201, as previously found with \cmc. Such analyses confirm the accuracy of both mass-orbit modelling and Monte Carlo $N$-body techniques, which together provide more robust predictions on the DCM of globular clusters (core-collapsed or not). This opens possibilities to understand a vast range of interesting astrophysical phenomena in clusters, such as fast radio bursts, compact object mergers, and gravitational waves. 
\end{abstract}













\section{Introduction}
\label{sec: intro}
Globular star clusters (GCs) are among the most dynamically active environments in the Universe. These roughly spherical, dense clusters are composed of $\sim 10^5-10^6$ stars tightly packed, with the densest GCs having inner densities as much as $\sim10^6$ times greater than what is observed in our solar neighbourhood \citep{McKee+15}. In such dense environments, stellar evolution is shaped by the internal dynamics by means of phenomena such as runaway mergers, as well as mass segregation: a consequence of dynamical friction and energy equipartition that leads more massive stars to locate closer to the cluster's centre, while less massive ones are moved towards the outskirts. This makes GCs excellent laboratories to study compact objects -- including white dwarfs \citep[e.g.,][]{Richer1997}, neutron stars \citep[e.g.,][]{Lyne1987}, stellar-mass black holes \citep[e.g.,][]{Giesers+18}, and intermediate-mass black holes (IMBHs; e.g. \citealt{Greene+20}) -- by means of simulations (e.g., \citealt{Wang2016dragon,Askar2017,Kremer+20,Rodriguez+21b}) and observational data analysis (e.g., \citealt{vanderMarel&Anderson&Anderson10,Vitral&Mamon21,Haberle+21}).

One of the most interesting phenomena related to GCs is the process of cluster core collapse, which is intrinsically related to the exchange of energy due to dynamical interactions in the cluster. In self-gravitating systems like GCs, the virial theorem reveals that the centres of GCs have a negative heat capacity, i.e., an energy input in the system triggers a decrease of the kinetic energy, which can be regarded as the system's ``temperature'' (e.g., \citealt{Binney2008}). Such a counter-intuitive relation tends to evolve in a typical cluster's interior, as it exchanges energy with its outer regions, naturally from the former to the latter. This process leads inevitably to the ``collapse'' of stars to the cluster's inner-most regions (e.g., \citealt{Henon61,LyndenBell&Wood&Wood68}).

The core-collapse process has been studied through many aspects (e.g., \citealt{Heggie79,Cohn80,Makino&Hut&Hut91,Goodman93}), but an important disparity arises when analysing the timescales expected for core-collapse from classic dynamical arguments and the ages of Galactic GCs: Many of Milky Way GCs are sufficiently dense to have experienced enough relaxation for core-collapse to occur in their lifetimes\footnote{GCs are particularly old systems with ages ranging up to 13 Gyrs \citep{MarinFranch+09}.} (e.g., \citealt{Spitzer87,Quinlan96} and equation~5 from \citealt{PortegiesZwart&McMillan02}). However, there is a clear bi-modality of core-collapsed and non core-collapsed clusters among the roughly 150 GCs observed in our galaxy, with only a fifth of them presenting a core-collapse structure \citep{Djorgovski&King86,Harris10}, characterised by a steep increase in the density profile at very inner radii. Many works argue that such bi-modality is related to three-body encounters, a process called \emph{binary burning}, where dynamical interactions of binaries with other stars cause tight (`hard') binaries to harden while the third star (not necessarily the original one) is kicked out at a higher speed than the initial third star came in with \citep{Heggie1975}. This process effectively pumps energy to the cluster's inner regions, thus preventing core-collapse from continuing indefinitely 
\citep{Hills75}.

More recently, \cite{Chatterjee+13} showed that the bi-modality between core-collapsed and non core-collapsed clusters could be associated with clusters having reached or not, respectively, this binary-burning phase. However, this study needed to assume relatively low initial cluster densities \citep[with respect to recent observations of young massive star clusters -- the expected local universe analogues of GC progenitors;][]{Bastian2005,Scheepmaker2007,PortegiesZwart2010} in order to obtain correct timescales of core-collapse, which still seemed to arrive too fast. Thus, the question remains: What mechanism is able to effectively delay core-collapse, in order to explain the relatively small core-collapsed GC population in the Milky Way?





The answer to this question has been gradually shaped in the last decade, especially thanks to the improvement of our knowledge of  
black hole populations in GCs (e.g., \citealt{Strader2012,Giesers+18,Giesers+19}). In fact, by means of realistic $N$-body simulations 
\citep[e.g.,][]{Morscher2015,Wang2016dragon, Askar2017,Kremer+20,Rodriguez+21}, black holes are now suggested to be behind the observed delay of core-collapse in many GCs (\citealt{Merritt2004,Mackey2007,BreenHeggie2013,Askar2018,Kremer+18,Kremer2019}). Black holes in GCs form and sink early (on $\lesssim100\,$Myr timescales) to the cluster's centre due to a combination of their high masses and energy equipartition. Once in the inner regions of GCs, black holes dynamically interact with one another and with luminous stars. Those living in hard binaries thus provide a similar energy exchange towards the cluster's interior as in the classical stellar binary-burning scenario, but amplified due to the relative high masses of black holes compared to stars. This phenomenon has been referred to as \textit{black hole binary-burning} \citep{Kremer+20p}.

This new theoretical comprehension of the physics governing GCs suggests that the ones without the characteristic inner cuspy structure of core-collapse probably harbour a segregated black hole population, responsible for the delay of core-collapse. Nevertheless, black holes are expected to eventually leave the cluster,
primarily from repeated dynamical encounters between black hole binaries and other black holes. These encounters harden the black hole binaries \citep[e.g.,][]{Heggie1975} and also pump linear momentum into the black hole binaries and single black holes, ultimately leading to ejection of black holes from their host clusters \citep[e.g.,][]{Kulkarni1993,Morscher2015,Kremer+20}.
Also, when binary black holes merge, the massive amount of energy released in the form of gravitational waves is in general anisotropic \citep[e.g.,][]{Barausse2009,Lousto2012,Gerosa2016}, and conservation of linear momentum leads to gravitational kicks \citep{Peres62}, whose amplitudes \citep{Lousto+10} should be usually sufficient to eject the resulting black hole from its host cluster. Ultimately ($\gtrsim10\,$Gyr timescales), the clustered black hole population becomes negligible, allowing other luminous stellar components to sink, as well as less massive compact objects such as neutron stars and white dwarfs. When these more luminous components collapse in the centre, forming the characteristic core-collapse inner cusp, \textit{stellar} and \textit{white dwarf} binary-burning effectively halts further shrinking of the core \citep{Kremer+21}.
Complementary observational constraints are required in order to validate these various theoretical predictions. Conveniently, we find ourselves in a prosperous moment to perform such observational analysis, since new releases from the \gaia\ mission \citep{GaiaHelmi+18,GaiaCollaboration+21} and long baselines from the \textit{Hubble Space Telescope} (\hst, \citealt{Bellini+14, Bellini+18} and \citealt{Libralato+18, Libralato+19}) are available, with proper motion data having unprecedented precision and completeness for nearby GCs. In addition, new mass-orbit modelling algorithms, such as \mpo\ (\citealt*{Mamon+13}; Mamon \& Vitral in prep.) and others (see \citealt{Read+21} for a comparison of different approaches), have become available to analyse large discrete kinematic datasets. 
These algorithms come to complement previous Jeans modelling techniques \citep[e.g.,][]{Binney&Mamon82,vanderMarel94,Cappellari08,Mamon&Boue10}, as well as distribution function modelling (e.g., \citealt{Wojtak+09}), in particular multi-mass distribution function based models\footnote{These methods provide, in particular, good constraints on which kind of stellar remnants (i.e., white dwarfs, neutron stars and black holes) compose dark central masses in GCs.}
dating from \cite{DaCosta&Freeman76,Illingworth&King76,Gunn&Griffin79}, with recent applications highlighted in \cite{Sollima+12,Gieles+18,Zocchi+19,HenaultBrunet+20}.

Such mass-orbit modelling of the kinematics can answer fundamental questions concerning the cores of globular clusters: 
1) Do globular clusters contain excess matter in their cores?
2) If yes, is the excess mass point-like, implying the presence of the long sought-after IMBHs, or are they extended?
Recent analyses of GC kinematics  
indicate that segregated compact objects inhabit the inner regions of GCs  (\citealt{Zocchi+19,Mann+19,Vitral&Mamon21}, hereafter, VM21).
Beyond the realm of mass-orbit modelling, we also ask
3) If the excess mass is extended, what dominates its mass: white dwarfs, neutron stars, or stellar-mass black holes?

In this work, we provide the first comparative analysis
of the inner unseen excess-mass of globular clusters, based on both observations and simulations, 
in a non core-collapsed cluster (NGC~3201) and a classic core-collapsed one (NGC~6397). 
We divide our paper as follows: In Section~\ref{sec: data-overview}, we overview the data we use and the two clusters we analyse. Sections~\ref{sec: data-clean} and \ref{sec: methods} explain the data cleaning procedure and the methods we use to analyse the data. We present our results and robustness checks in Section~\ref{sec: results}. Finally, we discuss and summarise our work in Section~\ref{sec: discussion}. 

\section{Data overview} \label{sec: data-overview}

\subsection{Proper motions}

We perform our mass-modelling fit with proper motion data from \hst\ and \gaia\ EDR3. We briefly describe the main aspects of these new data below.

\subsubsection{\gaia\ EDR3}


Among the main qualities of the \gaia\ EDR3 data set that impact our work (for a summary of the main aspects of this mission, see \citealt{GaiaCollaboration+21,Lindegren+21}), we highlight, compared to the previous \gaia\ DR2 data, the $\sim 2$ times better precision on proper motion measurements and better photometric precision, rendering more homogeneous colour-magnitude diagrams (CMDs). In practice, this yielded not only more reliability to our data, but also improved completeness,  especially for nearby clusters such as the ones we analyse. 

\gaia\ EDR3 data presents an inconvenient issue related to spatially correlated systematic errors \citep[e.g.,][]{Lindegren+21}, which is usually associated to the telescope scan directions, even though there has been significant improvement from DR2 to EDR3. The modelling and correction of these systematics in our data is beyond the scope of this work, and we  only use the statistical errors provided in the catalogue. In fact, the impact of these systematics on GCs is not yet very clear, with recent works focusing more on describing them rather than presenting a method to correct for them \citep[e.g.,][]{Fardal+21}. The most robust correction for these systematics in GCs is perhaps the one given in \cite{Vasiliev&Baumgardt&Baumgardt21} where the authors calculate an uncertainty floor of $\epsilon_{\mu} \sim 0.026 \, \masyr$, which remains considerably below the smallest statistical errors of our cleaned \gaia\ EDR3 data, of the order of $\epsilon_{\mu} \sim 0.06 \, \masyr$ for both of the clusters we study. Notice that in their figure~6, \cite{Vasiliev&Baumgardt&Baumgardt21} do test the impact of these systematics in the velocity dispersion profile of NGC~3201, and one can see that for the magnitude range of our \gaia\ EDR3 cleaned data for this cluster (i.e., $G \lesssim 18.5$, as seen in our Figure~\ref{fig: err-mag-relation}), the impact of these systematics is small, especially when considering the spatial range of our \gaia\ data (i.e., $\sim 2$ to 8 arcmin in projected radii, such as seen in Figure~\ref{fig: sky-view}, provided as online material). We checked with E.~Vasiliev that a similar trend was found for NGC~6397, in a \gaia\ EDR3 magnitude range close to ours (i.e., $G \lesssim 19$). Figure~\ref{fig: scaling}, provided as online material, also corroborates the negligible effect of such systematics in our mass modelling.

\subsubsection{\hst}
\label{sssec: hst}


The \hst\ data reduction and proper-motion computation were performed following the prescriptions of \citet{Bellini+14, Bellini+18} and \citet{Libralato+18, Libralato+19}. In this Section, we briefly summarize the salient points. The detailed description of the workflow will be provided in an upcoming paper (Libralato et al., in preparation).

We made use of all suitable {\tt \_flc} exposures taken before 2019 with the Wide-Field Channel (WFC) of the Advanced Camera for Surveys (ACS) and with the Ultraviolet-VISible (UVIS) channel of the Wide-Field Camera 3 (WFC3). In a first pass, an initial set of positions and fluxes for the brightest and most isolated sources in each exposure was estimated via fits of the point-spread-function (PSF) to the brighter sources. The PSF model varies across the frame and depends on the frame.
These sources, in combination with the \gaia\ Data Release 2 catalogue \citep{GaiaCollaboration+16, GaiaCollaboration+18B}, were then used to setup a common reference-frame system. Once onto the same reference system, all images were used at once to re-determine position and flux of all detectable sources, this time PSF-subtracting all close-by neighbours prior to the final fit.
This second-pass-photometry stage is designed to enhance the contribution of faint sources, and yields better measurements in crowded regions (by subtracting all detected close-by neighbours before estimating position and flux of an object)\footnote{NGC~6397 was analysed by \citet{Vitral&Mamon21} using the proper-motion catalogue made by \citet{Bellini+14}. The data reduction carried out in this manuscript mainly differs from that of \citet{Bellini+14} by the addition of the second-pass-photometry stage. As exhaustively described in \citet{Bellini+18} and \citet{Libralato+18, Libralato+19}, second-pass photometry provides better results for faint sources and crowded environments than first-pass photometry.}.

Finally, proper motions were computed following \citet{Bellini+14}, i.e. by fitting geometric-distortion-corrected positions transformed onto the same reference system as a function of epoch with a least-squares straight line.
The slope of the straight line provides an estimate of the proper motion. 
Spatial patterns in proper motions indicate systematic errors, which
were also corrected, both for low and high spatial frequency, with the prescriptions of \citet{Bellini+18} and \citet{Libralato+18, Libralato+19}.

Various \hst\ data sets were used to compute the astro-photometric catalogues of NGC~3201 and NGC~6397. In the following, we considered in the analysis only objects that were measured in both GO-10775 (ACS/WFC images in F606W and F814W filters; PI: Sarajedini) and GO-13297 (WFC3/UVIS exposures in F275W, F336W, and F438W filters; PI: Piotto) data.
Finally, as described in, e.g., \cite{Bellini+17} and \citet{Libralato+18}, the procedure used to compute proper motions removes any signature of the systemic rotation of the GC in the plane of the sky. Thus, we cannot infer rotation directly from our \hst\ proper motions.

\subsection{NGC~3201 \& NGC~6397} 

The choice of which clusters to study depends on the availability of good quality data as well as on structural characteristics that facilitate our modelling. For instance, strong imprints of rotation or non-spherical sources are not ideal, as our mass-modelling routine considers a spherical system with no rotation when solving the Jeans equation (see Section~\ref{sssec: mpopm-formalism}). Similarly, sources that are located too far away (e.g., $\gg 5$~kpc) usually have characteristic uncertainties much higher than the local velocity dispersion, which could induce an error underestimation that undermines our study. For those reasons, we choose to work with NGC~3201 and NGC~6397, whose main features we comment below.

\subsubsection{NGC~3201}

NGC~3201 is a $10.4$ Gyr old cluster \citep{MarinFranch+09} that orbits the Milky Way in a retrograde orbit and recedes from the Sun with a velocity of nearly $500~$km~s$^{-1}$ \citep{GaiaHelmi+18}. It is located at 4.74~kpc from the Sun \citep{Baumgardt&Vasiliev21}, it had its dynamics studied many times (e.g., \citealt{Bianchini+19,Wan+21}) and its ellipticity is $0.12$, according to \cite{Harris96,Harris10}\footnote{Ellipticity is defined in this catalogue as $e = 1 - b/a$, where $a$ and $b$ are the semi-major and minor axis of the isophote projected ellipse, respectively.}. The median and maximum \hst\ proper motion baselines among the stars are 4 and 8 years, for this cluster.

Rotation in this cluster can be overestimated when disregarding perspective rotation (see \citealt{vandeVen+06,Wan+21}, for details) due to its high line-of-sight velocity, but recent studies that treat this issue tend to agree that a rotation signal of amplitude $\sim 10$  per cent of the velocity dispersion is present in its innermost regions \citep{Sollima+19}, a feature that is erased, by construction, on our \hst\ subset. The outskirts of our \gaia\ data however, could have an increasing rotation pattern, 
but recent studies using \gaia\ proper motions found that its rotation is much smaller than its velocity dispersion \citep{Bianchini+18,Sollima+19,Vasiliev19c}.
Along with the fact that our \gaia\ data actually represents only $\sim 14$  per cent of our NGC~3201 subset, we thus ignore this cluster's rotation and assume it to have an weak effect on our mass-modelling.

Among the many interesting features of NGC~3201, we stress that it is far from a core-collapse state \citep{Djorgovski&King86}, which is most likely related to the black hole population thought to inhabit its inner regions \citep{Kremer2019,Weatherford+20}. Indeed, \cite{Giesers+18} recently provided solid evidence for a stellar-mass\footnote{They measured a mass of $4.36\pm0.41$~M$_{\odot}$.} black hole dynamical detection near the cluster's centre and follow-up observations revealed additional black holes \citep{Giesers+19}. Such a black hole population could provide enough energy, by means of dynamical interactions, to halt the cluster core-collapse.

\subsubsection{NGC~6397}

NGC~6397 is the second closest GC to our Sun, at only $2.48$~kpc away \citep{Baumgardt&Vasiliev21}, and is a very metal-poor ([M/H]$=-1.54$), old (12.87 Gyr) cluster \citep{MarinFranch+09}. 
It is very spherical ($e = 0.07$, \citealt{Harris96,Harris10}) and its rotation is negligible relative to its velocity dispersion \citep{Bianchini+18,Sollima+19,Vasiliev19c}.
As a matter of fact, recent mass-modellings of this cluster have neglected rotation and argued that such assumption did not affect its overall modelling (e.g., \citealt{Kamann+16},VM21). The median and maximum \hst\ proper motion baselines
among the stars are 9.7 years (both statistics), for this cluster.

Claims of a central dark mass in this cluster were recurrent: \cite{Larson84} first proposed that a central $1600$~M$_{\odot}$ component formed by compact remnants could reside int the cluster's core, while \cite{Kamann+16} fitted line-of-sight data of this cluster to claim a $600$~M$_{\odot}$ intermediate-mass black hole (IMBH) detection. Recently, VM21 showed, with proper motions from \hst, \gaia\ DR2 and the velocities from \cite{Kamann+16} that such a fit was indeed consistent with the data, but an extended component of roughly $1000$~M$_{\odot}$, composed by stellar remnants, was actually much favoured by statistical indicators and goodness-of-fit comparisons. As the authors did not account for a thorough cluster evolution analysis that took into account dynamical interactions and black hole ejection from the cluster, they proposed that such population could be dominated by stellar-mass black holes, in mass, and by white dwarfs, in number.

However, NGC~6397 is a core-collapsed cluster \citep{Djorgovski&King86}, which is not consistent with it presenting many black holes. As explained before, core-collapse is thought to occur once the black hole original population of the cluster has been almost entirely ejected, and therefore no strong energy input can be provided to delay the core from collapsing (e.g., \citealt{Merritt2004,Mackey2007,BreenHeggie2013,Wang2016dragon,Askar2018,Kremer2019,Kremer+20}). Recent studies have then proposed that the dark population detected by VM21 is most likely composed of segregated massive white dwarfs, which form a sub-cluster in the internal regions of the GC \citep{Rui+21,Kremer+21}.



\subsection{\parsec\ isochrones}

\begin{figure}
\centering
\includegraphics[width=0.95\hsize]{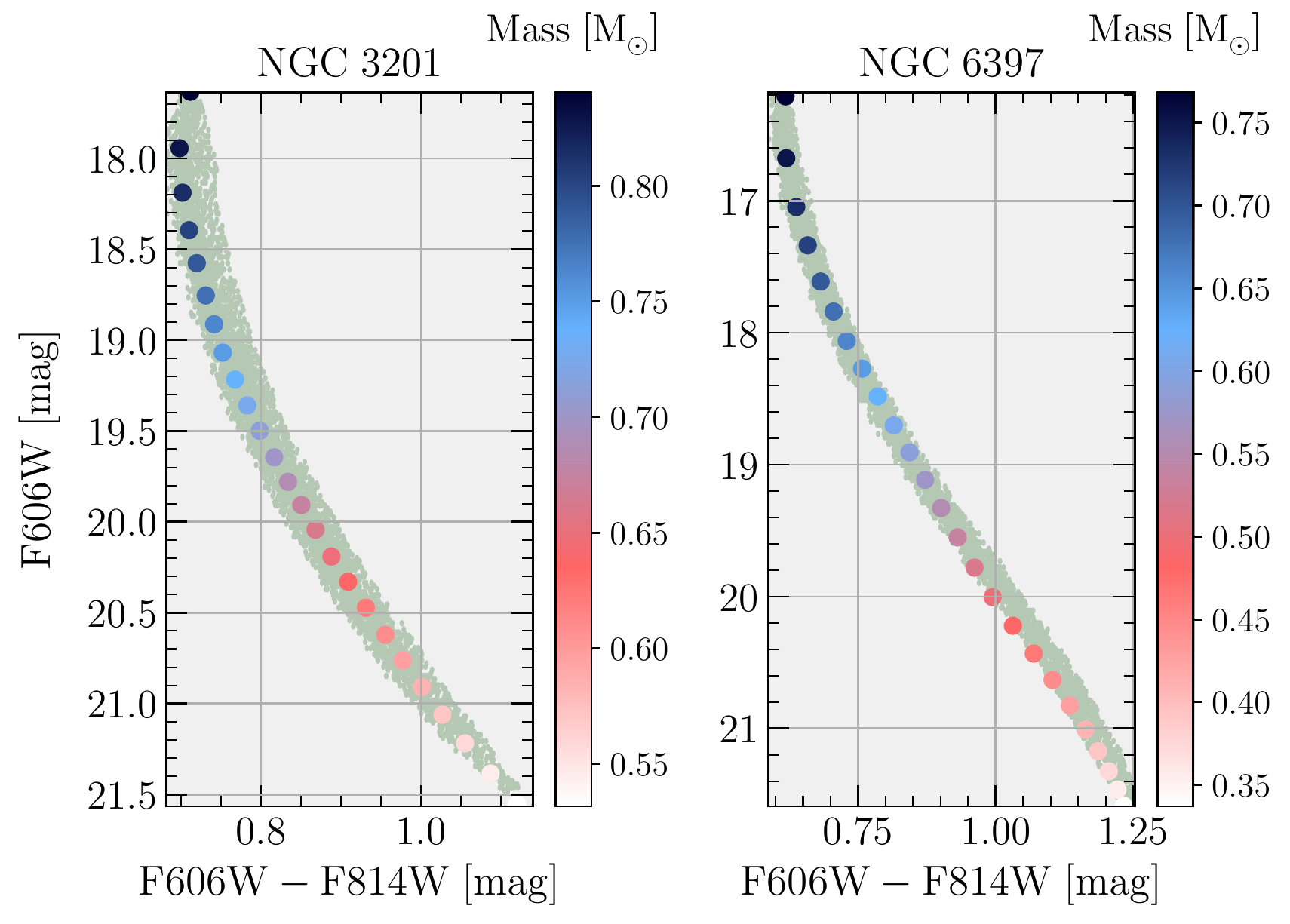}
\caption{\textit{Colour-magnitude diagrams of the two globular clusters:}
The \emph{small gray-green points} are the \hst\ data, cleaned according to Sect.~\ref{sec: data-clean}, while the \emph{filled circles} are the predictions from the \parsec\ code,
colour-coded by stellar mass.}
\label{fig: parsec-iso}
\end{figure}

\begin{table*}
\centering
\renewcommand{\arraystretch}{1.8}
\tabcolsep=8pt
\footnotesize
\caption{Main assumptions.}
\begin{tabular}{c|cccccr}
\hline\hline             
\multicolumn{1}{c}{Cluster ID} &
\multicolumn{1}{c}{Distance} &
\multicolumn{1}{c}{$\eta_{\rm R}$} &
\multicolumn{1}{c}{Age} &
\multicolumn{1}{c}{$A_{\rm v}$} &
\multicolumn{1}{c}{[M/H]} &
\multicolumn{1}{c}{($\alpha_0, \, \delta_0$)} \\
\multicolumn{1}{c}{} &
\multicolumn{1}{c}{[kpc]} & 
\multicolumn{1}{c}{} & 
\multicolumn{1}{c}{[Gyr]} & 
\multicolumn{1}{c}{} & 
\multicolumn{1}{c}{} &
\multicolumn{1}{c}{[deg]} \\ 
\hline
NGC~3201 & $4.74$ & $0.492$ & $10.40$ & $0.8215$ & $-1.02$ & ($154\fdg40346$, $-46\fdg41249$) \\
\hline
NGC~6397 & $2.48$ & $0.468$ & $12.87$ & $0.558$ & $-1.54$ & ($265\fdg17540$, $-53\fdg67441$) \\
\hline
\end{tabular}
\parbox{\hsize}{\textit{Notes}: Columns are: 
(1) Cluster ID; 
(2) Distance to the Sun, in kpc (\protect\citealt{Baumgardt&Vasiliev21});  
(3): Reimers scaling factor (\protect\citealt{McDonald&Zijlstra15}); 
(4) Age, in Gyr (\protect\citealt{MarinFranch+09}); 
(5) Total extinction, considering $R_{\rm v} = 3.1$ (\protect\citealt{Harris96,Harris10} and \protect\citealt{VandenBerg+13} for NGC~3201 and just \protect\citealt{Harris96,Harris10} for NGC~6397); 
(6) Metallicity, in log solar units (\protect\citealt{MarinFranch+09});
(7) Cluster centre, in degrees (calculated with \balrogo, \protect\citealt{Vitral21}).
For the total extinction of NGC~3201, we selected values between the different ones provided in the literature, so that we had a better adjustment of \parsec\ isochrones, as discussed in Section~\ref{sec: data-overview}. 
}
\label{tab: assumptions}
\end{table*}


In order to handle mass-magnitude conversions, as well as to relate the magnitude systems from \gaia\ EDR3 and \hst, we used \parsec\ isochrones\footnote{\url{http://stev.oapd.inaf.it/cgi-bin/cmd}} (e.g., \citealt{Bressan+12,Chen+14,Chen+15,Marigo+17,Pastorelli+19}). The input parameters we used and their references are displayed in Table~\ref{tab: assumptions} (along with a few other assumptions from our modelling). 

For the value of total extinction of NGC~3201, we used a value between the ones presented in \cite{Harris10} and \cite{VandenBerg+13}, which yielded a better fit to our data (as well as to the combination of other parameters). Figure~\ref{fig: parsec-iso} displays the isochrones of the two analysed clusters, with respect to the cleaned \hst\ data (according to Section~\ref{sec: data-clean}).

\section{Data cleaning} \label{sec: data-clean}

\subsection{Maximum projected radius} \label{ssec: max-r}

Passages close to the Milky Way's disk, as well as possible amounts of dark matter in its outskirts can provide an important source of dynamical heating to the outer regions of GCs. As our modelling does not include dark matter components, neither encompasses the influence of the Milky Way tidal field, we chose to analyse our data up to a maximum allowed radius, where we expect such effects to be negligible.

The mean plane-of-sky velocities in the frame aligned with the position of the star, $\overline {v_\posr}$ and $\overline{v_\post}$, are equal up to $2\,R_{\rm e}$, where  $R_{\rm e}$ is the \emph{effective radius} containing half the projected number of stars. The mean velocity profiles diverge further out, for both NGC~6397 (see VM21, fig. 6) and NGC~3201. Also, for both NGC~6397 (fig. 6 of VM21) and NGC~3201, the velocity dispersion profiles $\sigma_\posr$ and $\sigma_\post$ decrease up to $5\,R_{\rm e}$ and increase further out.
We therefore set the 
maximum allowed projected radius as $2\,R_{\rm e}$.

\subsection{Quality indicators}

The first step in our data cleaning was to remove stars with poor photometric and astrometric measurements. We detail below this cleaning procedure for both \gaia\ EDR3 and \hst.

\subsubsection{\gaia\ EDR3}

We retained \gaia\ stars that satisfied:
\begin{itemize}
\itemsep 0.5\baselineskip
    \item {\bf Astrometric accuracy}: ${\tt RUWE} < \eta_{90}({\tt RUWE})$, where {\tt RUWE} is \gaia's Renormalised Unit Weight Error and   $\eta_{n}(x)$ is the $n$-th percentile of the $x$ data
    ($\eta_{90}({\tt RUWE}) \simeq 1.1$, close to the threshold of 1.15 chosen by \citealt{Vasiliev&Baumgardt&Baumgardt21}).
    \item {\bf Photometric accuracy}: 
    \begin{equation}
            C(r) - f(G_{\mathrm{BP}} - G_{\mathrm{RP}}) < 3\,\sigma_C \ ,
    \label{eq: Gaia_flag2}
    \end{equation}
 where
    $C(r)$ is \gaia's 
     \texttt{phot\_bp\_rp\_excess\_factor},  $f(x) = \sum_i a_i \, x^i$,  with the polynomial coefficients $a_i$ taken from Table~2 of \cite{Riello+21}, and 
     $G_\mathrm{BP} - G_\mathrm{RP}$ is given by {\tt bp\_rp}.
\end{itemize}
%
Eq.~(\ref{eq: Gaia_flag2}) performs an additional filter for unreliable astrometric solutions (mainly in the cases of blended stars), affecting mainly faint sources in crowded areas. 

\subsubsection{\hst}\label{ssec: hst}


Our \hst\ astro-photometric catalogues include several diagnostic parameters to select trustworthy objects for the analysis. For each cluster, a sample of well-measured objects in the \hst\ data is defined with the following criteria 
(similarly to \citealt{Libralato+19}, but with some small changes labelled as `new' or `changed').
\begin{itemize}
    \item the star is unsaturated;
    \item the number of single exposures used to compute the magnitude of a star in the second-pass-photometry stage differs by less than 15  per cent from the number of images in which a star was actually found (new);
    \item 
    the star flux is greater than the flux from neighbours within the PSF fitting radius of the star.
    \item the photometric rms uncertainty is lower than 0.2 mag (changed);
    \item the quality, {\tt QFIT}
    of the PSF fit is greater than 0.8 (changed);
    \item the absolute value of the shape parameter \texttt{RADXS} \citep{Bedin+2008} is lower than 0.15 (changed). 
    The \texttt{RADXS} parameter represents the excess/deficiency of flux outside of the fitting radius with respect to the PSF prediction and helps discerning stars from other objects like galaxies or cosmic rays;
    \item all photometry-based selections above must be fulfilled in both (ACS/WFC) F606W and F814W filters;
    \item the acceptance rate in the proper-motion fit (number of measurements used to compute the proper motion of a star before and after all outlier rejections; see \citealt{Bellini+14}) is greater than 85\,per cent (changed);
    \item an a~posteriori correction was applied to the proper motion to account for spatial and magnitude-dependent systematics (see Sect.~\ref{sssec: hst} and Libralato et al., in prep.). 
\end{itemize}



\subsection{Proper motion error threshold} \label{ssec: pm-ilop}

Even though our mass-modelling routine takes into account the distribution of errors by convolving it with the local velocity distribution, it is wise to limit the tracers to a maximum error threshold, for robustness.
Indeed, if the proper motion errors are underestimated, which can be the case for stars with very high errors, an artificial increase of the velocity dispersion (and thus, of the mass) can take place.
We therefore removed stars with error greater than or equal to a constant times the local (mass and position) velocity dispersion of stars of given mass. The local velocity dispersion (for each star) was computed empirically, without relying on a particular model, according to the following steps:

\begin{enumerate}
    \item We converted F606W magnitudes
    ($G$ magnitudes
    for \gaia) into mass by interpolating the magnitude with respect to the respective \parsec\ isochrone of the cluster, disregarding the horizontal branch, which could cause a degeneracy in the interpolation and would be eventually removed further on. 
    \item We first selected the 100 closest stars in both $\log{M_{\star}}$ 
    and $\log{R}$, where $R$ is the projected distance to the cluster centre (hereatfer \emph{projected radius}).
The distance $\xi$ in the $\log{M_{\star}}$ vs.  $\log{R}$ plane was calculated as:
    \begin{equation} \label{eq: new-dist-final}
        \xi = \sqrt{\left(\Delta x_{\rm new}\right)^2 + \left(\Delta y_{\rm new}\right)^2} \ ,
    \end{equation}
    with $x \equiv \log{M_{\star}}$, $y \equiv \log{R}$, and
    \begin{subequations}
    \begin{flalign}
        \Delta x_{\rm new} &= \Delta x / [\eta_{84}(x) - \eta_{16}(x)]  \ ,\\
        \Delta y_{\rm new} &= \Delta y / [\eta_{84}(y) - \eta_{16}(y)] \ ,
    \end{flalign}
       \label{eq: new-dist-first}
    \end{subequations}
\noindent$\!\!$where, once again, $\eta_{n}(x)$ designates the $n$-th percentile of the variable $x$.
    \item We computed the velocity dispersion of this subset according to:
    \begin{equation} \label{eq: disp}
        \sigma_{\mu} = \sqrt{\sigma^2_{\rm POSr} + \sigma^2_{\rm POSt}} \ ,
    \end{equation}
    where POSr and POSt stand for \textit{plane of sky radial (tangential)} directions, respectively. Moreover, the proper motion in the radial direction is corrected for perspective rotation (causing apparent expansion) according to eq.~(4) of \cite{Bianchini+18}, by using the line-of-sight velocity displayed on the website of H.~Baumgardt\footnote{\url{https://people.smp.uq.edu.au/HolgerBaumgardt/globular/}, \copyright\ H.~Baumgardt, A.~Sollima, M.~Hilker, A.~Bellini \& E.~Vasiliev. \label{fn: Baumgardt}}
\end{enumerate}

\nocite{Baumgardt17}
\nocite{Baumgardt+20}
\nocite{Sollima&Baumgardt17}
\nocite{Baumgardt+19}

We finally applied $\epsilon_{\mu} < \sigma_{\mu}$ (we test this assumption in Section~\ref{ssec: error-limit}), where the proper motion error $\epsilon_{\mu}$ was calculated according to eq.~B2 from \cite{Lindegren+18}:
\begin{subequations} \label{eq: err_lind18}
\begin{flalign}
    \epsilon_\mu &= \sqrt{\frac{1}{2}(C_{33}+C_{44}) + \frac{1}{2}\sqrt{(C_{44}-C_{33})^2 + 4 C_{34}^2}} \ , \\
    C_{33} &= \epsilon_{\mu_{\alpha*}}^2 \ ,\\
    C_{34} &= \epsilon_{\mu_{\alpha*}}\, \epsilon_{\mu_\delta}\,\rho \ ,\\
    C_{44} &= \epsilon_{\mu_\delta}^2 \ ,
\end{flalign}
\end{subequations}
where $\rho$ is the correlation coefficient between $\mu_{\alpha*}$\footnote{We use the standard notation $\mu_{\alpha*} = \cos \delta\,{\rm d}\alpha/{\rm d}t$,
$\mu_\delta={\rm d}\delta/{\rm d}t$.} and $\mu_\delta$. Notice that $\rho$ is zero for \hst\ stars since $\mu_{\alpha*}$ and $\mu_\delta$ were independently calculated for this catalogue.

The steps above were repeated iteratively for each star, until no more star was discarded from the cluster. This ensures that at least the most discordant stars will be removed, so that they will not affect the dispersion of their network. The procedure usually consisted of $5-10$ iterations.

\subsubsection{Caveats}

During the procedure described above, it was possible (mostly for \gaia\ data) that a strong amount of Milky Way interlopers could bias the dispersion measurements. That is why, at each iteration, when picking the 100 closest stars in order to compute the velocity dispersion, we considered only a naively filtered subset, with less interlopers. We did this by first selecting stars whose errors were smaller than the previously computed $\sigma_{\mu}$ and whose proper motion moduli\footnote{We define the proper motion modulus as in eq.~(19) of \cite{Vitral&Mamon21}.} 
were smaller than five times the cluster velocity dispersion fitted jointly with the Milky Way contaminants by \balrogo\
\citep{Vitral21}.\footnote{\url{https://gitlab.com/eduardo-vitral/balrogo}}

\subsection{Proper motion interloper filtering} \label{ssec: pm-ilop-cut}

\mpo\ can handle the presence of interlopers in proper motion space. However, we have noticed that the best-fit \mpo\ parameters that linked to the visible components appear more physically realistic when the interloper fraction is much less than one-half.

This model assigns a fat-tailed Pearson~VII \citep{Pearson16} distribution to the Milky Way contaminants (as discovered by VM21), and a Gaussian to the cluster members, which allows us to compute membership probabilities to each star.\footnote{Following \cite{Vitral21} we allow asymmetric Pearson VII profiles.}
We then filter out stars whose membership probabilities are smaller than $90$  per cent. We provide the plot of stars having passed that test for both \gaia\ EDR3 and \hst\ as online material (Figure~\ref{fig: pm-cut}).

\subsubsection{\hst\ bulk proper motion} \label{sssec: hst-bulk-pm}


In contrast  with the \gaia\ data, the original \hst\ PMs are relative to the bulk motion of the clusters and do not provide information about the absolute motions of stars on the sky. Before applying our mixture model mentioned above, we registered the \hst\ relative PMs onto an absolute reference frame by cross matching well-measured \gaia\ and \hst\ stars and computing the proper motion offset between them (we test this assumption in Section~\ref{ssec: bulk-pm-robust}).

Well-measured stars in the \gaia\ catalogue were defined as those with ${\tt RUWE} < 1.3$ and $\epsilon_{\mu} < 0.1$ $\masyr$. For \hst\ data, we selected only unsaturated stars with $\texttt{QFIT}>0.99$ and magnitude rms lower than 0.1 mag in both F606W and F814W data. We refined this sample by including only objects whose proper motions have a rejection rate lower than 20 per cent, $\chi^2_x < 2$ and $\chi^2_y < 2$, and error lower than $0.1 \,\masyr$. Some of these quality selections are less severe than those described in Sect.~\ref{ssec: hst} and represent a good compromise between the need of a statistically-significant sample of objects to compute the offset, and the rejection of poorly-measured stars in both catalogues. Our final estimates had separations, with respect to those computed in \cite{Vasiliev&Baumgardt&Baumgardt21}, of the order of the \gaia\ systematics (i.e., $\sim 0.025\, \masyr$). 


\subsection{Colour-magnitude filtering}

Filtering the colour-magnitude diagram (CMD) not only removes field stars whose PMs coincide by chance with those of GC stars, but also removes GC members that are unresolved binary stars and lie in the edges of the stellar Main Sequence, as well as particular Blue Stragglers, which are believed to be associated with GC mergers and binaries (\citealt{Leonard89,Davies15}).
Removing binaries and stars that have gone through mergers is wise because their kinematics could be dominated by two-body interactions (i.e., their motions might be more affected by the companion or by previous encounters than by the cluster's potential), while our modelling (Sect.~\ref{ssec: mpo}) assumes that stellar motions are dominated by the global gravitational potential of the GC. 

Mass-orbit modelling of line-of-sight data is biased by the presence of binaries whose velocities are more affected by their mutual interaction than by the gravitational potenetial of the GC (e.g., \citealt{Rastello+20}).
In contrast, mass-orbit modelling based on PM-based fits are not very affected by them \citep{Bianchini+16}.
Our study is entirely based on PMs, and therefore, the influence of those binaries should be almost negligible, but we still filter them, for the reasons mentioned above.

We filter these outliers according to the \balrogo\ Kernel Density Estimation (KDE) confidence limits explained in \cite{Vitral21}, by keeping stars inside a $2-\sigma$ confidence contour. The KDE bandwidth was set as half the one derived by the Silverman rule \citep{Silverman86}, to access a better resolution.

\subsection{Data stitching}

In order to stitch our cleaned \gaia\ EDR3 and \hst\ data sets, we performed a similar approach to the one from VM21: First, we removed \gaia\ stars which had an angular separation in the sky from \hst\ stars smaller than one arcsec. Second, we selected only \gaia\ stars with magnitudes within the range of magnitudes of our cleaned \hst\ data. 

The conversion of \gaia\ $G$ magnitude into the \hst\ F606W filter, for comparison purposes, was done by interpolating the output of \parsec\ isochrones for a same cluster, with different filters, in a similar fashion as what is described in Section~\ref{ssec: pm-ilop} for the mass-magnitude interpolation. 

The final cleaned data sets from NGC~3201 and NGC~6397 contained $7422$ stars ($1064$ from the \gaia\ EDR3 catalogue, and $6358$ from \hst) and $12271$ stars ($5772$ from the \gaia\ EDR3 catalogue, and $6499$ from \hst), respectively. We provide, as online material, the plot of the the distribution of \gaia\ and \hst\ stars in the sky (i.e., in $\alpha \times \delta$ coordinates).

\section{Methods} \label{sec: methods}

\subsection{Mass modelling}
\label{ssec: mpo}
\subsubsection{General formalism} \label{sssec: mpopm-formalism}

We perform our mass modelling with the Bayesian code \mpo \ (Mamon \& Vitral in prep.), which is an extension of {\sc MAMPOSSt} \citep{Mamon+13} to handle PMs in addition to line-of-sight velocities. \mpo\ is briefly described in sect.~2 of  VM21, and was tested by \cite{Read+21}, who showed that \mpo\ reproduced well the radial profiles of mass density and velocity anisotropy of mock dwarf spheroidal galaxies. 

\mpo\ fits models for the radial profiles of total mass and the velocity anisotropy of the visible stars to the distribution of these stars in projected phase space. 
The velocity anisotropy (`anisotropy' for short) is defined as in \citep{Binney80}:
\begin{equation}    \label{eq: anisotropy}
    \beta(r) = 1 - \displaystyle{\frac{\sigma_{\theta}^{2}(r) + \sigma_{\phi}^{2}(r)}{2 \,\sigma_{r}^{2}(r)}} \ ,
\end{equation}
where $\theta$ and $\phi$ are the tangential components of the coordinate system, while $\sigma_{i}^2$ stands for the velocity dispersion of the component $i$ of the coordinate system. In spherical symmetry, $\sigma_\phi = \sigma_\theta$.
In these fits, \mpo\ assumes  that the \emph{local} velocity ellipsoid is an anisotropic Gaussian, whose major axis is aligned with the spherical coordinates.

The expression of the local velocity distribution function involves the observable components of the velocity dispersion. In the sky frame aligned with the position of a given star, one has \citep{Strigari+07}
\begin{subequations}
\begin{flalign}
\sigma_{\rm POSr}^2(R\,|\,r) &= \left[1-\beta (r) + \beta(r)\,\frac{R^2}{r^2}\right]\,\sigma_r^2(r) \ ,\\
\sigma_{\rm POSt}^2(R\,|\,r) &= \left[1-\beta(r)\right]\,\sigma_r^2(r) \ .
\end{flalign}
\end{subequations}
If  line-of-sight velocities are available:
\begin{equation}
   \sigma_{\rm LOS}^2(R\,|\,r) = \left[1-\beta (r) \,\frac{R^2}{r^2}\right]\,\sigma_r^2(r) 
\end{equation}
\citep{Binney&Mamon82}.
The required radial squared velocity dispersion $\sigma_r^2(r)$ is obtained by solving  
the spherical Jeans equation with no streaming motions
\citep{Binney80}    
\begin{equation} 
  \frac{{\rm d}\left (\rho\sigma_r^2\right)}{{\rm d}r} + 2\,\frac{\beta(r)}{
    r}\,\rho(r)\sigma_r^2(r) = -\rho(r) \frac{G\,M(r)}{r^2} \ ,
\label{eq: jeans}
\end{equation}
assuming a given mass profile $M(r)$ and anisotropy profile $\beta(r)$, for a previously determined mass density profile 
$\rho(r)$ for the kinematic tracers (here stars). The term $\rho\,\sigma_r^2$ is the dynamical pressure that counteracts gravity.\footnote{In fact, the Jeans equation~(\ref{eq: jeans}) is a consequence of the Collisionless Boltzmann Equation, which considers the incompressibility in phase space of the six-dimensional (6D) distribution function (DF). Expressing the DF  in terms of 6D number, mass or luminosity density, implies that the term $\rho$ in the Jeans equation is the number, mass or luminosity density. For the present case of a globular cluster made of stars, it makes more physical sense to reason with mass density. In the absence of mass segregation, the mass density is proportional to the number density, so the mass density profile is obtained from deprojecting the observed surface number density profile.}
We discuss our choices for $M(r)$ and $\beta(r)$ in Sect.~\ref{sssec: massbeta}, and our estimate of $\rho(r)$ in Sect.~\ref{sssec: surfdens}.

\subsubsection{Mass and velocity anisotropy profiles}
\label{sssec: massbeta}

In \mpo, the mass profile can be the sum of several components.
Here, the GCs always include a component for the Main Sequence stars, assuming that the mass density profile of this component follows the number density profile, i.e. that the mean stellar mass is independent of radius. 
However, Main Sequence stars of different mass should follow different distributions in projected phase space, because of exchanges of energy between stars of different masses by two-body relaxation, and also because stars of different masses are expelled differently in three-body encounters with hard binaries. This \emph{mass segregation} is difficult to quantify, because the selection effects from confusion vary with stellar magnitude and local surface density in a complicated way.
We shall re-discuss mass segregation in Sect.~\ref{sec: results}.

We consider four different mass scenarios: 1) no central dark component (i.e. just the Main Sequence stars); 2) central IMBH; 3)  cluster of unseen objects (CUO), and 4) central IMBH plus a CUO.

While GCs are often modelled with isotropic velocities, isolated  core-collapsed GCs should have radial anisotropy in their envelopes \citep{Takahashi95,Tiongco+16,Zocchi+16,Bianchini+17}. 
We  therefore performed many runs of \mpo\ with freedom in the anisotropy profile. 

The anisotropic runs of \mpo\ used the generalisation (hereafter gOM) of the Osipkov-Merritt model \citep{Osipkov79, Merritt85} for the velocity anisotropy profile:
\begin{equation}    \label{eq: gOM}
    \beta_{\mathrm{gOM}}(r) = \beta_{0} + (\beta_{\infty} - \beta_{0}) \ \displaystyle{\frac{r^2}{r^2 + r_{\beta}^2}} \ ,
\end{equation}
where $r_{\beta}$ is the anisotropy radius, which can be fixed to the scale radius of the luminous tracer by \mpo.\footnote{\cite{Mamon+19} and \cite{Vitral&Mamon21} found no significant change in models of galaxy clusters and globular clusters, respectively, when using this model for $\beta(r)$ compared to one with a softer transition: $\beta(r) = \beta_{0} + (\beta_{\infty} - \beta_{0})\,r/(r+r_{\beta})$, first used by \cite{Tiret+07}.} 

\subsubsection{Likelihood}
In \mpo, the likelihood is written
\begin{equation}
    {\cal L} = \prod_i p({\bf v}_\mathbf{i}\,|R_i) \ ,
\end{equation}
where the conditional probability of measuring a velocity ${\bf v}_\mathbf{i}$ is the mean of the local  velocity distribution function, $h({\bf v}\,|\,R,r)$, integrated along the line of sight\footnote{While the mass density $\rho$ enters the Jeans~equation~(\ref{eq: jeans}), it is the number density $\nu$ that enters the local velocity distribution function.}
\begin{equation}
    p({\bf v}\,|\,R) = \frac{2}{\Sigma(R)}\, \int_R^\infty h({\bf v}\,|\,R,r)\,\nu(r) \,\frac{r}{\sqrt{r^2-R^2}}\,{\rm d}r \ .
    \label{pvofR}
\end{equation}
\mpo\ determines the marginal distributions of the free parameters and their covariances by running the Markov Chain Monte Carlo (MCMC) routine ({\sc CosmoMC}\footnote{\url{https://cosmologist.info/cosmomc/}.}, \citealt{Lewis&Bridle02}).  
We generally use flat priors on the mass and anisotropy parameters, and Gaussian priors on the pre-determined surface density profile parameters (Sect.~\ref{sssec: surfdens}) and on the bulk motions (Sect.~\ref{sssec: bulkprior}).

\subsubsection{Priors on the surface density}
\label{sssec: surfdens}

In the absence of mass segregation, the tracer mass density profile, $\rho(r)$ is  proportional to the number density profile $\nu(r)$, which
is deprojected from the surface mass density profile, $\Sigma(R)$, using spherical symmetry.
But the surface distribution of kinematic tracers is usually incomplete, in particular in the inner regions of the GCs. We therefore estimate the surface density profile in two steps.
\begin{enumerate}
    \item First, we use Gaussian priors based on MCMC fits of the S\'ersic  \citep{Sersic63,Sersic68} model surface density profile plus a uniform field surface density to the distribution of projected radii. For this, we do not restrict our sample to the stars whose kinematics we later analyse with \mpo, because we noticed in VM21 that kinematic data are increasingly incomplete towards the GC centre. We therefore consider all stars in a magnitude range, but we also emphasise that spatial incompleteness can still strongly affect our results in the cluster's centre due to crowdness \citep{Arenou+18}, especially for NGC~6397, which is much denser. We try several maximum allowed projected radii, $R_{\rm max}^{\rm allow}$ and adopt the MCMC means for the log scale radius and S\'ersic index  in the middle of the plateau of $R_{\rm max}^{\rm allow}$ where the means are roughly constant (see left panel of fig.~10 of VM21).
    \item Since \mpo\ takes into account not only the distribution of projected radii, but also kinematic data, a second pass that uses it should add more constraints on our fits (and future priors, consequently), and decrease the spatial incompleteness intrinsic to our \gaia\ and \hst\ data (see previous step). Hence, we run \mpo\ with these means for the particular case of no central mass and isotropic velocities\footnote{We also tested other mass-anisotropy models to check if the derived priors were too different, which was not the case.}, using Gaussian priors with fairly wide uncertainties (0.1 dex in effective radius and 0.5 in S\'ersic index), centred on the means of the S\'ersic fits of the first step. 
\end{enumerate}
For each mass model, we then run \mpo\ with the mean and uncertainties on surface density parameters returned by \mpo\ in that first run.
The data used for the first step combines the \hst\ and \gaia\ data, stitched together in a given magnitude range: we used stars fainter than the brightest \hst\ star, while brighter than the 95th percentile of magnitudes associated with stars whose proper motion error was below a fixed threshold, set as the maximum proper motion error of our cleaned sample.

\subsubsection{Bulk motion and distance priors}
\label{sssec: bulkprior}

We set Gaussian priors for the bulk proper motion of our data, centred on the \hst\ offset discussed in Section~\ref{sssec: hst-bulk-pm}, with a width equivalent to the uncertainty of the offset calculation.
The current version of \mpo\ does not allow for a kinematics distance fit. We therefore fix the distance of each GC to those derived by \cite{Baumgardt&Vasiliev21}.

\subsubsection{Marginal distributions and covariances}

We explored the parameter space to determine marginal distributions and parameter covariances as in VM21.
In particular, we used 6 MCMC chains run in parallel and stopped the exploration of parameter space after one of the chains reached a number of steps $N_{\rm steps} = 10\,000 \,N_{\rm free}$, where $N_{\rm free}$ is the number of free parameters of the model\footnote{$N_{\rm free}$ varied from 6 (isotropic model with no central component) to 11 (anisotropic model with central IMBH plus CUO).}.
We discard the 3000\,$N_{\rm free}$ first steps of each MCMC chain, which are associated with a burn-in phase.

\subsection{Mock data} \label{ssec: mock-build}

\begin{figure}
\centering
\includegraphics[width=1.0\hsize]{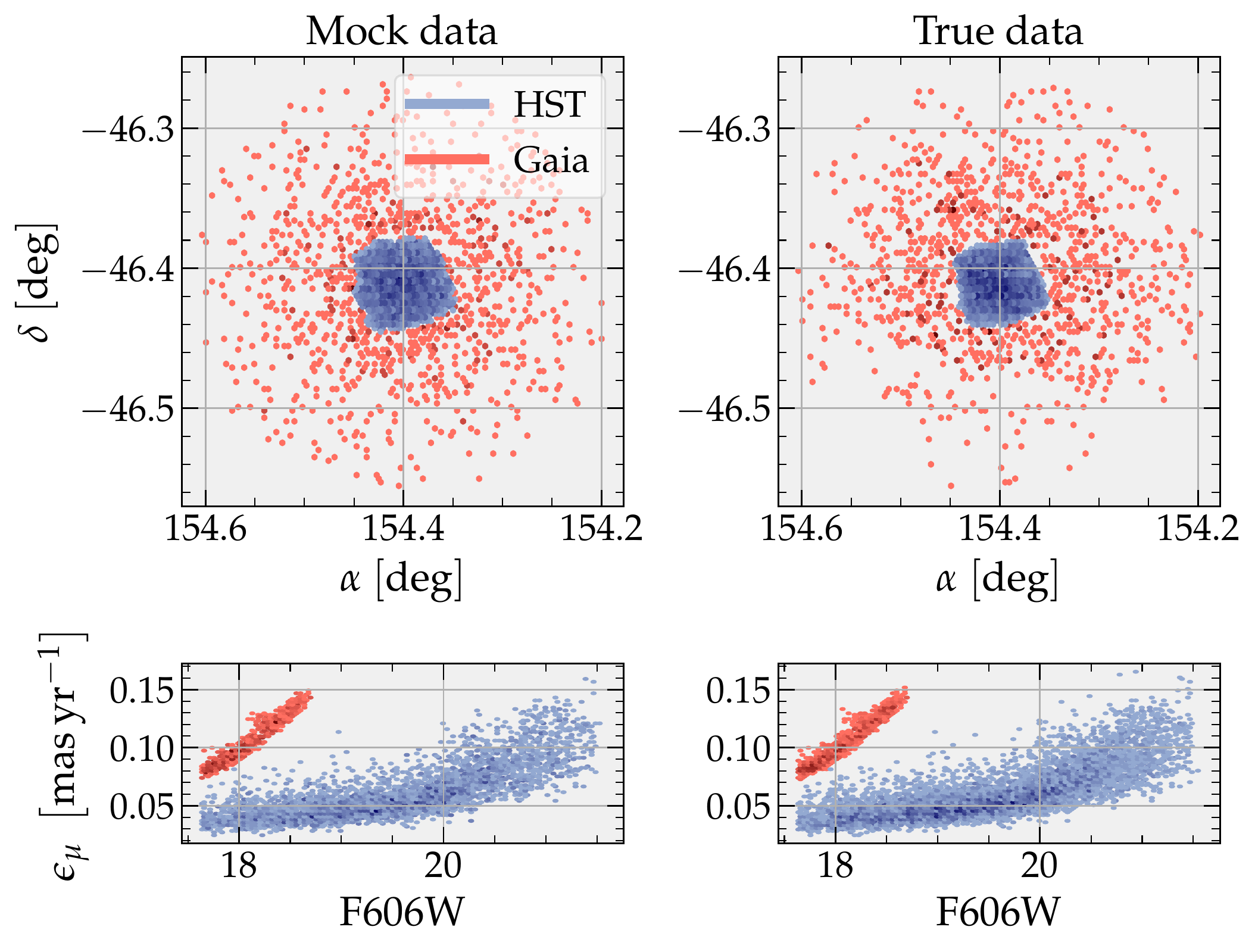}
\caption{\textit{Mocks:} Comparison of our mock data ({\it left}) and the true data set ({\it right}), for NGC~3201. Stars associated with \hst\ are in {\it blue}, while the ones associated with \gaia\ EDR3 are in {\it red}. The \emph{upper panels} show the positions on the sky, while the \emph{lower panels} show the dependence of proper motion errors with magnitude. The axis limits are the same between mock and data. The colours go from \emph{faint tones} in less populated  regions to \emph{darker tones} in  more populated regions).}
\label{fig: mock}
\end{figure}

We built mock datasets to test the ability of \mpo\ to recover central mass excesses, either point-like (IMBH) or extended (CUO).
For each GC, we generated four mocks, one with no central excess mass, one with an IMBH, one with an extended central mass, and one with both IMBH and an extended central mass.

\subsubsection{Positions and velocities in a cartesian frame}

The mocks were constructed with the \agama\ software \citep{Vasiliev19a}. For each GC, we used information obtained from our \mpo\ runs (see Sect.~\ref{sec: results} below): i.e. with the same mass profiles as determined by \mpo. 
Therefore, the GC followed a S\'ersic profile and the CUO, if present, followed a Plummer profile, while the GC Main Sequence and CUO stars had isotropic velocities.
The parameters are displayed in the online version of Table~\ref{tab: results-full}. 
The mean tracer mass of each population (i.e., GC and CUO) is required by \agama\ for the construction of the mock. 
\begin{itemize}
    \item For GC tracers, we estimated the mean mass assuming a power-law mass function (MF), $\d N/\d M$:
    \begin{equation}
        \bar{m}_{\star} = \frac{\int_{m_{\rm min}}^{m_{\rm max}} m \, [\d N /\d m] \d m}{\int_{m_{\rm min}}^{m_{\rm max}} [\d N /\d m] \d m} = \frac{m_{\rm max}^{\alpha+2} - m_{\rm min}^{\alpha+2}}{m_{\rm max}^{\alpha+1} - m_{\rm min}^{\alpha+1}}
    \end{equation}
    where
   we used the MF slopes $\alpha$ available at the website of H. Baumgardt, see footnote \ref{fn: Baumgardt}), while $m_{\rm min}$ and $m_{\rm max}$ are the minimum and maximum stellar masses of our data, derived from \parsec\ isochrones (see Fig.~\ref{fig: parsec-iso}).
    \item For CUO tracers, we used the mean mass of the compact objects from the Monte Carlo models described in Section~\ref{ssec: cmc}, up to twice the 3D half mass radius we derived from the real data with \mpo\ (Table~\ref{tab: results-full}).
\end{itemize}

\subsubsection{Sky membership}

We transformed the Cartesian coordinates into astrometric data (i.e., $\alpha, \, \delta, \, \mu_{\alpha,*}, \, \mathrm{and} \ \mu_{\delta}$) positioned similarly to the studied clusters with the routine \textsc{angle.cart\_to\_radec} from \balrogo. We used the same values of cluster centre, mean bulk motion\footnote{We considered the \hst\ bulk motion as calculated in Section~\ref{sssec: hst-bulk-pm} for \hst-like stars, and the bulk motion values from \cite{Vasiliev&Baumgardt&Baumgardt21} for the \gaia-like stars.
Having two different bulk motions may surprise the reader, and we test in Sect.~\ref{ssec: bulk-pm-robust} the robustness of our results to the choice of bulk motions.},
line-of-sight velocity and distance we considered for the true data.

Next, for each star, we assigned \gaia\ EDR3 and \hst-like memberships by mimicking the membership of the closest star in the true data set.
The top panels of Figure~\ref{fig: mock} show the mock data used for NGC~3201 next to the true data for this cluster.

\subsubsection{Proper motion errors}

We estimated proper motion errors in our mock GC stars, separately for the \hst\ and \gaia\ EDR3 data sets, according to the following steps:

\begin{enumerate}
    \item We constructed an empirical cumulative distribution function (CDF) of magnitudes (the original F606W for \hst\ data, and the converted $G_{\rm mag}$ to F606W for \gaia) for the true data, by sorting and arranging it from zero to one.
    \item We interpolated the CDF with the respective magnitudes for a uniformly distributed array (from zero to one) of length equal the mock \gaia\ EDR3 or \hst-like data set, which is greater than the length of the true data set. This creates a random distribution of magnitudes following the same shape as the true data.
    \item From those magnitudes, we associated an error (both in $\alpha$ and $\delta$) by picking the same $\epsilon_{\alpha,*}$ and $\epsilon_{\delta}$ from the star with the closest magnitude from the true data set.
   In this way, the proper motion errors follow the same trend and scatter with magnitude as the observed ones, as can be seen from the similarity of the bottom panels of Fig.~\ref{fig: mock}.
    \item Having those errors, we add them up to the proper motions by sorting random Gaussian variables with zero mean and standard deviation equal to the respective error. The original errors are saved and taken into account during the mass modelling, when convolving the velocity distribution function of the tracers.
\end{enumerate}

Finally, we randomly selected, from each data set, a number of tracers equal to the amount of respective \gaia\ and \hst\ stars from our true data set, which artificially (and intentionally) adds incompleteness to our subset.

\subsection{Statistical tools} \label{ssec: stats}

We first use Bayesian evidence to compare our four basic models for each GC: no excess inner mass, a central IMBH, a CUO, and a combination of IMBH and CUO.  This model selection
involves comparing the maximum log posteriors using a Bayesian information criteria.
We then measure how well the posterior distributions obtained by \mpo\ on the observations match those obtained on mock data constructed to mimic these observations.

\subsubsection{Bayesian inference} \label{sssec: aicc}

We use the corrected Akaike Information Criterion (derived by \citealt{Sugiyara78} and independently by \citealt{HurvichTsai89} who demonstrated its utility for a wide range of models)
\begin{equation}
      \mathrm{AICc} =\mathrm{AIC} + 2 \, \frac{N_{\mathrm{free}} \,  (1 + N_{\mathrm{free}})}{N_{\mathrm{data}} - N_{\mathrm{free}} - 1} \ ,
\end{equation}
where AIC is the original 
{Akaike Information Criterion}
\citep{akaike1973information}
\begin{equation}
 \label{eq: AIC}
    \mathrm{AIC} = - 2 \, \ln \mathcal{L_{\mathrm{MLE}}} + 2 \, N_{\mathrm{free}} \ ,
\end{equation}
and where ${\cal L}_{\rm MLE}$ is the maximum likelihood estimate found when exploring the parameter space, $N_{\rm free}$ is the number of free parameters, and $N_{\rm data}$ the number of data points. 
We prefer AICc to the other popular simple Bayesian evidence model, the Bayes Information Criterion (BIC, \citealt{Schwarz78}), because  AIC(c) is more robust for situations where the true model is not among the tested ones (for example our choice of S\'ersic density profiles is purely empirical and not theoretically motivated), in contrast with BIC
\citep{Burnham&Anderson02}.


The likelihood (given the data) of one model relative to a reference one is 
\begin{equation}
\rm \exp\left(-\frac{AIC-AIC_{\rm ref}}{2}\right)
\label{eq: pAIC}
\end{equation}
\citep{Akaike83} and we assume strong evidence for one reference model over another whenever $95$  per cent confidence is attained (i.e., $\rm AICc > AICc_{\rm ref}+6$). We consider AICc differences smaller than $4.5$ (i.e., less than $90$  per cent confidence) are usually not enough to consistently distinguish two models, based on purely statistical arguments (thus, no astrophysics involved).

\subsubsection{Distance on parameter space} \label{sssec: dist-mcmc}

To correctly compare the mass-modelling outputs of the mock data and the true, observed data, we also compute the distance on parameter space from the maximum likelihood\footnote{Since some of our priors are Gaussians, our maximum likelihood parameter vectors are really maximum posteriors, but we will refer to these as `maximum likelihood' to avoid confusion with the modes of the marginal parameter distributions.} solutions of the real data and the mock data set. For each free parameter $k$, we define the distance between the maximum likelihood solutions $\lambda_{ik}^{\rm MLE}$ and $\lambda_{jk}^{\rm MLE}$ from the set of chains from the data and mock, $\mathcal{C}_i$ and $\mathcal{C}_j$ as:

\begin{equation} 
\Delta_{ij}(k) = 
|\lambda_{ik}^{\rm MLE} - \lambda_{jk}^{\rm MLE}| \ .
\label{eq: dist-mcmc}
\end{equation}

With this information, we follow the iteration below:

\begin{enumerate}
    \item For each parameter $k$, select a random value from the chain $\mathcal{C}_i$, and another one from the chain $\mathcal{C}_j$.
    \item Evaluate if the modulus of the difference between these two values is greater than $\Delta_{ij}(k)$.
    \item Repeat it $10^6$ times.
\end{enumerate}
Then, we compute the fraction $\phi_{\Delta k}$ of times where the absolute difference between the random values from $\mathcal{C}_i$ and $\mathcal{C}_j$ is greater than $\Delta_{ij}(k)$. If this fraction is high, it means that the distance between the fits from the mock and true data is small when compared to the overall difference of MCMC chain values. On the opposite, small fractions (e.g., $\lesssim 50\%$) indicate a disagreement between the fits of mock and true data. We test this statistic for the free parameters of a dark central component fit (i.e., mass and scale radius of the dark component), in order to better evaluate its composition.

\subsubsection{AD and KS statistics} \label{sssec: ad-ks}

In some cases, the marginal distribution of the posterior might be very broad, indicating higher uncertainties for the maximum likelihood solutions. In that case, it is interesting to compare the \emph{shapes} of the marginal distributions obtained by \mpo\ on the mock and observed datasets, to probe the expected contrast between different mass models. For this purpose, we used Kolmogorov-Smirnov (\citealt{Kolmogorov1933,Smirnov1939}, hereafter KS) as well as Anderson-Darling (\citealt{Anderson&Darling52}, hereafter AD)  statistics to quantify the disagreement between mock and observed marginal distributions of mass and scale radius of a dark central component.

Since the KS and AD tests quantify whether two 1-D distributions arise form a single parent distribution, they will be sensitive to any shift between them. We adapt these statistics to compare the distribution of shapes without being sensitive to any offset.
For this, we translated (shifted) the two marginal distributions by a proxy of their respective median. In practice, we performed the following iteration three times:
\begin{enumerate}
    \item We first consider only the intersection of the two chains $\mathcal{C}_i$ and $\mathcal{C}_j$, to be compared. 
    \item We assign this intersection to an auxiliary pair of chains $\widetilde{\mathcal{C}}_i$ and $\widetilde{\mathcal{C}}_j$.
    \item We translate the distributions of $\mathcal{C}_i$ and $\mathcal{C}_j$ by a respective amount of M$\left[\widetilde{\mathcal{C}}_i\right]$ and M$\left[\widetilde{\mathcal{C}}_j\right]$, where M[$x$] is the median of a distribution $x$.
\end{enumerate}

This iteration removes undesirable effects on the borders of the distributions, where there may be  artefacts of our choice of priors. This iteration thus allows a more honest comparison between the shapes of the distributions than just a single shift by the difference of medians, because our distributions can be skewed non-Gaussians (the median does not necessarily follow the mode of a skewed distribution). We considered the intersection of the translated mock and observed marginal distributions of a specific parameter, and computed the KS and AD statistics associated with them. We remind that smaller KS and AD statistics relate to a better match of distributions.

\subsection{Monte Carlo $N$-body models} \label{ssec: cmc}

\begin{table}
\centering
\renewcommand{\arraystretch}{1.3}
\tabcolsep=9pt
\footnotesize
\caption{Number of compact objects in our \texttt{CMC} Monte Carlo $N$-body models.}
\begin{tabular}{l|rrrrr}
\hline\hline             
\multicolumn{1}{c}{ID} &
\multicolumn{1}{c}{BH} &
\multicolumn{1}{c}{NS} &
\multicolumn{1}{c}{WD} &
\multicolumn{1}{c}{WD} &
\multicolumn{1}{c}{WD} \\
\multicolumn{1}{c}{} &
\multicolumn{1}{c}{} &
\multicolumn{1}{c}{} &
\multicolumn{1}{c}{[ONeMg]} &
\multicolumn{1}{c}{[CO]} &
\multicolumn{1}{c}{[He]} \\
\multicolumn{1}{c}{(1)} &
\multicolumn{1}{c}{(2)} &
\multicolumn{1}{c}{(3)} &
\multicolumn{1}{c}{(4)} &
\multicolumn{1}{c}{(5)} & 
\multicolumn{1}{c}{(6)}  \\ 
\hline
NGC~3201 & 108\, & 334 & 1954\ \ \ \ \ \  & 78501 & 315\ \  \\
NGC~6397 & 0\, & 191 & 620\ \ \ \ \ \  & 31941 & 73\ \  \\
\hline
\end{tabular}
\parbox{\hsize}{\textit{Notes}: Columns are: (1) Cluster ID; (2) Number of black holes; (3) Number of neutron stars; (4) Number of [ONeMg] white dwarfs; (5) Number of [CO] white dwarfs; (6) Number of [He] white dwarfs.
}
\label{tab: cmc-numbers}
\end{table}

To facilitate the interpretation of our results, we use Monte Carlo $N$-body cluster models of NGC 3201 and NGC 6397, computed using the cluster dynamics code \texttt{CMC} \citep{Kremer+20,Rodriguez+21}. \texttt{CMC} is a H\'{e}non-type Monte Carlo code that includes various physical processes relevant to the dynamical evolution of clusters including two-body relaxation, tidal mass loss, and direct integration of small-$N$ resonant encounters. For strong binary-mediated encounters, \texttt{CMC} computes the energy exchange between binaries and stars directly, by performing direct $N$-body integrations using the \texttt{Fewbody} code \citep{Fregeau&Rasio07}, now updated to include post-Newtonian effects for black hole encounters \citep{Rodriguez2018}. Although individual distant encounters (with pericentre distances much larger than the characteristic hard-soft boundary) are not modelled directly, the cumulative effect of many distant encounters is computed as a single \textit{effective} encounter at each time step using the scheme described in \citet{Stod1982,Joshi2000}. This method captures the effect of distant encounters on the cluster as a whole, in particular upon the two-body relaxation process. For a detailed and current explanation of the methods implemented in \texttt{CMC}, see \citeauthor{Rodriguez+21} (\citeyear{Rodriguez+21}, specifically Section~2.1 for discussion of the treatment of weak encounters and two-body relaxation, and Section~2.2 for the treatment of strong encounters). The Monte Carlo approach employed in \cmc\ has been shown to agree well with direct $N$-body models, especially pertaining to dynamical evolution of black holes \citep[e.g.,][]{Rodriguez2016dragon}.
Finally, by employing the \texttt{COSMIC} single/binary star evolution code \citep{Breivik+20}, \texttt{CMC} tracks various evolution features (including stellar type, mass, radius, luminosity, etc.) for all $N$ stars as the model cluster evolves dynamically. This makes it straightforward to compute standard observed cluster features from the \texttt{CMC} snapshots, in particular surface brightness and velocity dispersion profiles, binary fractions, and colour-magnitude diagrams \citep[e.g.,][]{Rui+21b}.

Previous studies have identified specific  \texttt{CMC} models that match accurately both NGC 3201 and NGC 6397 (using observed surface brightness and velocity dispersion profiles as the key diagnostics to evaluate goodness of fit;
for details, see~\citealt{Rui+21b}). For NGC 3201, we use the \texttt{CMC} model presented in \cite{Kremer2019}. For NGC 6397, we use the models published in \cite{Kremer+21} and also compute a few additional models in effort to more accurately match the compact object distributions inferred from our analysis.\footnote{The \cmc\ N-body models are intended to provide a basic numerical supplement that complements the Jeans modelling constraints. We are not claiming to have performed an exhaustive match between the models and observed cluster properties, which may include other diagnostics such as mass segregation measurements \citep[e.g.,][]{Weatherford+20}, blue straggler populations, cataclysmic variables, etc. Such a comparison would be a much more intensive endeavour than is intended here.} 
In both clusters, \cmc\ starts with isotropic stellar orbits \citep[e.g., assumes standard King profiles as initial conditions;][]{King66}, and despite 3-body encounters and natal kicks, the models remain roughly isotropic over time.
In Table~\ref{tab: cmc-numbers}, we list various features of our best-fit models for both of these clusters. Figure~\ref{fig: mass-seg}, provided as online material, also argues in favour of the good agreement between our \cmc\ models and the \hst\ data we use.

\section{Results \& Robustness} \label{sec: results}

We ran a total of 48 \mpo\ fits on the \hst\ plus \gaia\ EDR3 data, using different mass models, data cuts, and prior assumptions, and we present the main outcome of those runs in the online version of 
Table~\ref{tab: results-full}.
In the following, we present 
our results on velocity anisotropy and on the excess of mass in the centre.

\subsection{Velocity anisotropy}

\begin{figure}
\centering
\includegraphics[width=0.99\hsize]{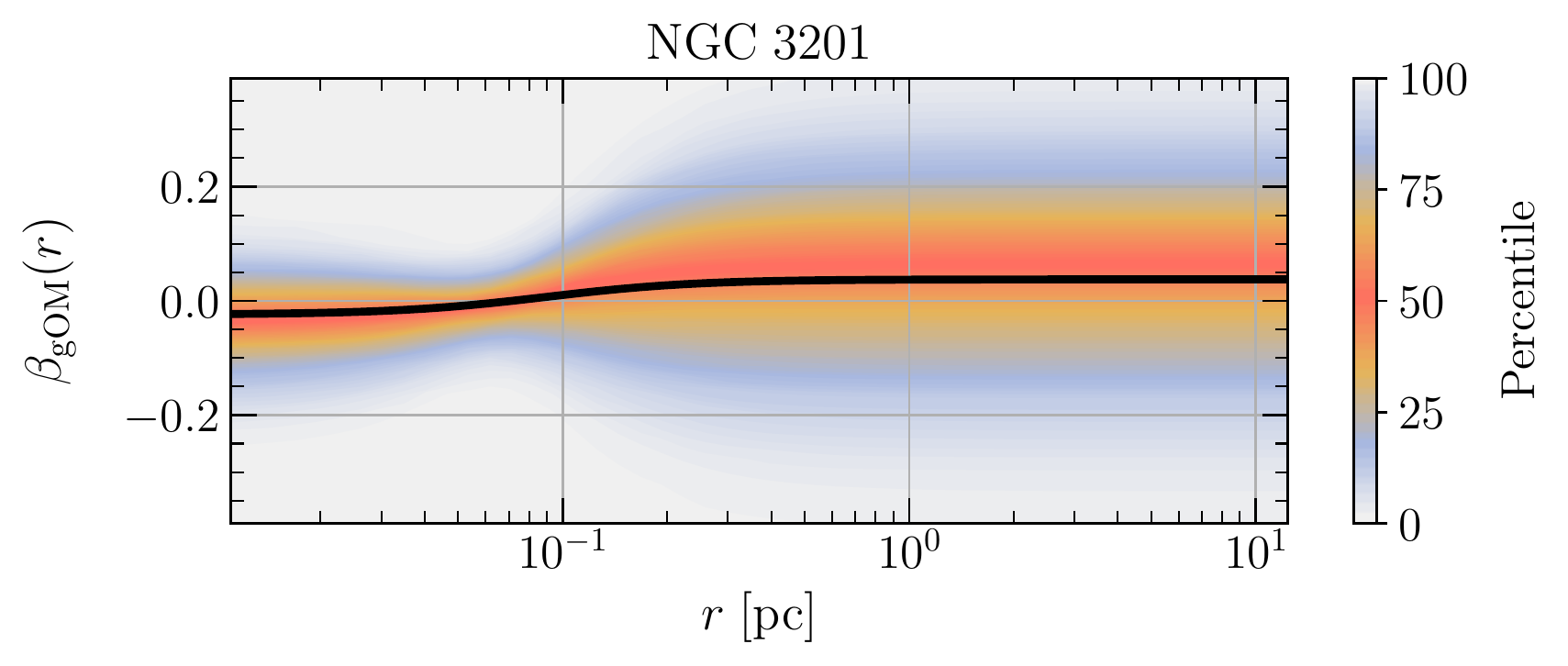} \\
\includegraphics[width=0.99\hsize]{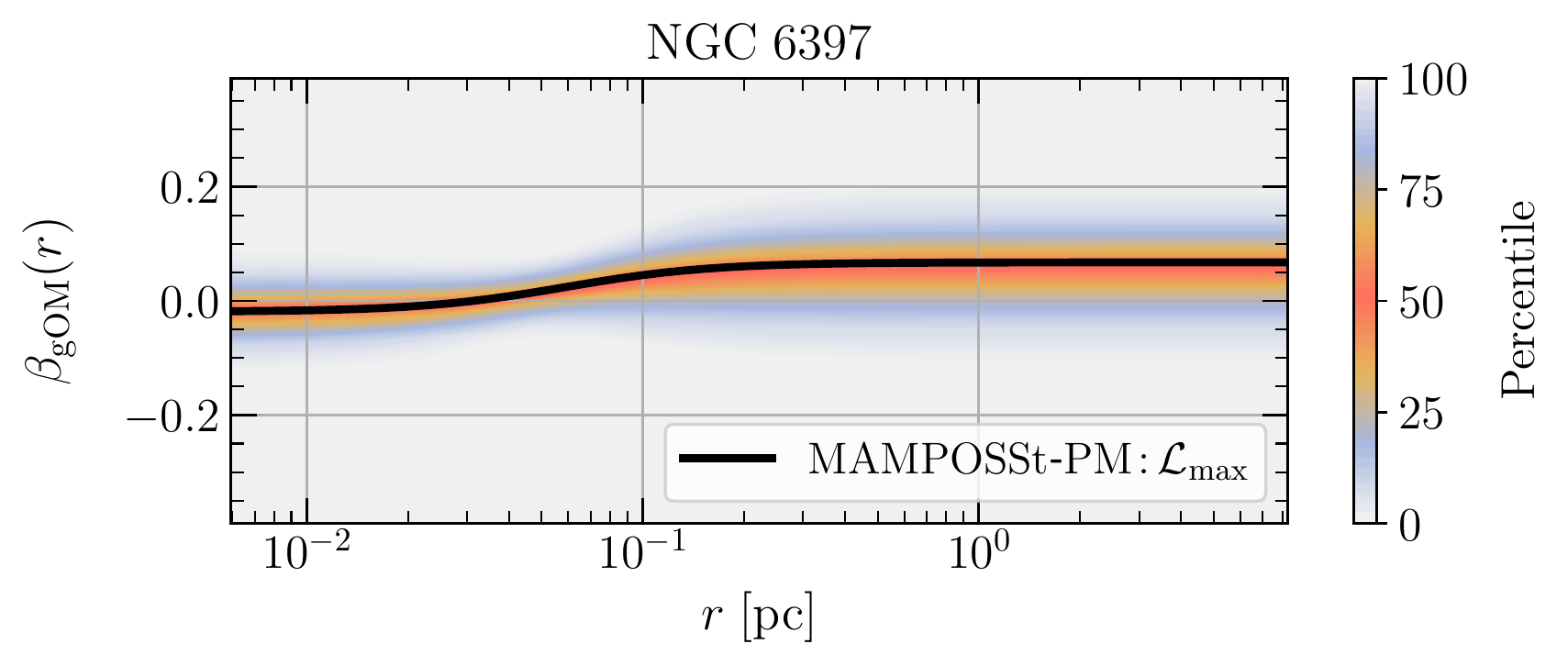}
\caption{{\it Velocity anisotropy:} \mpo\ fits of the velocity anisotropy, using the gOM parameterization (Eq.~[\ref{eq: gOM}]), as a function of the distance (in pc) to the respective cluster's centre. The colour bar indicates the percentile of the MCMC chain post burn-in phase. The \emph{black curves} represent the maximum likelihood solution of our fit. The \emph{upper} plot displays the fits for NGC~3201, while the \emph{bottom} plot presents the fits for NGC~6397.
The range of physical radii is set to the range of projected radii in the data we analysed. 
}
\label{fig: anis-mpo}
\end{figure}

\begin{figure*}
\centering
\includegraphics[width=0.12\hsize]{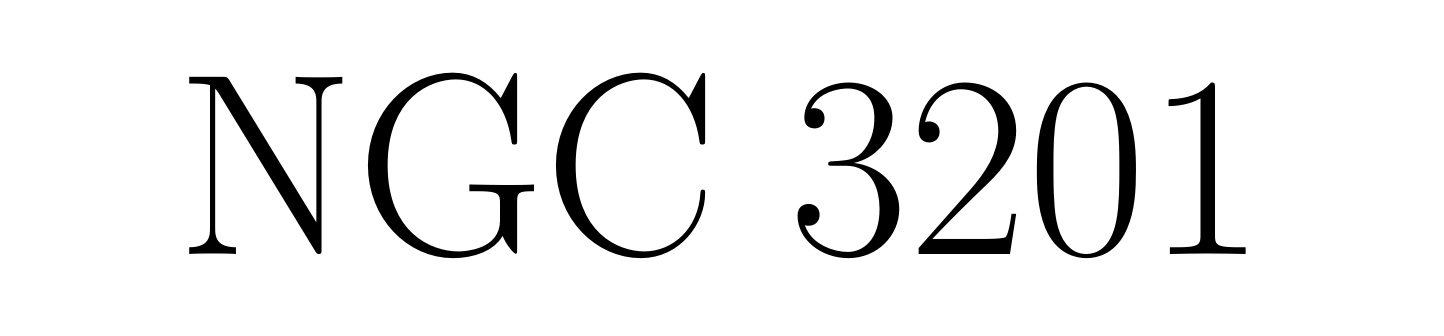} \\
\includegraphics[width=0.247\hsize]{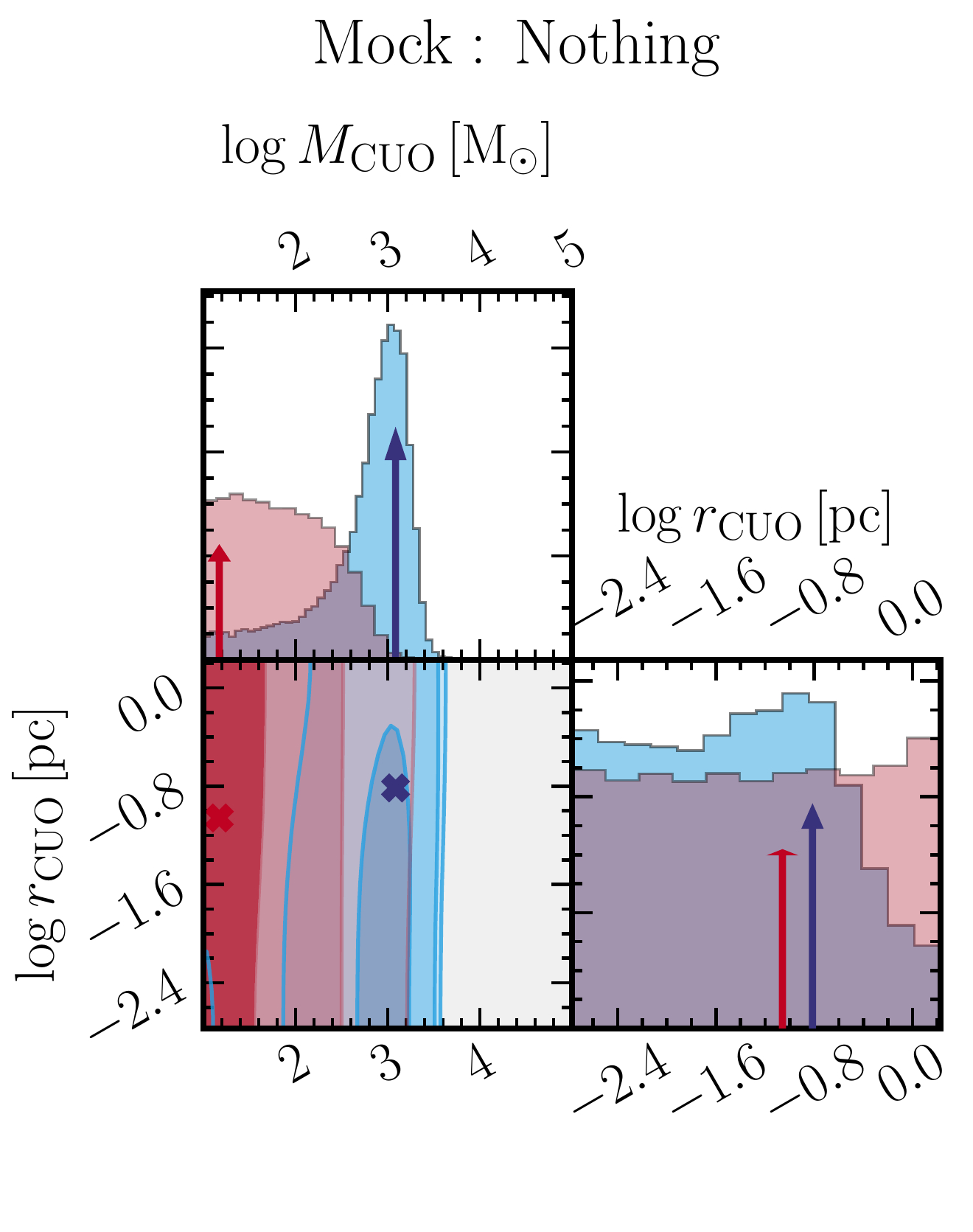}
\includegraphics[width=0.247\hsize]{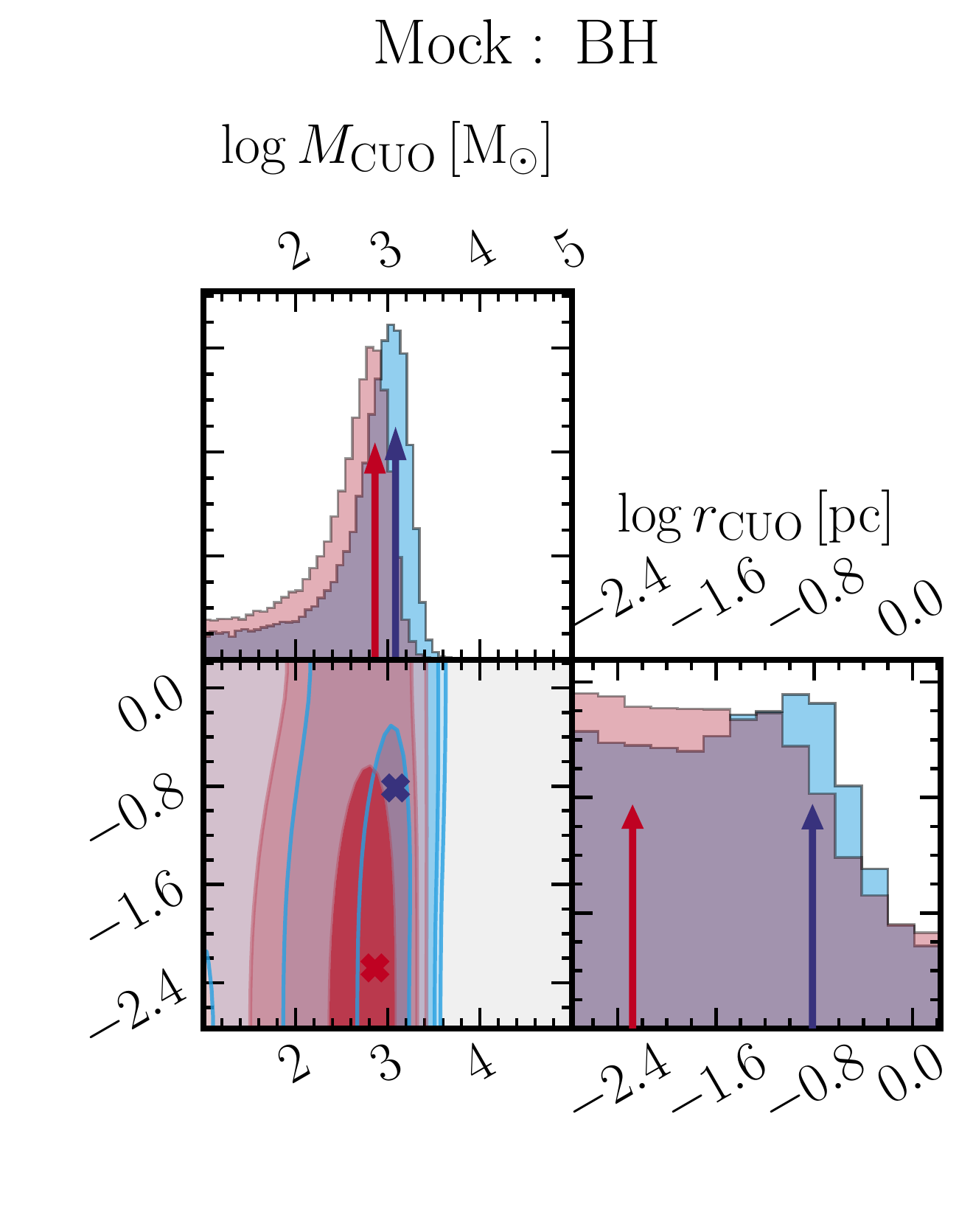}
\includegraphics[width=0.247\hsize]{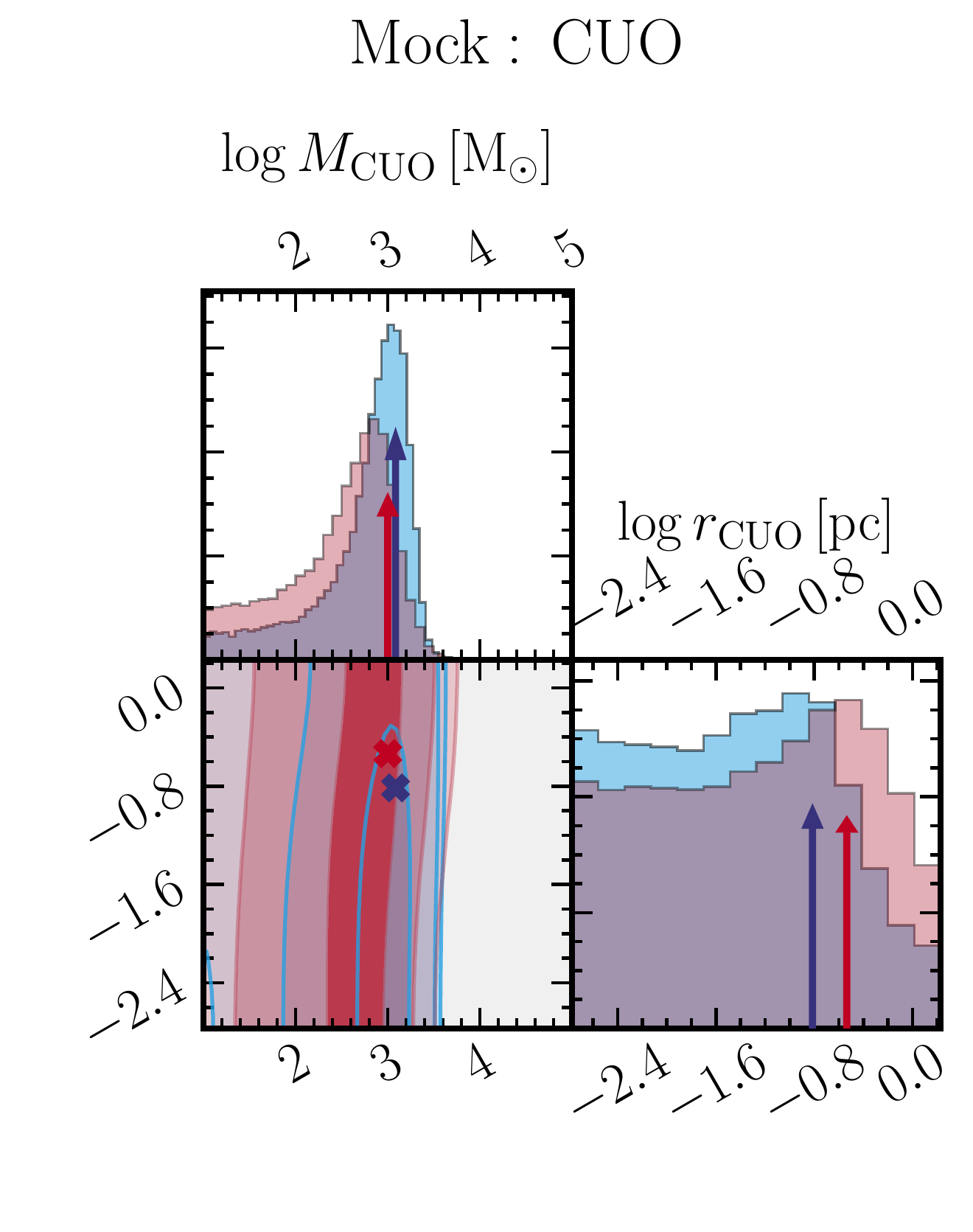}
\includegraphics[width=0.247\hsize]{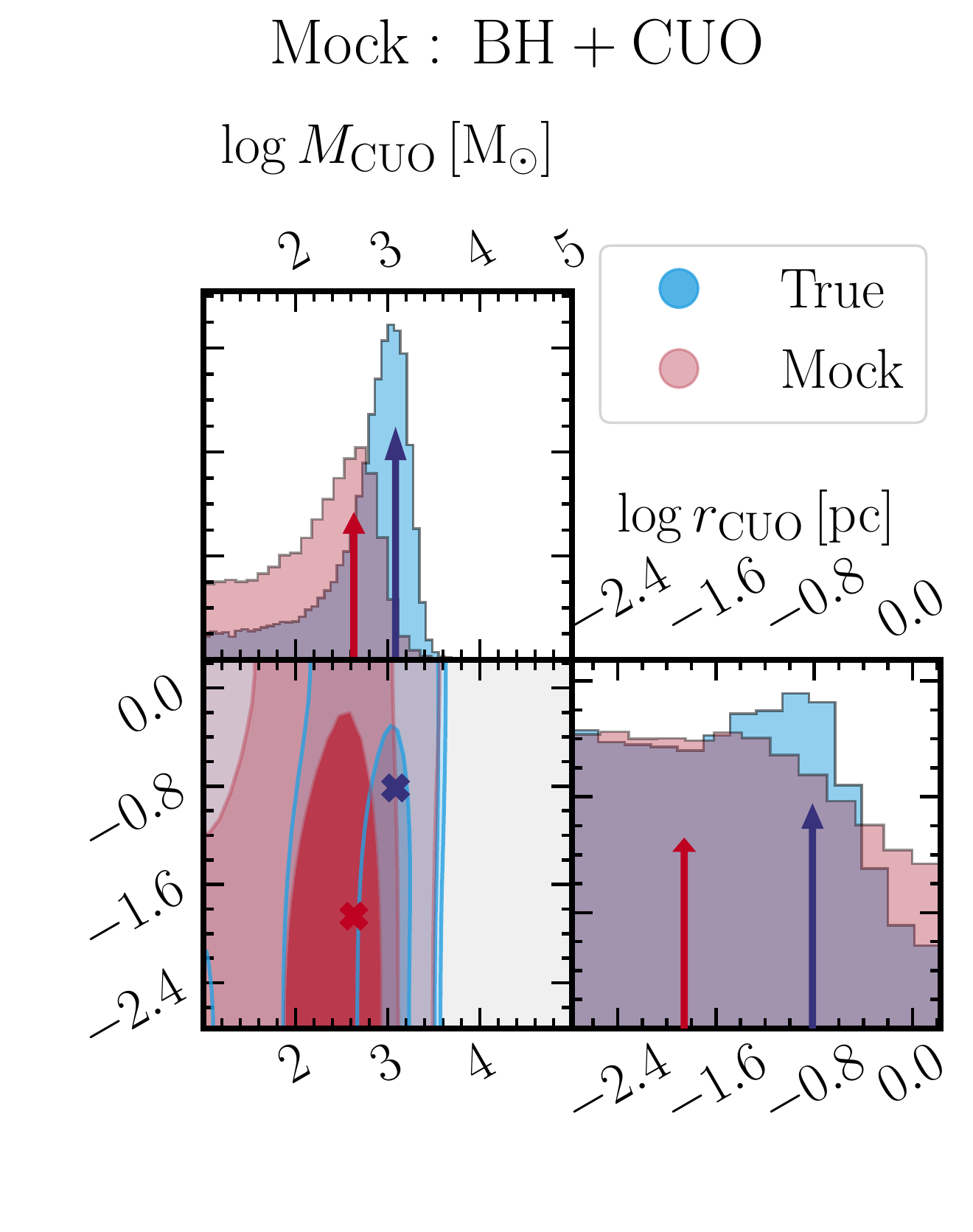} \\
\vspace{0.2cm}
\includegraphics[width=0.12\hsize]{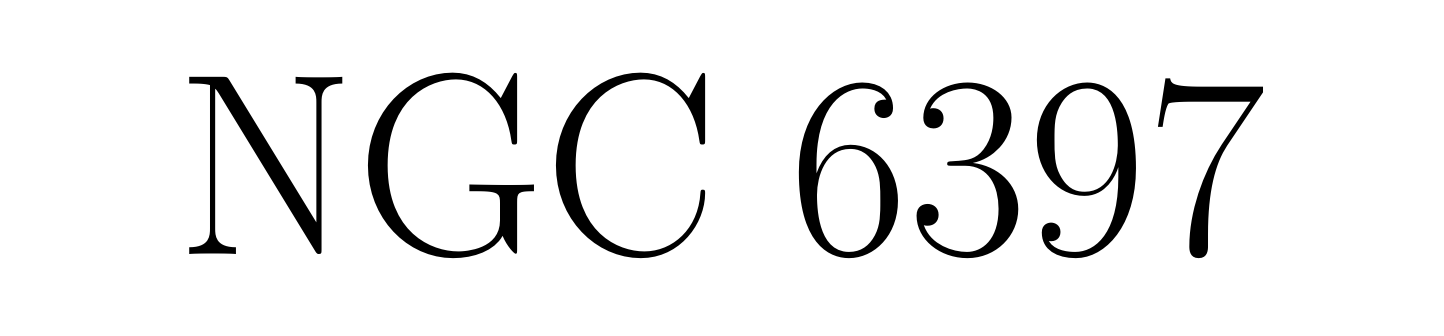} \\
\includegraphics[width=0.247\hsize]{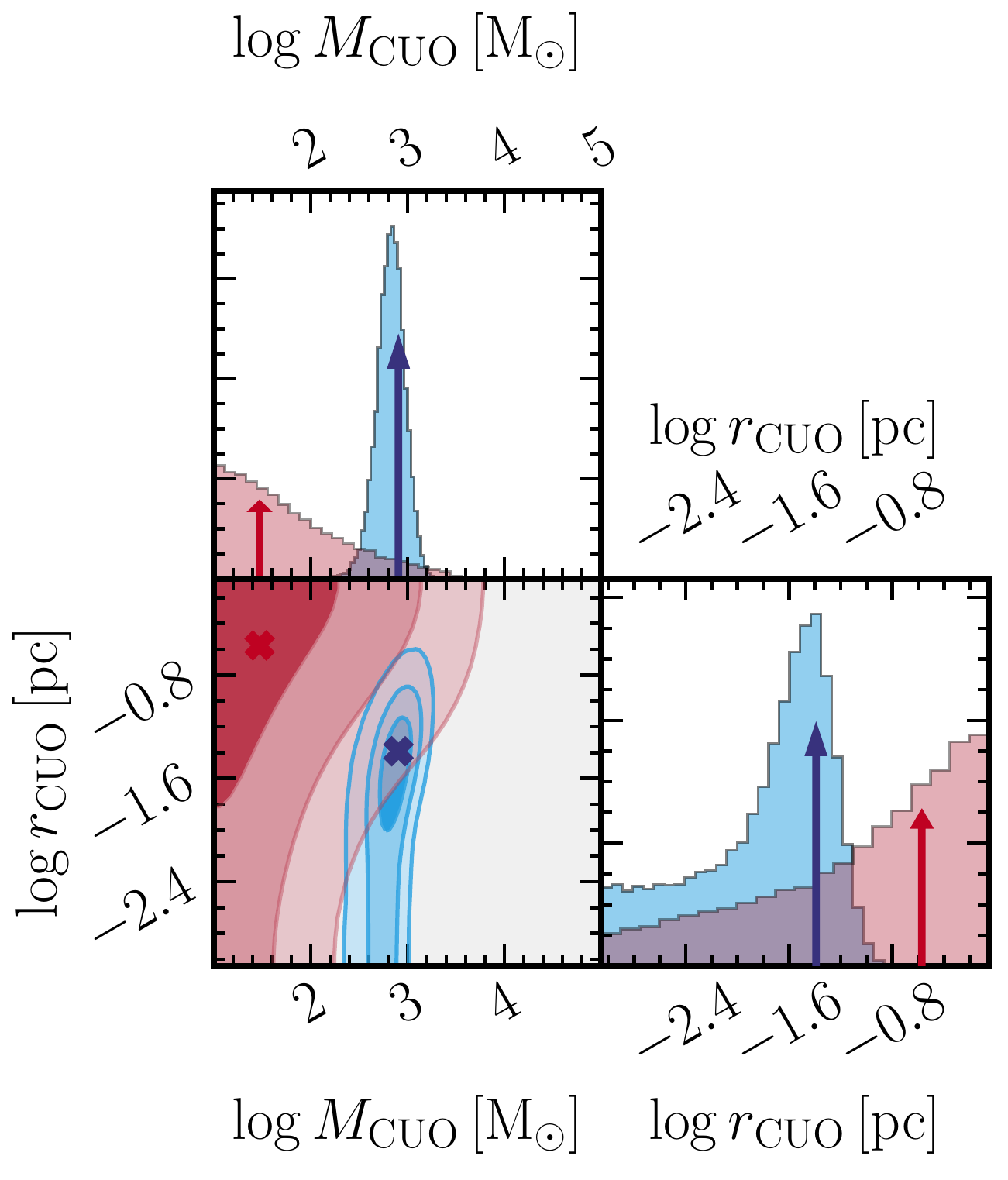}
\includegraphics[width=0.247\hsize]{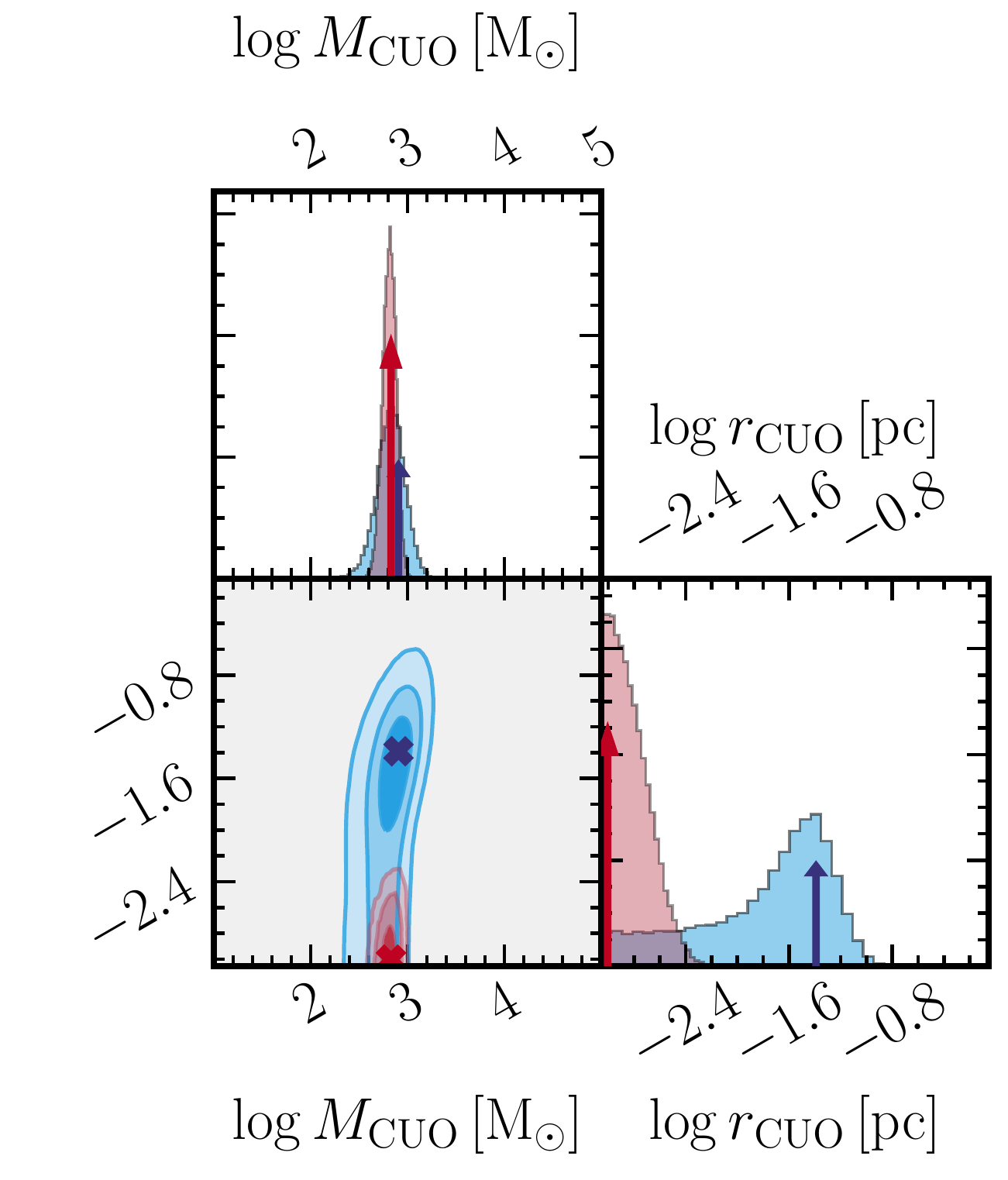}
\includegraphics[width=0.247\hsize]{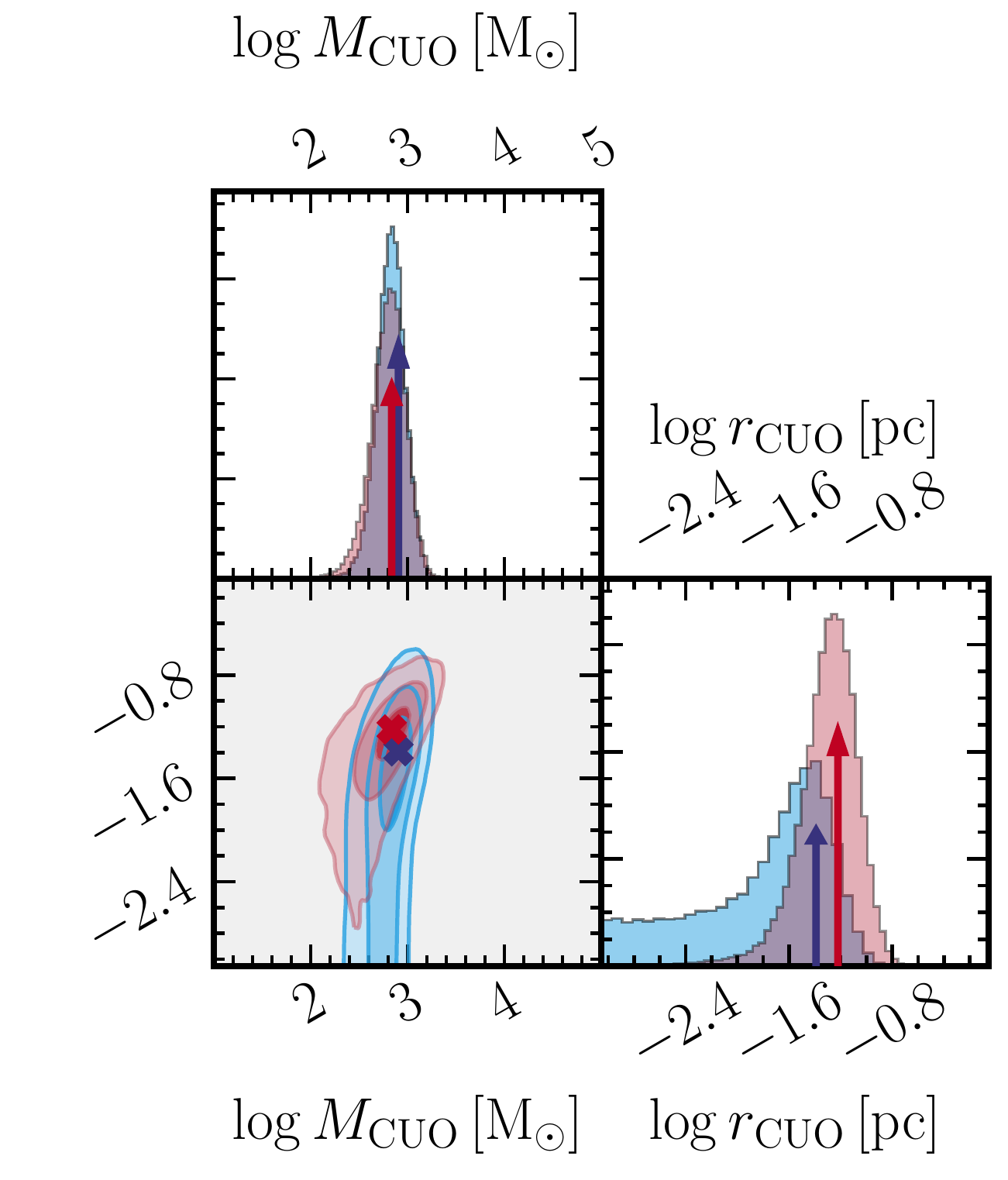}
\includegraphics[width=0.247\hsize]{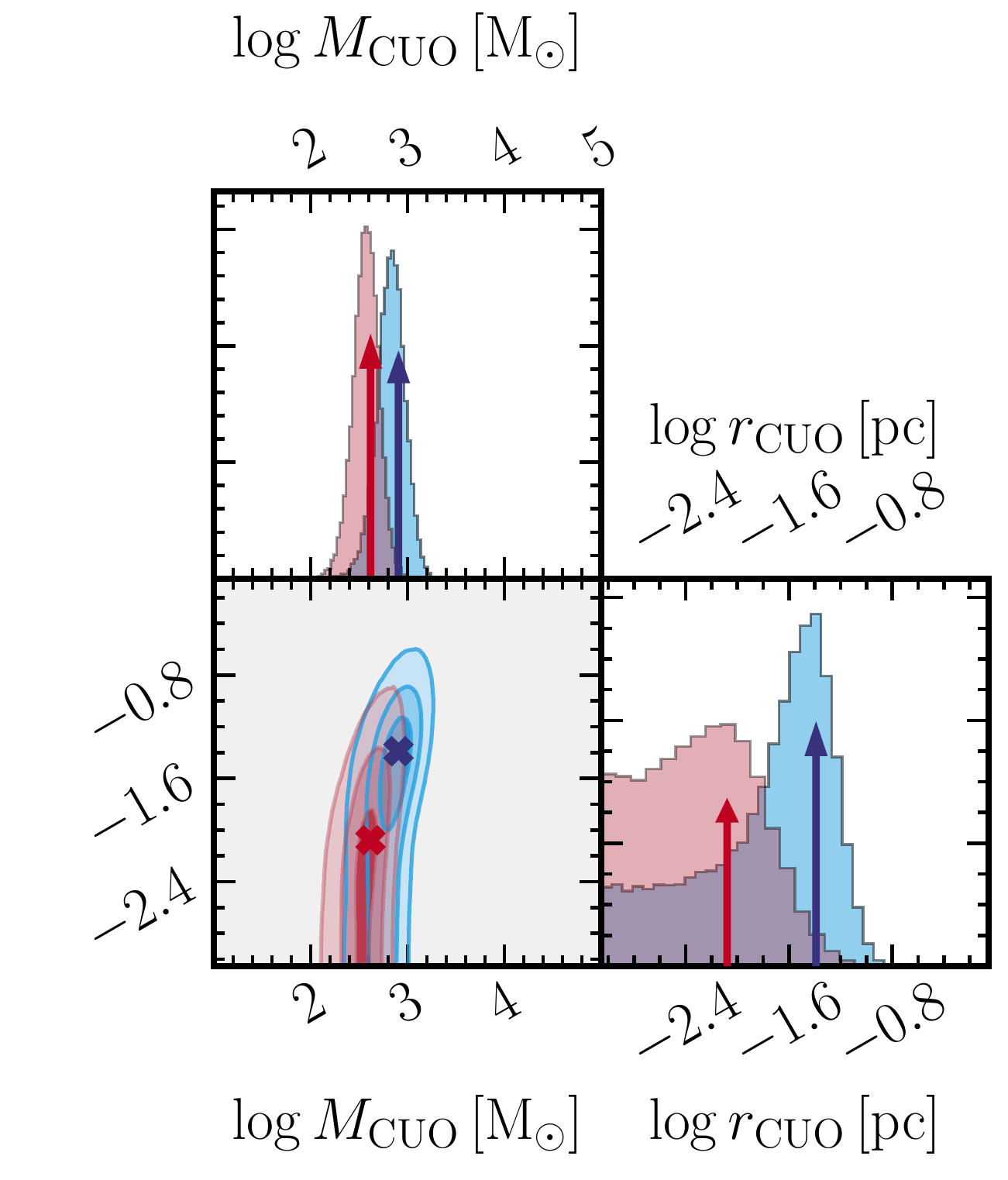}
\caption{{\it \mpo\ outputs on mock data compared to those on real data:} 
Marginal distributions of  the cluster of unresolved objects (CUO) mass and 2D Plummer half mass radius and their covariances for the true data (\hst\ and \gaia\ EDR3) in \textit{blue} and the mock data (constructed with \agama) in \textit{red}, for NGC~3201 \textbf{top} and NGC~6397 \textbf{bottom}, respectively. 
The priors are flat for $M_{\rm CUO}$ within the plotted range and zero outside, while they are Gaussian for the scale radii, centred on the middles of the panels and extending to $\pm3\,\sigma$ at the edges of the panels, and zero beyond. The arrows indicate the respective best likelihood solutions of the Monte Carlo chains.
The mock data prescription is, from \textbf{left} to \textbf{right}: No central dark component (Nothing); a central black hole alone (BH); a central CUO (CUO) and both a central black hole and CUO (BH$+$CUO). The mocks were constructed with the best values of each respective isotropic mass model from the online version of 
Table~\ref{tab: results-full}. The fits alone indicate preference for a CUO in NGC~6397 and for a central mass excess in NGC~3201, without strong distinction between extended and point-like scenarios (see text for details).}
\label{fig: corner-comp-partial}
\end{figure*}

Our \mpo\ fits to the kinematic data allow us to constrain the radial profile of velocity anisotropy of the visible (Main Sequence) stars. For each mass model, we performed \mpo\ fits to the data using,  isotropic and gOM velocity anisotropy (Eq.~\ref{eq: gOM}]) in turn.
%
The online version of
Table~\ref{tab: results-full}
shows that, for both GCs, all the free anisotropy runs (labelled as ``$\beta(r)$'', in the third column) present roughly isotropic shapes, with a slight 
tendency for radial anisotropy in the outskirts and tangential anisotropy in the centre. 
However, the uncertainties in the inner and outer anisotropies encompass the isotropic solution.
This can be seen in Figure~\ref{fig: anis-mpo}, which displays the radial profiles of velocity anisotropy for the two clusters with the CUO mass model. Admittedly, the uncertainties of the outer anisotropy profile of NGC~3201 are quite large.
Yet, AICc Bayesian evidence prefers the isotropic solution, although the preference is only moderate ($\Delta \rm AICc \approx 3.5$ in both GCs, with $\Delta\rm AICc = 6.9$ for the anisotropic model with both IMBH and CUO compared to the isotropic one). Thus, given the present quantity of kinematic data in both GCs, there is no compelling evidence for anisotropic motions in either cluster.

We thus now assume velocity isotropy to better explore other free parameters such as the central unseen mass.


\subsection{Central dark component}

\subsubsection{\mpo\ results on observed data}
We now compare, for both clusters, our four mass models using \mpo\ fits to the kinematic data assuming isotropic velocities. We first use AICc to compare our mass models. 
The online version of Table~\ref{tab: results-full}
indicates that there is only marginal
evidence of an excess mass in NGC~3201 ($\Delta\rm AICc = 3.53$). On the other hand, there is very strong evidence for an excess inner mass in NGC~6397, $\Delta\rm AICc = 24$, yielding a probability of no mass excess of less than $10^{-5}$ (from Eq.~[\ref{eq: pAIC}]).

Among the remaining three mass models with a central mass excess, the Akaike information criterion given in the online version of Table~\ref{tab: results-full} indicates a very weak preference for a central black hole in NGC~3201 compared to the second-best model being a CUO ($\Delta\rm AICc = 1.88$).
This difference is too small to distinguish between IMBH and CUO models.
While the preferred mass of the central black hole is $825\,\msun$, the 16th percentile of the central black hole mass is only $209\,\msun$.
In summary, there is marginal evidence of an excess inner mass in NGC~3201, but while there is weak evidence in favor of an IMBH (of mass between $200\,\msun$ and $1050\,\msun$) relative to a CUO, it is too small to be taken in consideration.
This suggests that we need to consider other indicators to compare the models. We will discuss in Sect.~\ref{sssec: test_mocks} whether the CUO model fits produce marginal distributions of the CUO scale radius that match those of \mpo\ fits to a mock with an IMBH in their centre.

For NGC~6397, the best fit IMBH model gives a more constrained IMBH mass between 400 and $700\,\msun$.
But  AICc leads to a very weak preference for the CUO over the IMBH model ($\Delta\rm AICc = 1.08$), which, again, is too small to consider.
In summary, there is very strong evidence for an excess inner mass in NGC~6397, but it is difficult to tell whether it is extended or not.

\subsubsection{Tests with mock data}
\label{sssec: test_mocks}

We ran \mpo\ on our 4 mock datasets, with isotropic velocities, realistic proper motion errors, for our 4 mass models. In all these \mpo\ runs, we assumed a CUO mass model, for reasons that will be clear below. Figure~\ref{fig: corner-comp-partial} compares the marginal distributions  of CUO mass and scale radius and their covariance for the \mpo\ fit to mock data (light red) to the \mpo\ fit to the observed data (light blue), both assuming the CUO mass model. The figure also compares the values of the maximum likelihood estimates.

First, the marginal distributions of CUO mass fit on the mock with no central mass excess (left top panels of left corner plots, in light red) show a significantly different pattern from the true data marginal distribution of CUO mass (light blue in same panels), spanning significantly lower masses. 
On the other hand, for both clusters, the marginal distributions of the CUO mass obtained on the three other mocks with extra inner mass (second, third and fourth columns of panels) show general agreement on the shape of the marginal CUO mass distributions obtained from the observed data, with some fairly small shifts of the peaks.

The marginal distribution of CUO scale radius is even more interesting. For the fit on the mock with no excess mass, the marginal distribution of CUO scale radius is flat for NGC~3201, suggesting a CUO scale radius that cannot be determined. For NGC 6397, this distribution is rising, suggesting a large CUO scale radius of the order of the radius of the visible stars. For the black hole mocks, the CUO scale radius marginal distributions also show a different pattern than what is observed: for both clusters, it fails to reproduce the  peak in the marginal distribution of CUO scale radius obtained on the observed data: this is very striking for NGC~6397, but is also visible for NGC~3201. On the other hand, as expected, the marginal distributions of the CUO scale radius obtained on the CUO mocks are consistent with those obtained on the data.

Whether or not the CUO model is the correct mass model, one expects that the marginal distributions of the CUO 
scale radii should have similar shapes when comparing those obtained on a mock that represents the observed data and those directly obtained from the same data.
The marginal distribution of CUO scale radii obtained from our CUO fits 
should therefore be a sensitive discriminator between single BH and a CUO, with no peak or a clear peak in the distribution of CUO scale radii, respectively.
Although AICc (based on likelihood) provides a global score for a particular model, it misses the differences in the  marginal distributions of CUO scale radii, which appear to be the critical aspect to differentiate black hole and CUO scenarios. In AICc, the differences of these marginal distributions are blurred by small differences in the marginal distributions of the structural properties of GC stars.
We therefore favour the comparison of CUO marginal distributions between mock and observed data to using AICc, and will hereafter omit model comparisons based on AICc.

The large discrepancy noted between the CUO 
mass and scale radii marginal distributions between the mock data
with no-inner-excess-mass and the observed data indicates that this model is ruled out by these \mpo\ fits. Similarly, the lack of a peak in the marginal distributions of CUO scale radii for the black hole mock, for both GCs, suggests that the data prefers an extended extra mass (CUO) compared to a point-like central mass (BH) for both GCs.
This indicates that the comparison of marginal distributions is a more sensitive tool for model selection than the comparison of likelihoods with AICc Bayesian evidence.

\begin{table}
\caption{Main statistical tests used for model selection.}
\label{tab: statistics}
\centering
\renewcommand{\arraystretch}{1.2}
\tabcolsep=2.5pt
\footnotesize
\begin{tabular}{l@{\hspace{3mm}}lccrrrr}
\hline\hline   
\multicolumn{1}{c}{Cluster ID} &
\multicolumn{1}{c}{Mock} &
\multicolumn{1}{c}{$\phi$} &
\multicolumn{1}{c}{$\phi$} &
\multicolumn{1}{c}{AD} &
\multicolumn{1}{c}{AD} &
\multicolumn{1}{c}{KS} &
\multicolumn{1}{c}{KS} \\
\multicolumn{1}{c}{} &
\multicolumn{1}{c}{model} &
\multicolumn{1}{c}{$M_{\rm dark}$} &
\multicolumn{1}{c}{$r_{\rm dark}$} &
\multicolumn{1}{c}{$M_{\rm dark}$} &
\multicolumn{1}{c}{$r_{\rm dark}$} &
\multicolumn{1}{c}{$M_{\rm dark}$} &
\multicolumn{1}{c}{$r_{\rm dark}$} \\
\multicolumn{1}{c}{(1)} &
\multicolumn{1}{c}{(2)} &
\multicolumn{1}{c}{(3)} &
\multicolumn{1}{c}{(4)} &
\multicolumn{1}{c}{(5)} &
\multicolumn{1}{c}{(6)} &
\multicolumn{1}{c}{(7)} &
\multicolumn{1}{c}{(8)} \\ 
\hline
NGC~3201 & Nothing  & 6\%  & 84\% & 18346 & 1709  & 0.229 & 0.070 \\
NGC~3201 & BH       & 73\% & 20\% & 1307  & 317   & 0.045 & 0.027 \\
NGC~3201 & CUO      & 90\% & 82\% & 7778  & 208   & 0.127 & 0.026 \\
NGC~3201 & BH$+$CUO & 63\% & 38\% & 13979 & 979   & 0.179 & 0.058 \\
\hline
NGC~6397 & Nothing  & 33\% & 59\% & 56540 & 29034  & 0.412 & 0.290 \\
NGC~6397 & BH       & 63\% & 11\% & 15783 & 50704 & 0.148 & 0.361 \\
NGC~6397 & CUO      & 76\% & 77\% & 1678  & 22912 & 0.050 & 0.219 \\
NGC~6397 & BH$+$CUO & 44\% & 47\% & 333   & 8459  & 0.022 & 0.150 \\
\hline
\end{tabular}
\parbox{\hsize}{\textit{Notes}: 
Columns are 
\textbf{(1)} Cluster ID; 
\textbf{(2)} Mass model assigned to the mock data; 
\textbf{(3)} Fraction of chain elements that present absolute dark mass distances greater than the distance between the mock and true data fit's best solutions -- Higher values indicate good agreement between the mock and true data fits;
\textbf{(4)} Same than (3), but considering only the dark radius;
\textbf{(5)} AD statistic, for $M_{\rm dark}$ -- High values indicate poor matches; 
\textbf{(6)} AD statistic, for $r_{\rm dark}$; 
\textbf{(7)} KS statistic, for $M_{\rm dark}$ -- High values indicate poor matches; 
\textbf{(8)} KS statistic, for $r_{\rm dark}$; 
}
\end{table}

We quantify the preferences mentioned above in Table~\ref{tab: statistics}, with the statistical indicators presented in Section~\ref{ssec: stats}. 
We first notice that in both clusters, models without a central dark mass (``Nothing'') are quickly ruled out, given their very different maximum likelihood solutions (small percentages in column 3) and disagreeing marginal shapes (high values in columns~5 and 7) concerning the mass fit.

We compare the remaining three dark mass models by
analysing column~4 to see how distant the maximum likelihood scale radius solutions are, and also columns~6 and 8 to compare the scale radius marginal distributions.
The two clusters display extremely unsatisfactory agreements for the case with a central black hole alone (``BH'') in column~4, and poor agreements for the case with both a CUO and a black hole (``BH$+$CUO''). The best match of shapes is for the CUO model in NGC~3201, followed by a ``BH$+$CUO'' model. In NGC~6397, this trend is inverted, but the reader should keep in mind that the ``BH$+$CUO'' case for this cluster consists of a $\sim750 \, \msun$ CUO, with a black hole of only $\sim20 \, \msun$, hence nearly a CUO case actually.

Therefore, our comparisons between \mpo\ fits of mock and observed data yield robust evidence for a dark central mass in both clusters. While in NGC~3201 we have reasonable, but not strong arguments to defend that this mass is extended, the case for NGC~6397 is more straightforward, with robust evidence for an extended mass, in agreement with previous fits from VM21\footnote{We notice however, that in VM21, their extended mass was roughly twice more massive and twice more diffuse. This difference could be related to a less complete data set in VM21 and a less  conservative data cleaning.}. Among the online material we provide, we display once again the comparison of marginal distributions for mock and observed data, now adding the structural parameters of the GCs (i.e., S\'ersic index and scale radius, and total cluster mass).
One also sees an excellent agreement between the marginal distributions obtained by \mpo\ on the two datasets (the GC effective radii differ by less than 0.02 dex).

\begin{figure}
\centering
\includegraphics[width=0.6\hsize]{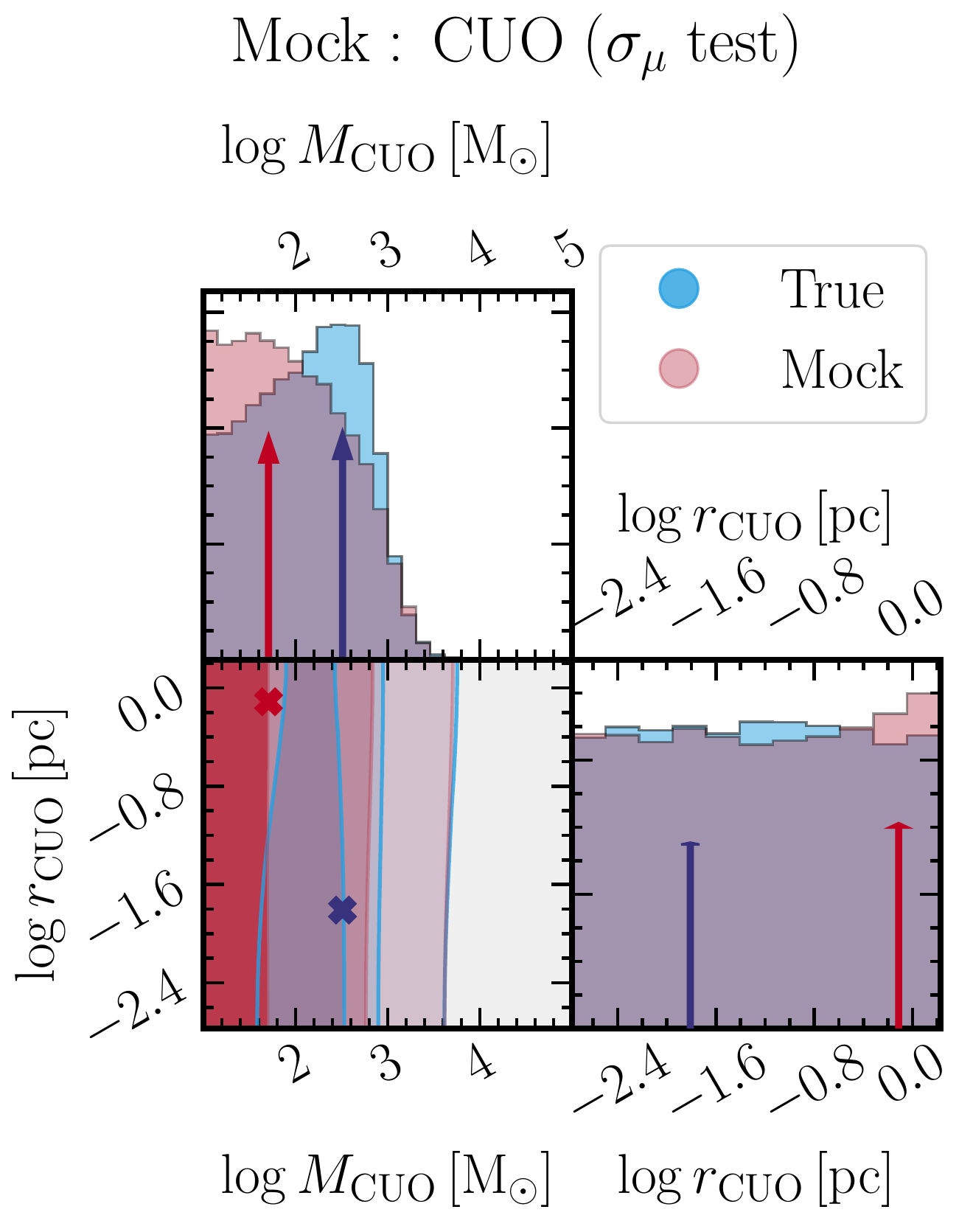}
\caption{{\it Error threshold test:} Similarly to Figure~\ref{fig: corner-comp-partial}, we present the corner plot of the logarithm of the cluster of unresolved objects (CUO) mass and 2D Plummer half mass radius for the true data (\hst\ and \gaia\ EDR3) in \textit{blue} and the mock data (constructed with \agama) in \textit{red}, for NGC~3201. In this case, the true data's maximum error threshold was set as half of our standard choice (see Section~\ref{ssec: pm-ilop}) and the mock data was constructed with a CUO prescription based on the standard case, but with the error budget similar to the new true data. This test helps one to check if the eventual non-detection of a central component is likely due to the lacking of such component or to a too conservative data cleaning (here, limiting too much the error threshold, thus forcing incompleteness).
}
\label{fig: corner-comp-sigmu}
\end{figure}

\subsection{Robustness}

\subsubsection{Error threshold} \label{ssec: error-limit}


The choice of error threshold (Sect.~\ref{ssec: pm-ilop}) when cleaning the observational dataset can affect the conclusions of our analysis. The advantage of adopting a liberal (high) maximum allowed proper motion error is to increase the size of the dataset analysed by \mpo, which handles the observed proper motion errors.
However, if the maximum allowed error is large, in comparison with the proper motion dispersion (i.e. POS velocity dispersion, after incorporating the distance), then over-estimated errors will lead to underestimated true proper motion velocity dispersions and masses. Conversely, if the proper motion errors are underestimated, then we would overestimate the true dispersions. If the high density of the central areas leads to confusion, the systematic errors may indeed be underestimated in the inner regions,  and may lead one to conclude to a spurious extended central mass (i.e. CUO). 
We therefore also ran \mpo\ using datasets filtered more conservatively, where  the data is cleaned with half of our standard error threshold.

The online version of Table~\ref{tab: results-full}
also displays the \mpo\ fits when the data is cleaned with half of our standard error threshold.
The best-fit black hole and CUO masses, as well as CUO radii of NGC~6397 remain roughly the same, attesting the robustness of the presence of a CUO. On the other hand, the analogous best-fit parameters of NGC~3201 predict lower central masses (black hole or CUO) and lower CUO scale radii,  with a lower mass limit also equivalent to the mass of a single stellar-mass black hole.
The difference between the two clusters is predictable, since NGC~3201 is a more distant cluster, leading to higher proper motion errors (and therefore more affected by a lower error threshold).

Are the differences in maximum likelihood CUO parameters for NGC~3201 between the standard and lower proper motion error thresholds caused by shot noise or by the lack of an inner dark component?
We investigated this more closely with extra mock data sets. We constructed them with using the fits from models~1, 2, 3 and 4 of the online version of Table~\ref{tab: results-full} (i.e., the fits of the data with the standard error threshold), but with an error budget and completeness similar to the \hst\ plus \gaia\ EDR3 data set with half of our standard error threshold (the procedure was the same as the one described in Section~\ref{ssec: mock-build}). 
This more conservative error threshold leaves us with 3176 \hst\ stars for NGC~3201 and with  5510 \hst\ stars, and 1551 \gaia\ stars for NGC~6397.
We next ran \mpo\ on these mocks with a CUO 
model and compared 
the marginal distributions of the CUO parameters for the mock data and the true data (again, both with these conservative error thresholds).

In Figure~\ref{fig: corner-comp-sigmu}, one clearly sees that the more conservative data cleaning of NGC~3201 affects considerably the detection of a central dark component in two ways.
First, while the marginal distribution of true CUO mass retains a peak, with a mass 3 times lower than obtained with the liberal threshold, that of the mock CUO mass is shifted to lower masses without a peak (we observed the same behaviour for the black hole mass from a black hole mock).
Second, the marginal distributions of CUO radius are flat for both mock and observed data, while the corresponding marginal distributions in the analyses of the mock and observed data with the liberal proper motion error threshold had clear peaks (Fig,~\ref{fig: corner-comp-partial}).
This suggests that the conservative data is too sparse to conclude on the mass and extent of a central component.

Figure~\ref{fig: corner-comp-sigmu} helps to argue that a non-detection of a central mass in the more conservative data cleaning cannot be attributed to the lack of such a  CUO component in NGC~3201, as mock data shows that an existing $\sim 1200$~M$_{\odot}$ component has its signatures erased by an incompleteness similar to the one of our more conservative data set. In fact, setting this smaller error threshold for NGC~3201 reduces its number of tracers (stars) to less than half of the less conservative case, and effectively removes all \gaia\ stars from the subset.

Finally, we tested if the mass peak observed for the data could be due to underestimated proper motion errors in the cluster's inner regions, caused by systematic errors from confusion in these dense regions. Indeed, an underestimation of proper motion errors will lead to overestimated plane-of-sky velocities, and therefore an overestimated, extended inner mass. We constructed mocks similarly to the ones in Figure~\ref{fig: corner-comp-partial}, but set the actual proper motion uncertainties to be 10\% higher than the values that were previously provided to \mpo, inside a projected radius of two times the length of the CUO scale radius we fitted. The outcome of these extra runs are shown in Figure~\ref{fig: underestimated-err} (online material). The figure shows that a central mass overestimation due to 10\% underestimated proper motion errors is insufficient to mimic a signal of a $\sim 1000 \, \msun$ central mass, with a peak in $M_{\rm CUO}$ as clear as the one observed in the data. We repeated the analysis for 20\% underestimated errors and retrieved the same results. This indicates that our standard data cleaning is not so liberal as to allow a false detection of an inner, possibly extended, dark mass.


\subsubsection{Mass segregation}

\begin{figure}
\centering
\includegraphics[width=0.95\hsize]{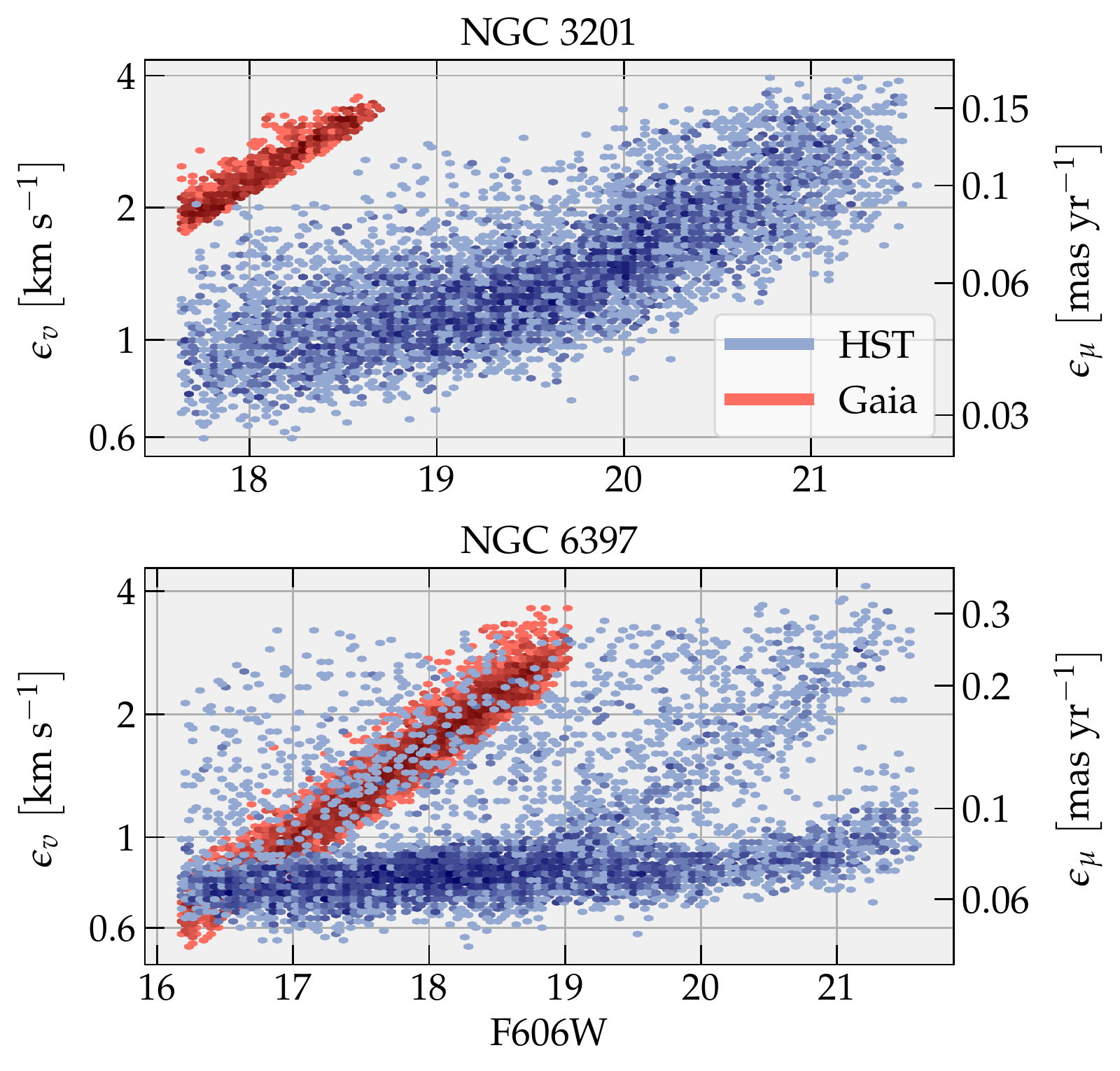}
\caption{
\textit{Error-magnitude relation:}
Same as the lower-right panel of Fig.~\ref{fig: mock}, adding NGC~6397, a log $y$-axis scale and physical velocity uncertainties given together with proper motion uncertainties.
}
\label{fig: err-mag-relation}
\end{figure}

A possible worry when modelling GCs is the presence of mass segregation, which implies different dynamics and stellar distributions for stars having different mass ranges. However, the presence of systematics and uncertainties in the data might erase some imprints from this segregation, so one should balance if the effort to consider mass segregation when solving the Jeans equation is worth it. 

For instance, due to energy equipartition, one can naively estimate the amplitude of mass segregation effects on the velocity dispersion, by assuming the relation $\sigma_i / \sigma_j \approx \sqrt{m_j / m_i}$, where $\sigma$ is the velocity dispersion and $m$ is the mass of two classes of objects ($i$ and $j$) of different masses.  With the masses derived from our \parsec\ fits, we can thus use typical mass ratios within our data set, for each cluster,  to probe the expected change in the velocity dispersion profiles of our clusters. 

Figure~\ref{fig: mass-mag-relation} shows extreme mass ratios in our data of 1.58 and 2.28 for NGC~3201 and NGC~6397, respectively.
The typical mass ratio of a bright and faint component would roughly correspond to the ratio of 75th to 25th percentiles. These are 1.22 and 1.18 for NGC~3201 and NGC~6397, respectively.
The corresponding mass segregation effects on the velocity dispersions will amount to
$\left(\sigma_2/\sigma_1-1\right)\,\sigma = \left[(m_1/m_2)^{1/2}-1\right]\,\sigma$.
Considering typical values of $0.2$ $\masyr$ and $0.4$ $\masyr$ for the velocity dispersion profiles of NGC~3201 and NGC~6397, respectively (see Figure~A1 from \citealt{Vasiliev&Baumgardt&Baumgardt21}), this yields of the order 
0.022\,$\masyr$ and 0.036\,$\masyr$ effects from mass segregation. More sophisticated analyses, from $N$-body simulations,
lead to
even weaker mass segregation, with a shallower relation between velocity dispersion and stellar mass, with a typical slope shallower than $-0.2$ instead of $-1/2$ \citep{Trenti&vanderMarel13,Bianchini+16_equipartition}, inducing even smaller changes in PMs.

Figure~\ref{fig: err-mag-relation} shows the proper motion errors as a function of apparent magnitude for the two GCs. The shifts in proper motion caused by mass segregation are so low that they are not included in the figure panels. Therefore, the mass segregation effects are  smaller than all the proper motion errors.

As a matter of fact, it has been shown in \cite{Vitral&Mamon21} with similar data that, when modelling NGC~6397 with a single mass component and two mass populations, the estimated masses agreed within the 1-$\sigma$ error bars and the respective density profiles related such that the prescriptions of the single component fits were alike the ones of the brightest component of the two population fits. Based on these results and on the numbers given above, we argue that for NGC~6397, our results should not drastically change if accounting for mass segregation in our fits.
NGC~3201, which has a shallower inner surface density profile compared to NGC~6397, should be less dynamically evolved and show even less segregation.

\subsubsection{Different centres}

Our standard \mpo\ fits assumed a GC centre calculated from \gaia\ EDR3 with \balrogo, but other measurements of centre exist (e.g. \citealt{Goldsbury+10,GaiaHelmi+18}). In particular a wrong centre could for instance privilege a CUO over a single black hole, as the velocity dispersion would have an increase towards higher radii, a characteristic sign of a CUO component.

Thus, we fitted the same data set using the centres of \cite{Goldsbury+10}, and as it can be seen in the online version of Table~\ref{tab: results-full}, the CUO prescriptions are roughly similar, within the error bars. We therefore argue that our fits are robust considering our choice of GC centre.

\subsubsection{\hst\ bulk proper motion} \label{ssec: bulk-pm-robust}

We finally address the question if the bulk proper motion we set for \hst\ could have affected our modelling. Since the \hst\ data bulk proper motion was slightly different from the bulk motion of our \gaia\ EDR3 data, one could reasonably wonder that strange effects could be observed in our fits.

Hence, we also performed fits with the bulk \hst\ proper motion being set as the same \gaia\ EDR3 bulk motion shared in \cite{Vasiliev&Baumgardt&Baumgardt21} for NGC~3201 and NGC~6397. These fits are displayed in the online version of Table~\ref{tab: results-full} and the reader can see that the fits show no significant difference with respect to the standard ones. Thus, once again we argue that our fits are robust, now with regard to the choice of bulk proper motion of \hst\ stars.

\begin{figure*}
\centering
\includegraphics[width=0.9\hsize]{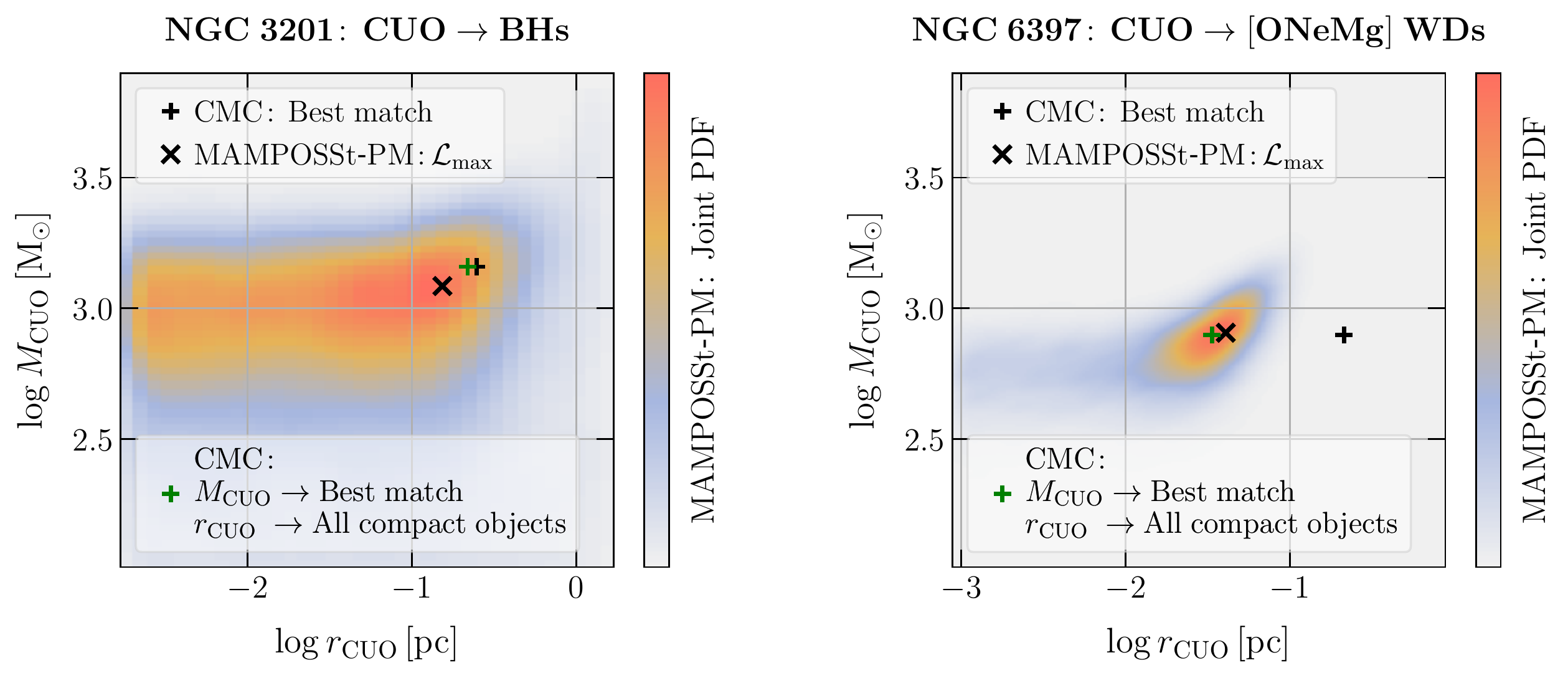}
\caption{{\it Comparison between \mpo\ fits and  {\tt CMC} dynamical simulations:} Mass vs. half-mass radius  of the sub-cluster of unresolved objects (CUO) for NGC~3201 (\textbf{left}) and for NGC~6397 (\textbf{right}). In each panel, the \textit{black cross} indicates the position of the maximum likelihood solution of \mpo, while the \mpo\ joint probability distribution function (PDF) 
is linearly colour-coded from grey to orange. Also, in each panel, the \textit{black plus sign} indicates 
the CUO mass and scale radius (containing half the projected mass) of the \texttt{CMC} simulation snapshot whose surface brightness and velocity dispersion profiles match best the observations, when the CUO is that of the dominant compact component (black holes for NGC~3201 and [ONeMg] white dwarfs for NGC~6397).
Other compact objects can also contribute to the sub-cluster, especially in NGC~6397, but they end up mixing themselves within the stellar component (thus, not forming a sub-cluster), so the \mpo\ fits of scale radius might be affected. For that reason, we also show the 
case considering all compact remnants (\textit{green plus sign}): the mass is still that of the main CUO component (black holes or [ONeMg] white dwarfs), 
but the scale radius corresponds to where the mass contribution of all compact remnants from the \texttt{CMC} simulation snapshot reaches half of the CUO mass found by \mpo\ (the radii are projected half-mass radii, estimated from the 3D half-mass radius, converted to projected assuming a Plummer model). 
}
\label{fig: cmc-mpo}
\end{figure*}

\subsection{Comparison with \cmc\ models}

The analysis of the internal kinematics of core-collapsed NGC~6397, with a code such as \mpo, may miss important priors set by the full, complex dynamics of the clusters. Dynamical simulations incorporating the small-range dynamical processes should therefore be used to complement our Jeans modelling. 
We then selected Monte Carlo simulations from \texttt{CMC} according to Section~\ref{ssec: cmc}, and 
picked the snapshot whose
surface brightness and velocity dispersion profiles best match 
the observed ones. For NGC~6397, this best snapshot was in one of the new models that were run\footnote{The new model was similar to the ones presented in \cite{Kremer+21}, but with a slightly smaller initial size for the cluster.}.


Figure~\ref{fig: cmc-mpo} shows the distinctive match of the
CUO mass and scale radius, between our \mpo\ mass-modelling fits (black cross) and from the snapshot of the \texttt{CMC} Monte Carlo simulation that best matches the observed surface brightness and velocity dispersion profiles (black plus sign) for NGC~3201 (left) and NGC~6397 (right). 
We are able to find {\tt CMC} simulations that predict a CUO mass that agrees well with those found by \mpo, attesting simultaneously the good performances of both \mpo\ and \texttt{CMC}, which 
are completely different methods
arriving at roughly the same result. The match of CUO scale radius is also very good for the black hole population in NGC~3201, but less good for the [ONeMg] white dwarf population in NGC~6397, with the simulations predicting a scale radius roughly five times greater. 

\begin{figure*}
\centering
\includegraphics[width=0.9\hsize]{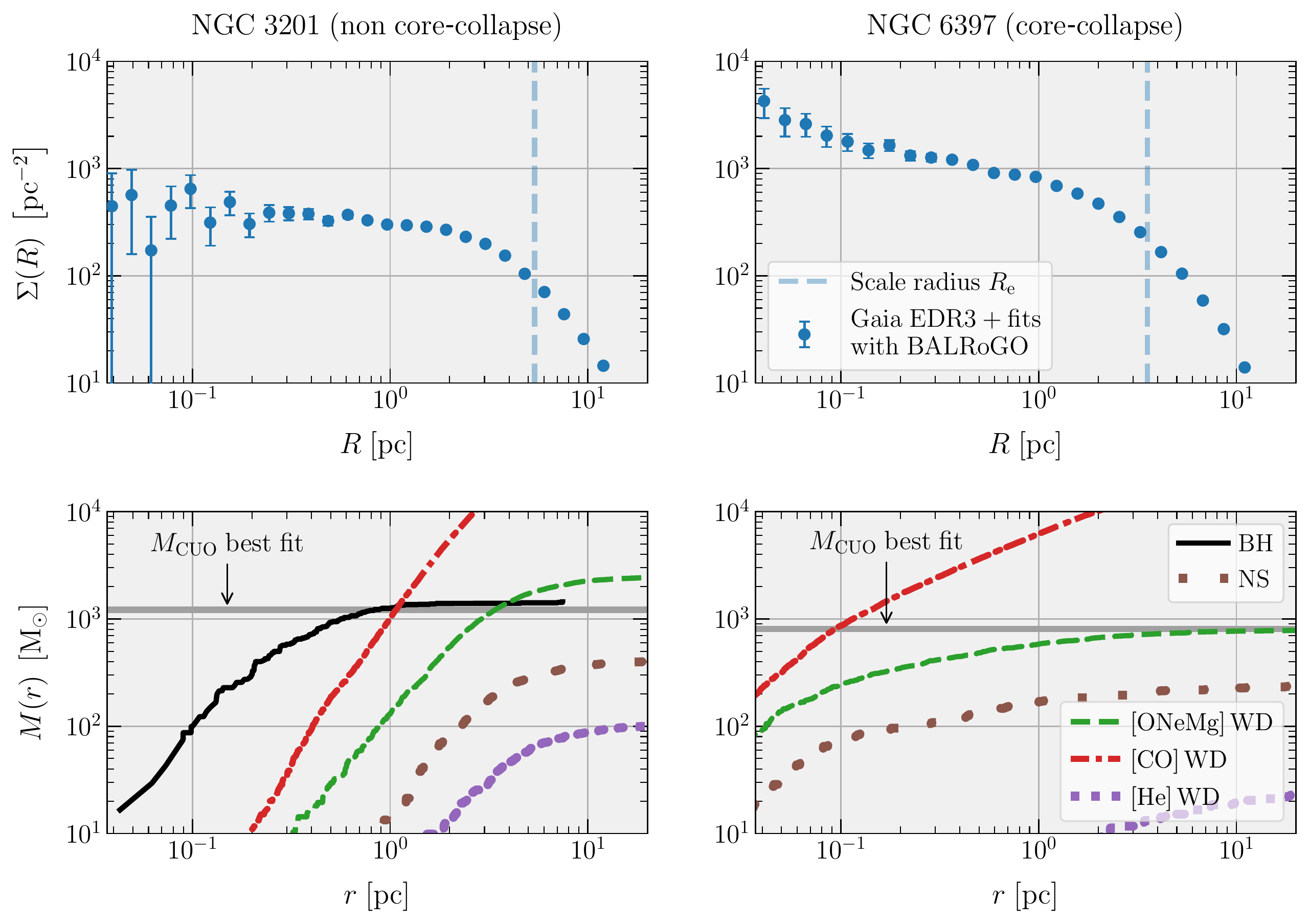}
\caption{{\it The effects of core-collapse:} Comparison between NGC~3201 (\textbf{left}), a cluster which has not yet experienced core-collapse, and NGC~6397 (\textbf{right}), a classic example of a cluster in a post core-collapse state. The \textbf{upper} panels display the surface density profiles of both clusters, from \gaia\ EDR3, using the same range of absolute magnitudes.
The angles are converted  to distances with the values displayed in Table~\ref{tab: assumptions} as a function of projected radius $R$. A uniform contribution from Milky Way interlopers (from fits with \balrogo) is subtracted. The error bars are a quadratic sum from the Poisson errors plus the errors on the fits. We also display a \textit{dashed} transparent-blue line at our best-fit $R_{\rm e}$ scale radius (cf., Table~\ref{tab: results-full}). The \textbf{lower} panels show the contribution in mass (M$_{\odot}$) from each compact object (remnant) as a function of the 3D distance from the cluster's centre (in pc), obtained from \texttt{CMC} models \protect\cite{Kremer2019,Kremer+21}, in addition to a grey line depicting our best \mpo\ mass fit of a sub-cluster of unseen objects (i.e., $M_{\rm CUO}$). The correspondences for each remnant are \textit{solid black}: black holes; \textit{loosely dotted brown}: neutron stars; \textit{dashed green}: [ONeMg] white dwarfs; \textit{dashed-dotted red}: [CO] white dwarfs; \textit{densely dotted purple}: [He] white dwarfs. This  illustrates that a collapsed population of stellar-mass black holes is a plausible explanation to the delay of core-collapse in globular clusters.
}
\label{fig: core-collapse}
\end{figure*}

This scale radius issue can be explained as follows.
As shown in the lower right panel of Figure~\ref{fig: core-collapse}, the unresolved population in NGC~6397 is
composed of 
two white dwarf  components ([CO] and [ONeMg]). 
The [CO] white dwarfs follow a density profile close to the resolved stars and eventually mix up with the stellar component.
In contrast, the [ONeMg] white dwarfs are much more concentrated in the inner regions. It is therefore natural that when \mpo\ fits a single clustered dark population, it gets confused
if the CUO is made of several components with different density profiles. For that reason, we also highlight in Fig.~\ref{fig: cmc-mpo} the CUO half projected-mass radius when considering all compact remnants (green plus sign) instead of the major one (black plus sign).
The match is indeed much better for NGC~6397. The difference is not significant for NGC~3201, whose black hole population strongly dominates the mass excess in the centre (see Figure~\ref{fig: core-collapse}), and tends to eject other less massive remnants through dynamical interactions. We also provide, as online material, the comparison of cumulative mass profiles from our \mpo\ fits of a dark sub-cluster and from their respective stellar remnant counterpart in our \cmc\ models (Figure~\ref{fig: mass-prof-cuo}).

All in all, our Monte Carlo simulations provide a remarkable agreement with our \mpo\ fits, and argue in favour of a central extended mass excess in both clusters, instead of a single central IMBH (or even an IMBH plus a CUO). While in NGC~6397, the CUO is likely to be formed by [ONeMg] white dwarfs with a still important contribution from [CO] white dwarfs (although these are more extended),
NGC~3201's CUO is mostly formed by stellar-mass black holes. 

\section{Discussion \& conclusions} \label{sec: discussion}

\subsection{Comparison to the analysis of Vitral \& Mamon (2021)} 

We now compare our \mpo\ results for NGC~6397 to those that were previously found by VM21.
In this comparison, we take into account the different adopted distances to NGC~6397: 2.39 kpc by VM21 and 2.48 kpc now.
Both physical size and plane-of-sky velocities scale as distance, for given observations in angular distance -- proper motion phase space. Since the mass at given radius scales approximately as radius times squared plane-of-sky velocity dispersion, it then scales as the cube of the distance. Therefore, one may expect that the new masses are  $(2.48/2.39)^3=1.12$ greater than the old ones.\footnote{Indeed, we ran extra fits assuming a lower distance to check that this scaling factor is applicable, within the error bars.}

The total mass and stellar effective radius of NGC~6397 are similar to what was previously found by VM21.
Indeed, VM21 found a Main Sequence mass of $1.16\times 10^5\,\msun$ for their 
best (lowest AICc) model (with two Main Sequence components) and 8 per cent lower for their single Main sequence best-fit model (again with a CUO).
We now find a Main Sequence mass of $1.13\times 10^5\,\msun$, which would be $1.01\times10^5\,\msun$ had we adopted the same distance to NGC~6397 as VM21, which is 6 per cent lower than the single Main Sequence mass of VM21.
The cluster effective radius has decreased from 5.31 arcmin in VM21 (for their single mass model) to 4.91 arcmin now, i.e. 8 per cent lower.

On the other hand, the CUO properties of NGC~6397 have changed appreciably in comparison  to those previously found by VM21. Previously, we  had found a mass of $1720^{+280}_{-670}\,\msun$ for the best two-component Main Sequence model, while we now find $807^{+123}_{-323}\,\msun$, i.e. 2.1 times lower, or 2.4 times lower when factoring in the different adopted distance.
The CUO radius of slope $-2$ was $0.11^{+0.02}_{-0.04}\,\rm arcmin$ in VM21, which corresponds to an effective radius (of half projected mass) $\sqrt{3/2}$ higher, i.e. $R_e = 0.135^{+0.025}_{-0.005}\,\rm arcmin$. The new CUO effective radius is
$R_e = 0.041^{+0.007}_{-0.037}\,\rm pc = 0.057^{+0.010}_{-0.051}\,arcmin$, for the adopted distance to NGC~6397.
Thus, the new CUO radius is 2.4 times lower than the old one, and now has a poorly constrained low-end tail (see light blue shaded regions of the lower panels of Fig.~\ref{fig: corner-comp-partial} for $r_{\rm CUO}$).
Combining the new mass and radius, leads to a new mean CUO density that is roughly 6 times denser than that of VM21.

These differences in CUO properties are the consequences in differences in the new data sets from \gaia, and especially \hst, whose much more precise PMs compared to \gaia\ (Fig.~\ref{fig: err-mag-relation}) are used to probe the GC inner mass profile. Although we now use \hst\ PMs with longer baselines (up to nearly 10 years instead of 5.5), which should lead to smaller proper motion errors, we turn out having 3 times higher proper motion errors for the brighter stars ($17 < \rm F606W < 19$) compared to VM21 (compare the lower panel of Fig.~\ref{fig: err-mag-relation} with figure 3 of VM21). These higher errors are caused by the correspondingly higher estimated systematic errors, produced by our careful re-calibration. This explains why we are no longer able to set a secure lower limit to the CUO radius of NGC~6397.

Nevertheless, our conclusions on NGC~6397 are qualitatively similar to those of VM21: the orbits are close to isotropic and the mass excess is very significant, with good evidence for it being extended (but not as strong as found by VM21).

\subsection{Core-collapse: White dwarfs vs. Black holes}

The mass-orbit modelling analyses of VM21 and in the present paper detect and fit a small extended inner mass excess in the centre of NGC~6397. On the basis of the initial mass function and simple stellar evolution, VM21 argued that the inner mass excess may be dominated by stellar-mass black holes if these avoided merging and escaping from the momentum acquired by the anisotropic emission of gravitational waves. However, they did not consider the importance of black hole ejection from dynamical interactions, nor that a large black hole population is inconsistent with the core-collapsed structure of NGC~6397 \citep{Rui+21}.

As discussed in Sect.~\ref{sec: intro}, once formed, black holes mass segregate to the centres of their host clusters, creating a black hole subsystem. Once a central black hole subsystem forms (typically on $\lesssim100\,$Myr timescales), black hole--black hole binaries\footnote{Black hole binaries can form from the evolution of primordial massive stellar binaries in the cluster or through the three-body binary formation mechanism involving three single black holes \citep[e.g.,][]{Binney2008,Ivanova2010,Morscher2015}.} within this subsystem begin to undergo binary--single and binary--binary dynamical encounters with single black holes and other binary black holes, respectively, on $\mathcal{O}(\rm{Myr})$ timescales. On average, these dynamical encounters lead to hardening of the black hole binaries \citep[e.g.,][]{Heggie1975}. Conservation of energy requires that this dynamical hardening is accompanied by an increase in kinetic energy of the single and binary black holes involved: the single and binary black holes receive dynamical ``kicks.'' Inevitably, the consequence of many of these dynamical encounters is for nearly the entire black hole population (both singles and binaries) to be ejected from their host cluster \citep[e.g.,][]{Kulkarni1993,PZ2000,Morscher2015,Wang2016dragon,Askar2018,Kremer2020}.

Although the dynamical ejection of all black holes is expected to be the ultimate fate of all GCs, not all clusters have evolved sufficiently long to have reach this state. \cite{Kremer+20} showed that the initial size at given mass of a GC impacts considerably its dynamical and evolutionary timescales, with denser clusters evolving faster and thus ejecting their black holes faster. Furthermore, while present in a cluster, the dynamical activity of stellar-mass black holes pump energy into their host cluster's luminous stellar population, 
preventing cluster core collapse \citep{Merritt2004,Mackey2007,BreenHeggie2013,Askar2018,Kremer2019,Kremer+20,Weatherford+20}. This implies that non-core-collapsed clusters have yet to eject their full black hole population, while core-collapsed clusters should contain a negligible number of stellar-mass black holes. In the absence of black holes in the latter case, the inner regions of core-collapsed clusters are expected to be dominated by white dwarfs, the next most massive stellar population.\footnote{Of course, neutron stars likely have comparable masses to white dwarfs (or even slightly larger masses). However white dwarfs dominate overwhelmingly by number (see Figure~\ref{fig: core-collapse}), and thus are expected to dominate the central regions.}

Since NGC~6397 is a core-collapsed GC \citep{Djorgovski&King86}, 
its inner mass should be dominated by white dwarfs instead of by stellar-mass black holes.
This idea was recently confirmed by \cite{Kremer+21}, who found that the observed surface brightness and velocity dispersion profiles of NGC~6397 were effectively reproduced by core-collapsed {\tt CMC} models that included an inner population of hundreds of [ONeMg] white dwarfs (with mean mass of $1.3\,\msun$).
On the other hand, the dark central mass in NGC~3201 
is consistent with a population of roughly
100 stellar mass black holes, with a mean mass of $13\,\msun$ \citep{Kremer+18,Kremer2019}, which is consistent with the non core-collapsed status of this cluster.

To illustrate this trend, we plot in the upper panels of Figure~\ref{fig: core-collapse}, for each GC, the surface number density of 
Gaia EDR3 stars obtained
after subtraction of the constant surface number density of Milky Way field stars of the same range of absolute magnitudes, obtained by \balrogo\ fits.
%
We clearly observe that NGC~6397 is a much denser cluster, with a steep inner slope at roughly $10^{-1}$~pc, characteristic of a post core-collapse cluster. In contrast,  NGC~3201 has a visible uniform, lower density core, with no signs of core-collapse.

Given these different surface density profiles, it is not surprising that the compact object population in each cluster is 
considerably different, as seen in the lower panels of Figure~\ref{fig: core-collapse}. For NGC~3201, the best-fitting  \cmc\ snapshot shows a dominance of black holes: This is also in agreement with the recent analyses of \citet{Giesers+19}, who predicted a population of roughly 50 black holes at present in NGC~3201, and \citet{Aros+21}, who showed (see their figure~D1) that the binary fractions in NGC~3201 are more consistent with a sub-cluster of stellar-mass black holes than with a single IMBH. The black holes in NGC~3201 prevent other components from sinking further to the cluster's centre, thus delaying gravitational collapse.
On the other hand, the compact object population in NGC~6397 is nearly devoid of black holes, with a predominance of [CO] white dwarfs, and a main sub-cluster formed of [ONeMg] white dwarfs, followed by a sub-cluster of neutron stars roughly five times less massive.

\subsection{Astrophysical implications of black hole/white dwarf sub-clusters}

The presence of black hole sub-clusters and, at late times after their host clusters have undergone core-collapse, white dwarf sub-clusters, leads naturally to a number of interesting astrophysical implications. In both scenarios,  compact object binaries form through both three-body encounters \citep[e.g.,][]{Morscher2015} and binary exchange encounters. Once formed, these compact object binaries harden through subsequent dynamical encounters \citep[e.g.,][]{Heggie1975} until, ultimately, they either merge or are ejected from their host cluster after attaining a sufficiently large dynamical recoil kick. For black hole sub-clusters, this process yields black hole--black hole binary mergers which are detectable as gravitational sources by instruments such as LIGO/Virgo \citep{LIGO1,LVC2021}. A number of recent analyses have demonstrated that the black hole binary mergers that occur in typical dense star clusters occur at rates comparable to the local universe rates predicted from the latest LIGO/Virgo results \citep[e.g.,][]{Rodriguez+21b}. Furthermore, the dynamical processes operating in black hole sub-clusters enable the formation of black hole mergers with components in the proposed pair-instability mass gap \citep[e.g.,][]{Rodriguez2019,DiCarlo2020,Kremer2020,GerosaFishbach2021}, which may be difficult to produce through alternative formation channels. Black hole sub-clusters are expected to also lead to the formation of compact black hole--luminous star binaries \citep[e.g.,][]{Kremer2018xrb} similar to those detected in a number of MW GCs \citep[e.g.,][]{Strader2012,Giesers+19} as well as stellar-mass tidal disruption events \citep[e.g.,][]{Perets2016,Kremer2019b,Kremer2022}, which may be detectable as bright electromagnetic transients by both current \citep[e.g., Zwicky Transient Facility;][]{Bellm2019} and upcoming \citep[e.g., Vera Rubin Observatory;][]{LSST2009} all-sky surveys.

In the case of core-collapsed clusters like NGC~6397 that are expected to have ejected nearly all of their black holes and host instead a compact sub-cluster of white dwarfs, the formation of inspiralling white dwarf--white dwarf binaries is the natural outcome \citep[e.g.,][]{Kremer+21}. As they inspiral, these binaries may be detectable as millihertz gravitational-wave sources by instruments such as LISA \citep{LISA2017}. At merger, they may be detectable at decihertz frequencies by proposed instruments such as DECIGO \citep[e.g.,][]{DECIGO2020}. Depending on the uncertain details of white dwarf merger physics, these mergers may plausibly lead to Type Ia supernovae \citep[e.g.,][]{Webbink1984}, rejuvenated massive white dwarfs \citep[e.g.,][]{Schwab2021}, or, in the event of collapse, young neutron stars \citep[e.g.,][]{NomotoIben1985}. Neutron stars formed through the latter scenario may be observable in old GCs as young pulsars \citep[e.g.,][]{Boyles2011,Tauris2013} and may potentially be the source of fast radio bursts similar to FRB20200120E in a GC in M81 \citep{Bhardwaj2021,Kirsten2021,Kremer2021frb,Lu2021}.

\subsection{Summary and prospects}

We provide the first comparative analysis, based on both observations (using \gaia\ EDR3 and \hst\ proper motions) and simulations, between the sub-clustering of compact objects in a non core-collapse cluster (NGC 3201) and a classic core-collapse one (NGC 6397). After confirming previous detections from a clustered dark population in NGC~6397, we associate this signal to hundreds of massive white dwarfs, instead of 
stellar mass black holes. Furthermore, our analysis of NGC~3201
is the first to provide compelling evidence of a dark central component of $\sim1000$~M$_{\odot}$ from mass-orbit Jeans modelling in this cluster, and although our fits alone yield no more than mild evidence for a sub-cluster of stellar-mass black holes instead of a central IMBH, we use Monte Carlo $N$-body simulations to robustly constrain this mass as the former case, by finding very good matches between the structural parameters from this dark component in our fits and in the simulations.

With the promising horizons of black hole searches in the next decade (see \citealt{Greene+19b}), we can expect that proper motion measurements from ground-based telescopes such as the \textit{Extremely Large Telescope (ELT)} will provide even tighter constraints on the nature of central dark components in GCs \citep{Davies+21}, and help to properly distinguish between the imprints of IMBHs and sub-clusters of compact objects.  This will help to verify our current understand of the physics of GC evolution, whose implications extend from our grasp on black hole physics up to our knowledge of galaxy formation.

\section*{Acknowledgements}

We thank the anonymous referee for the constructive report, with many insightful comments that have helped us to improve the quality of our results and clarify some descriptions in the manuscript.
We also acknowledge Eugene Vasiliev for great help with the \agama\ software, which allowed us to construct our mock data and for providing an unpublished analysis of the impact of \gaia\ EDR3 systematics in the velocity dispersion profile of NGC~6397. We thank Sebastian Kamann as well, for useful exchanges concerning mass segregation.
\\
Eduardo Vitral was funded by an AMX doctoral grant from \'Ecole Polytechnique.
Kyle Kremer is supported by an NSF Astronomy and Astrophysics Postdoctoral Fellowship under award AST-2001751. 
\\
Support for this work was provided by a grant for \hst\ program 13297 provided by the Space Telescope Science Institute, which is operated by AURA, Inc., under NASA contract NAS 5-26555. This work has made use of data from the European Space Agency (ESA) mission \gaia\ (\url{https://www.cosmos.esa.int/gaia}), processed by the \gaia\ Data Processing and Analysis Consortium (DPAC, \url{https://www.cosmos.esa.int/web/gaia/dpac/consortium}). Funding for the DPAC has been provided by national institutions, in particular the institutions participating in the \gaia\ Multilateral Agreement.
We greatly benefited from the public software {\sc Python} \citep{VanRossum09} packages 
{\sc BALRoGO} \citep{Vitral21},
{\sc Scipy} \citep{Jones+01},
{\sc Numpy} \citep{vanderWalt11} and
{\sc Matplotlib} \citep{Hunter07}. We also used the {\sc Spyder} Integrated Development Environment \citep{raybaut2009spyder}.


\section*{Data Availability}

The data that support the plots within this paper and other findings of this study are available from the corresponding author upon reasonable request.



\bibliographystyle{mnras}
\bibliography{src} 




\appendix

\section{\mpo\ outcome}


\begin{table}
\caption{Main results of the \mpo\ mass-modelling fits of NGC~3201 and NGC~6397.}
\label{tab: results-full}
\centering
\renewcommand{\arraystretch}{1.7}
\tabcolsep=2.0pt
\footnotesize
\begin{tabular}{ll@{\hspace{2mm}}crrrrr}
\hline\hline             
\multicolumn{1}{c}{ID} &
\multicolumn{1}{c}{$r_{\rm GC}$} &
\multicolumn{1}{c}{$n_{\rm GC}$} &
\multicolumn{1}{c}{$M_{\rm GC}$} &
\multicolumn{1}{c}{$r_{\rm CUO}$} &
\multicolumn{1}{c}{$M_{\rm CUO}$} \\
\multicolumn{1}{c}{[NGC]} &
\multicolumn{1}{c}{[pc]} & 
\multicolumn{1}{c}{} & 
\multicolumn{1}{c}{[$10^{5} $ M$_\odot$]} & 
\multicolumn{1}{c}{[pc]} & 
\multicolumn{1}{c}{[M$_\odot$]} \\ 
\multicolumn{1}{c}{(1)} &
\multicolumn{1}{c}{(2)} &
\multicolumn{1}{c}{(3)} &
\multicolumn{1}{c}{(4)} &
\multicolumn{1}{c}{(5)} & 
\multicolumn{1}{c}{(6)} \\ 
\hline
3201 & $  5.378^{+ 0.130}_{- 0.453} $ & $  1.03^{+ 0.02}_{- 0.06} $ & $  1.56^{+ 0.05}_{- 0.10} $ & $  0.153^{+ 0.087}_{- 0.149} $ & $  1217^{+ 254}_{- 1046} $ \\
6397 & $  3.544^{+ 0.116}_{- 0.213} $ & $  3.27^{+ 0.05}_{- 0.06} $ & $  1.13^{+ 0.03}_{- 0.04} $ & $  0.041^{+ 0.007}_{- 0.037} $ & $  807^{+ 123}_{- 323} $ \\
\multicolumn{1}{c}{\vdots} \\
\hline
\end{tabular}
\parbox{\hsize}{\textit{Notes}: Columns are 
(1) Cluster ID; 
(2) S\'ersic projected half mass radius $R_{\rm e}$, in pc, of the mass density profile of the globular cluster;
(3) S\'ersic index $n$ of the mass density profile of the globular cluster;
(4) Total globular cluster mass (without dark central component), in M$_{\odot}$;
(5) Plummer projected half mass radius $a_{\rm P}$, in pc, of the mass density profile of the central sub-cluster of unresolved objects (CUO);
(6) Total mass of the CUO, in M$_{\odot}$;
The uncertainties are from the 16th and 84th percentiles of the marginal distributions.
The complete version of this table with all \mpo\ results and AICc diagnostics is provided as online material.}
\end{table}


\begin{table*}
\caption{Main results of the \mpo\ mass-modelling fit of NGC~3201 and NGC~6397.
}
\label{tab: results-full2}
\centering
\renewcommand{\arraystretch}{1.7}
\tabcolsep=2.7pt
\footnotesize
\begin{tabular}{ll@{\hspace{2mm}}ccrrrrrrrrr}
\hline\hline             
\multicolumn{1}{c}{Model} &
\multicolumn{1}{c}{Cluster ID} &
\multicolumn{1}{c}{Test} &
\multicolumn{1}{c}{$R^{-1}$} & 
\multicolumn{1}{c}{$\beta_{0}$} &
\multicolumn{1}{c}{$\beta_{\rm out}$} &
\multicolumn{1}{c}{$r_{\rm GC}$} &
\multicolumn{1}{c}{$n_{\rm GC}$} &
\multicolumn{1}{c}{$M_{\rm GC}$} &
\multicolumn{1}{c}{$r_{\rm CUO}$} &
\multicolumn{1}{c}{$M_{\rm CUO}$} &
\multicolumn{1}{c}{$M_{\rm BH}$} &
\multicolumn{1}{c}{$\Delta \rm AICc$} \\
\multicolumn{1}{c}{} &
\multicolumn{1}{c}{} &
\multicolumn{1}{c}{} &
\multicolumn{1}{c}{} &
\multicolumn{1}{c}{} &
\multicolumn{1}{c}{} &
\multicolumn{1}{c}{[pc]} & 
\multicolumn{1}{c}{} & 
\multicolumn{1}{c}{[$10^{5} $ M$_\odot$]} & 
\multicolumn{1}{c}{[pc]} & 
\multicolumn{1}{c}{[M$_\odot$]} &
\multicolumn{1}{c}{[M$_\odot$]} &
\multicolumn{1}{c}{} \\ 
\multicolumn{1}{c}{(1)} &
\multicolumn{1}{c}{(2)} &
\multicolumn{1}{c}{(3)} &
\multicolumn{1}{c}{(4)} &
\multicolumn{1}{c}{(5)} & 
\multicolumn{1}{c}{(6)} & 
\multicolumn{1}{c}{(7)} & 
\multicolumn{1}{c}{(8)} & 
\multicolumn{1}{c}{(9)} & 
\multicolumn{1}{c}{(10)} & 
\multicolumn{1}{c}{(11)} &
\multicolumn{1}{c}{(12)} & 
\multicolumn{1}{c}{(13)} \\ 
\hline
1 & NGC~3201 & \multicolumn{1}{c}{--} & $  0.002 $ & \multicolumn{1}{c}{\darkg{\bf 0 }} & \multicolumn{1}{c}{\darkg{\bf 0 }} & $  5.070^{+ 0.233}_{- 0.273} $ & $  0.99^{+ 0.04}_{- 0.03} $ & $  1.55^{+ 0.07}_{- 0.08} $ & \multicolumn{1}{c}{--} & \multicolumn{1}{c}{--} & \multicolumn{1}{c}{--} &  0.00 \\
2 & NGC~3201 & \multicolumn{1}{c}{--} & $  0.007 $ & \multicolumn{1}{c}{\darkg{\bf 0 }} & \multicolumn{1}{c}{\darkg{\bf 0 }} & $  5.269^{+ 0.254}_{- 0.327} $ & $  1.01^{+ 0.04}_{- 0.04} $ & $  1.53^{+ 0.09}_{- 0.06} $ & \multicolumn{1}{c}{--} & \multicolumn{1}{c}{--} & $  1144^{+ 202}_{- 935} $ & -3.53 \\
\rowcolor{lavender} 3 & NGC~3201 & \multicolumn{1}{c}{--} & $  0.004 $ & \multicolumn{1}{c}{\darkg{\bf 0 }} & \multicolumn{1}{c}{\darkg{\bf 0 }} & $  5.378^{+ 0.130}_{- 0.453} $ & $  1.03^{+ 0.02}_{- 0.06} $ & $  1.56^{+ 0.05}_{- 0.10} $ & $  0.153^{+ 0.087}_{- 0.149} $ & $  1217^{+ 254}_{- 1046} $ & \multicolumn{1}{c}{--} & -1.65 \\
4 & NGC~3201 & \multicolumn{1}{c}{--} & $  0.005 $ & \multicolumn{1}{c}{\darkg{\bf 0 }} & \multicolumn{1}{c}{\darkg{\bf 0 }} & $  5.283^{+ 0.258}_{- 0.316} $ & $  1.01^{+ 0.04}_{- 0.04} $ & $  1.53^{+ 0.08}_{- 0.08} $ & $  0.122^{+ 0.265}_{- 0.118} $ & $  342^{+ 697}_{- 310} $ & $  825^{+ 230}_{- 783} $ &  0.35 \\
5 & NGC~3201 & $\beta(r)$ & $  0.002 $ & $  0.03^{+ 0.07}_{- 0.06} $ & $  0.02^{+ 0.13}_{- 0.20} $ & $  5.119^{+ 0.285}_{- 0.307} $ & $  1.00^{+ 0.04}_{- 0.03} $ & $  1.56^{+ 0.08}_{- 0.08} $ & \multicolumn{1}{c}{--} & \multicolumn{1}{c}{--} & \multicolumn{1}{c}{--} &  3.65 \\
6 & NGC~3201 & $\beta(r)$ & $  0.005 $ & $ -0.07^{+ 0.11}_{- 0.06} $ & $  0.07^{+ 0.16}_{- 0.18} $ & $  5.341^{+ 0.284}_{- 0.375} $ & $  1.03^{+ 0.03}_{- 0.05} $ & $  1.54^{+ 0.08}_{- 0.08} $ & \multicolumn{1}{c}{--} & \multicolumn{1}{c}{--} & $  1273^{+ 378}_{- 991} $ & -0.07 \\
7 & NGC~3201 & $\beta(r)$ & $  0.006 $ & $ -0.02^{+ 0.07}_{- 0.10} $ & $  0.04^{+ 0.19}_{- 0.15} $ & $  5.376^{+ 0.236}_{- 0.410} $ & $  1.01^{+ 0.04}_{- 0.03} $ & $  1.55^{+ 0.06}_{- 0.09} $ & $  0.107^{+ 0.111}_{- 0.103} $ & $  1289^{+ 478}_{- 1039} $ & \multicolumn{1}{c}{--} &  1.76 \\
8 & NGC~3201 & $\beta(r)$ & $  0.012 $ & $ -0.04^{+ 0.07}_{- 0.10} $ & $  0.05^{+ 0.18}_{- 0.15} $ & $  5.278^{+ 0.354}_{- 0.288} $ & $  1.01^{+ 0.04}_{- 0.04} $ & $  1.50^{+ 0.11}_{- 0.04} $ & $  0.089^{+ 0.312}_{- 0.084} $ & $ \darkor{ 1367^{+ 0}_{- 1334}} $ & $  138^{+ 1156}_{- 98} $ &  3.85  \\
9 & NGC~3201 & $(\alpha_{0}, \, \delta_{0})$ & $  0.002 $ & \multicolumn{1}{c}{\darkg{\bf 0 }} & \multicolumn{1}{c}{\darkg{\bf 0 }} & $  5.026^{+ 0.271}_{- 0.237} $ & $  1.00^{+ 0.04}_{- 0.04} $ & $  1.54^{+ 0.08}_{- 0.07} $ & \multicolumn{1}{c}{--} & \multicolumn{1}{c}{--} & \multicolumn{1}{c}{--} &  87009.01 \\
10 & NGC~3201 & $(\alpha_{0}, \, \delta_{0})$ & $  0.004 $ & \multicolumn{1}{c}{\darkg{\bf 0 }} & \multicolumn{1}{c}{\darkg{\bf 0 }} & $  5.224^{+ 0.293}_{- 0.298} $ & $  1.01^{+ 0.04}_{- 0.03} $ & $  1.53^{+ 0.09}_{- 0.07} $ & \multicolumn{1}{c}{--} & \multicolumn{1}{c}{--} & $  992^{+ 385}_{- 797} $ &  87005.52 \\
11 & NGC~3201 & $(\alpha_{0}, \, \delta_{0})$ & $  0.003 $ & \multicolumn{1}{c}{\darkg{\bf 0 }} & \multicolumn{1}{c}{\darkg{\bf 0 }} & $  5.192^{+ 0.318}_{- 0.264} $ & $  1.02^{+ 0.03}_{- 0.05} $ & $  1.51^{+ 0.10}_{- 0.05} $ & $  0.091^{+ 0.150}_{- 0.086} $ & $  1081^{+ 386}_{- 905} $ & \multicolumn{1}{c}{--} &  87007.14 \\
12 & NGC~3201 & $(\alpha_{0}, \, \delta_{0})$ & $  0.009 $ & \multicolumn{1}{c}{\darkg{\bf 0 }} & \multicolumn{1}{c}{\darkg{\bf 0 }} & $  5.291^{+ 0.264}_{- 0.325} $ & $  1.01^{+ 0.05}_{- 0.03} $ & $  1.54^{+ 0.08}_{- 0.08} $ & $  0.134^{+ 0.268}_{- 0.129} $ & $  1021^{+ 21}_{- 990} $ & $  59^{+ 1011}_{- 18} $ &  87009.26 \\
13 & NGC~3201 & $\sigma_{\mu}$ & $  0.002 $ & \multicolumn{1}{c}{\darkg{\bf 0 }} & \multicolumn{1}{c}{\darkg{\bf 0 }} & $  4.987^{+ 0.463}_{- 0.305} $ & $  1.00^{+ 0.05}_{- 0.04} $ & $  1.42^{+ 0.14}_{- 0.09} $ & \multicolumn{1}{c}{--} & \multicolumn{1}{c}{--} & \multicolumn{1}{c}{--} &  36566.09 \\
14 & NGC~3201 & $\sigma_{\mu}$ & $  0.005 $ & \multicolumn{1}{c}{\darkg{\bf 0 }} & \multicolumn{1}{c}{\darkg{\bf 0 }} & $  5.029^{+ 0.455}_{- 0.320} $ & $  1.01^{+ 0.05}_{- 0.05} $ & $  1.41^{+ 0.15}_{- 0.08} $ & \multicolumn{1}{c}{--} & \multicolumn{1}{c}{--} & $  320^{+ 125}_{- 298} $ &  36567.18 \\
15 & NGC~3201 & $\sigma_{\mu}$ & $  0.001 $ & \multicolumn{1}{c}{\darkg{\bf 0 }} & \multicolumn{1}{c}{\darkg{\bf 0 }} & $  5.065^{+ 0.416}_{- 0.368} $ & $  1.02^{+ 0.04}_{- 0.05} $ & $  1.42^{+ 0.13}_{- 0.10} $ & $  0.016^{+ 0.522}_{- 0.010} $ & $  323^{+ 211}_{- 298} $ & \multicolumn{1}{c}{--} &  36569.16 \\
16 & NGC~3201 & $\sigma_{\mu}$ & $  0.004 $ & \multicolumn{1}{c}{\darkg{\bf 0 }} & \multicolumn{1}{c}{\darkg{\bf 0 }} & $  5.081^{+ 0.426}_{- 0.350} $ & $  1.03^{+ 0.04}_{- 0.05} $ & $  1.42^{+ 0.13}_{- 0.10} $ & $  0.116^{+ 0.498}_{- 0.110} $ & $  363^{+ 67}_{- 342} $ & $ \darkor{ 15^{+ 351}_{- 0}} $ &  36571.14 \\
17 & NGC~3201 & Bulk $\mu$ & $  0.002 $ & \multicolumn{1}{c}{\darkg{\bf 0 }} & \multicolumn{1}{c}{\darkg{\bf 0 }} & $  5.022^{+ 0.277}_{- 0.231} $ & $  1.00^{+ 0.04}_{- 0.03} $ & $  1.54^{+ 0.08}_{- 0.07} $ & \multicolumn{1}{c}{--} & \multicolumn{1}{c}{--} & \multicolumn{1}{c}{--} &  87210.35 \\
18 & NGC~3201 & Bulk $\mu$ & $  0.004 $ & \multicolumn{1}{c}{\darkg{\bf 0 }} & \multicolumn{1}{c}{\darkg{\bf 0 }} & $  5.290^{+ 0.206}_{- 0.368} $ & $  1.02^{+ 0.03}_{- 0.04} $ & $  1.54^{+ 0.08}_{- 0.07} $ & \multicolumn{1}{c}{--} & \multicolumn{1}{c}{--} & $  1055^{+ 236}_{- 892} $ &  87207.12 \\
19 & NGC~3201 & Bulk $\mu$ & $  0.003 $ & \multicolumn{1}{c}{\darkg{\bf 0 }} & \multicolumn{1}{c}{\darkg{\bf 0 }} & $  5.370^{+ 0.131}_{- 0.440} $ & $  1.03^{+ 0.03}_{- 0.05} $ & $  1.56^{+ 0.06}_{- 0.09} $ & $  0.131^{+ 0.131}_{- 0.126} $ & $  1242^{+ 191}_{- 1082} $ & \multicolumn{1}{c}{--} &  87208.94 \\
20 & NGC~3201 & Bulk $\mu$ & $  0.008 $ & \multicolumn{1}{c}{\darkg{\bf 0 }} & \multicolumn{1}{c}{\darkg{\bf 0 }} & $  5.245^{+ 0.278}_{- 0.295} $ & $  1.00^{+ 0.05}_{- 0.03} $ & $  1.53^{+ 0.08}_{- 0.07} $ & $  0.118^{+ 0.286}_{- 0.113} $ & $  985^{+ 58}_{- 952} $ & $  133^{+ 825}_{- 99} $ &  87210.98 \\
\hline
\end{tabular}
\parbox{\hsize}{\textit{Notes}: Columns are 
(1) Model number; 
(2) Cluster ID; 
(3) Test type: "$\beta(r)$" for a free anisotropy model, "$(\alpha_{0}, \, \delta_{0})$" for the test of a different centre (\protect\citealt{Goldsbury+10}), "$\sigma_{\mu}$" for the test with half of the standard error threshold and "Bulk $\mu$" for the test setting the \hst\ bulk proper motion as the one from \protect\cite{Vasiliev&Baumgardt&Baumgardt21}.
(4) MCMC convergence criterion ($R^{-1} \leq 0.02$ is considered as properly converged);
(5) anisotropy value at $r=0$;
(6) anisotropy value at the data's most distant projected radius (usually around 10 arcmin);
(7) S\'ersic projected half mass radius $R_{\rm e}$ (in pc) of the mass density profile of the globular cluster;
(8) S\'ersic index $n$ of the mass density profile of the globular cluster;
(9) Total globular cluster mass (without dark central component), in M$_{\odot}$;
(10) Plummer projected half mass radius $a_{\rm P}$ (in pc) of the mass density profile of the central sub-cluster of unresolved objects (CUO);
(11) Total mass of the CUO, in M$_{\odot}$;
(12) Central black hole mass, in M$_{\odot}$;
(13) Difference in AICc (eq.~[\ref{eq: AIC}]) relative to model 1 for NGC~3201 and to model 23, for NGC~6397.
We highlight the maximum likelihood values in \emph{orange} when they were outside the 16-84 percentiles of the posterior distribution, and the convergence criterion $R^{-1}$ in \emph{red} when the MCMC convergence was poor.
The uncertainties are from the 16th and 84th percentiles of the marginal distributions.
The lines coloured in \textit{lavender} indicate our preferred models for each cluster. We did not consider the AICc diagnosis when the data set was different from the respective standard model.}
\end{table*}

\begin{table*}
\contcaption{}
\centering
\renewcommand{\arraystretch}{1.7}
\tabcolsep=2.7pt
\footnotesize
\begin{tabular}{ll@{\hspace{2mm}}ccrrrrrrrrr}
\hline\hline             
\multicolumn{1}{c}{Model} &
\multicolumn{1}{c}{Cluster ID} &
\multicolumn{1}{c}{Test} &
\multicolumn{1}{c}{$R^{-1}$} & 
\multicolumn{1}{c}{$\beta_{0}$} &
\multicolumn{1}{c}{$\beta_{\rm out}$} &
\multicolumn{1}{c}{$r_{\rm GC}$} &
\multicolumn{1}{c}{$n_{\rm GC}$} &
\multicolumn{1}{c}{$M_{\rm GC}$} &
\multicolumn{1}{c}{$r_{\rm CUO}$} &
\multicolumn{1}{c}{$M_{\rm CUO}$} &
\multicolumn{1}{c}{$M_{\rm BH}$} &
\multicolumn{1}{c}{$\Delta \rm AICc$} \\
\multicolumn{1}{c}{} &
\multicolumn{1}{c}{} &
\multicolumn{1}{c}{} &
\multicolumn{1}{c}{} &
\multicolumn{1}{c}{\white{$ -0.01^{+ 0.05}_{- 0.05} $}} &
\multicolumn{1}{c}{\white{$ -0.01^{+ 0.05}_{- 0.05} $}} &
\multicolumn{1}{c}{[pc]} & 
\multicolumn{1}{c}{} & 
\multicolumn{1}{c}{[$10^{5} $ M$_\odot$]} & 
\multicolumn{1}{c}{[pc]} & 
\multicolumn{1}{c}{[M$_\odot$]} &
\multicolumn{1}{c}{[M$_\odot$]} &
\multicolumn{1}{c}{} \\ 
\multicolumn{1}{c}{(1)} &
\multicolumn{1}{c}{(2)} &
\multicolumn{1}{c}{(3)} &
\multicolumn{1}{c}{(4)} &
\multicolumn{1}{c}{(5)} & 
\multicolumn{1}{c}{(6)} & 
\multicolumn{1}{c}{(7)} & 
\multicolumn{1}{c}{(8)} & 
\multicolumn{1}{c}{(9)} & 
\multicolumn{1}{c}{(10)} & 
\multicolumn{1}{c}{(11)} &
\multicolumn{1}{c}{(12)} & 
\multicolumn{1}{c}{(13)} \\ 
\hline
29 & NGC~6397 & \multicolumn{1}{c}{--} & $  0.008 $ & \multicolumn{1}{c}{\darkg{\bf 0 }} & \multicolumn{1}{c}{\darkg{\bf 0 }} & $  3.150^{+ 0.514}_{- 0.097} $ & $  3.32^{+ 0.60}_{- 0.04} $ & $  1.08^{+ 0.07}_{- 0.02} $ & \multicolumn{1}{c}{--} & \multicolumn{1}{c}{--} & \multicolumn{1}{c}{--} &  0.00 \\
30 & NGC~6397 & \multicolumn{1}{c}{--} & $  0.005 $ & \multicolumn{1}{c}{\darkg{\bf 0 }} & \multicolumn{1}{c}{\darkg{\bf 0 }} & $  3.461^{+ 0.133}_{- 0.169} $ & $  3.27^{+ 0.05}_{- 0.06} $ & $  1.12^{+ 0.03}_{- 0.03} $ & \multicolumn{1}{c}{--} & \multicolumn{1}{c}{--} & $  578^{+ 136}_{- 174} $ & -22.98 \\
\rowcolor{lavender} 31 & NGC~6397 & \multicolumn{1}{c}{--} & $  0.005 $ & \multicolumn{1}{c}{\darkg{\bf 0 }} & \multicolumn{1}{c}{\darkg{\bf 0 }} & $  3.544^{+ 0.116}_{- 0.213} $ & $  3.27^{+ 0.05}_{- 0.06} $ & $  1.13^{+ 0.03}_{- 0.04} $ & $  0.041^{+ 0.007}_{- 0.037} $ & $  807^{+ 123}_{- 323} $ & \multicolumn{1}{c}{--} & -24.06 \\
32 & NGC~6397 & \multicolumn{1}{c}{--} & $  0.017 $ & \multicolumn{1}{c}{\darkg{\bf 0 }} & \multicolumn{1}{c}{\darkg{\bf 0 }} & $  3.564^{+ 0.073}_{- 0.247} $ & $  3.29^{+ 0.03}_{- 0.07} $ & $  1.13^{+ 0.02}_{- 0.04} $ & $  0.034^{+ 0.064}_{- 0.031} $ & $ \darkor{ 749^{+ 0}_{- 703}} $ & $ \darkor{ 21^{+ 518}_{- 0}} $ & -21.91 \\
33 & NGC~6397 & $\beta(r)$ & $  0.012 $ & $  0.06^{+ 0.03}_{- 0.06} $ & $ -0.01^{+ 0.07}_{- 0.05} $ & $  3.197^{+ 0.397}_{- 0.124} $ & $  3.32^{+ 0.50}_{- 0.05} $ & $  1.09^{+ 0.05}_{- 0.03} $ & \multicolumn{1}{c}{--} & \multicolumn{1}{c}{--} & \multicolumn{1}{c}{--} &  2.32 \\
34 & NGC~6397 & $\beta(r)$ & $  0.004 $ & $  0.00^{+ 0.04}_{- 0.05} $ & $  0.08^{+ 0.03}_{- 0.09} $ & $  3.488^{+ 0.148}_{- 0.171} $ & $  3.28^{+ 0.05}_{- 0.06} $ & $  1.12^{+ 0.03}_{- 0.03} $ & \multicolumn{1}{c}{--} & \multicolumn{1}{c}{--} & $  598^{+ 146}_{- 183} $ & -19.27 \\
35 & NGC~6397 & $\beta(r)$ & $  0.010 $ & $ -0.02^{+ 0.04}_{- 0.05} $ & $  0.07^{+ 0.05}_{- 0.07} $ & $  3.538^{+ 0.162}_{- 0.179} $ & $  3.29^{+ 0.03}_{- 0.08} $ & $  1.12^{+ 0.03}_{- 0.03} $ & $  0.036^{+ 0.012}_{- 0.032} $ & $  711^{+ 275}_{- 201} $ & \multicolumn{1}{c}{--} & -20.57 \\
36 & NGC~6397 & $\beta(r)$ & $  0.016 $ & $ -0.01^{+ 0.05}_{- 0.05} $ & $  0.07^{+ 0.04}_{- 0.08} $ & $  3.202^{+ 0.083}_{- 0.236} $ & $  2.18^{+ 0.11}_{- 0.02} $ & $  1.07^{+ 0.01}_{- 0.04} $ & $  0.084^{+ 0.071}_{- 0.031} $ & $  1924^{+ 301}_{- 853} $ & $  28^{+ 462}_{- 6} $ & -14.97 \\
37 & NGC~6397 & $(\alpha_{0}, \, \delta_{0})$ & $  0.004 $ & \multicolumn{1}{c}{\darkg{\bf 0 }} & \multicolumn{1}{c}{\darkg{\bf 0 }} & $  3.151^{+ 0.517}_{- 0.098} $ & $  3.31^{+ 0.61}_{- 0.03} $ & $  1.08^{+ 0.07}_{- 0.02} $ & \multicolumn{1}{c}{--} & \multicolumn{1}{c}{--} & \multicolumn{1}{c}{--} &  144066.39 \\
38 & NGC~6397 & $(\alpha_{0}, \, \delta_{0})$ & $  0.008 $ & \multicolumn{1}{c}{\darkg{\bf 0 }} & \multicolumn{1}{c}{\darkg{\bf 0 }} & $  3.474^{+ 0.125}_{- 0.180} $ & $  3.27^{+ 0.05}_{- 0.06} $ & $  1.12^{+ 0.03}_{- 0.03} $ & \multicolumn{1}{c}{--} & \multicolumn{1}{c}{--} & $  590^{+ 127}_{- 183} $ &  144044.21 \\
39  & NGC~6397 & $(\alpha_{0}, \, \delta_{0})$ & $  0.010 $ & \multicolumn{1}{c}{\darkg{\bf 0 }} & \multicolumn{1}{c}{\darkg{\bf 0 }} & $  3.525^{+ 0.141}_{- 0.188} $ & $  3.27^{+ 0.05}_{- 0.06} $ & $  1.12^{+ 0.03}_{- 0.03} $ & $  0.033^{+ 0.017}_{- 0.029} $ & $  780^{+ 168}_{- 290} $ &\multicolumn{1}{c}{--} &  144042.75 \\
40  & NGC~6397 & $(\alpha_{0}, \, \delta_{0})$ & $  0.013 $ & \multicolumn{1}{c}{\darkg{\bf 0 }} & \multicolumn{1}{c}{\darkg{\bf 0 }} & $  3.540^{+ 0.105}_{- 0.217} $ & $  3.28^{+ 0.04}_{- 0.07} $ & $  1.13^{+ 0.03}_{- 0.04} $ & $  0.040^{+ 0.058}_{- 0.037} $ & $ \darkor{ 759^{+ 0}_{- 713}} $ & $  42^{+ 506}_{- 13} $ &  144045.05 \\
41 & NGC~6397 & $\sigma_{\mu}$ & $  0.002 $ & \multicolumn{1}{c}{\darkg{\bf 0 }} & \multicolumn{1}{c}{\darkg{\bf 0 }} & $  3.342^{+ 0.302}_{- 0.210} $ & $  3.93^{+ 0.22}_{- 0.22} $ & $  1.05^{+ 0.06}_{- 0.04} $ & \multicolumn{1}{c}{--} & \multicolumn{1}{c}{--} & \multicolumn{1}{c}{--} &  84517.03 \\
42 & NGC~6397 & $\sigma_{\mu}$ & $  0.002 $ & \multicolumn{1}{c}{\darkg{\bf 0 }} & \multicolumn{1}{c}{\darkg{\bf 0 }} & $  3.235^{+ 0.270}_{- 0.141} $ & $  3.27^{+ 0.09}_{- 0.08} $ & $  1.05^{+ 0.06}_{- 0.03} $ & \multicolumn{1}{c}{--} & \multicolumn{1}{c}{--} & $  569^{+ 151}_{- 195} $ &  84499.35 \\
43 & NGC~6397 & $\sigma_{\mu}$ & $  0.005 $ & \multicolumn{1}{c}{\darkg{\bf 0 }} & \multicolumn{1}{c}{\darkg{\bf 0 }} & $  3.462^{+ 0.103}_{- 0.322} $ & $  3.26^{+ 0.08}_{- 0.08} $ & $  1.09^{+ 0.03}_{- 0.07} $ & $  0.045^{+ 0.008}_{- 0.041} $ & $  851^{+ 154}_{- 372} $ & \multicolumn{1}{c}{--} &  84497.79 \\
44 & NGC~6397 & $\sigma_{\mu}$ & $ \darkr{ 0.046} $ & \multicolumn{1}{c}{\darkg{\bf 0 }} & \multicolumn{1}{c}{\darkg{\bf 0 }} & $  3.420^{+ 0.133}_{- 0.312} $ & $  3.25^{+ 0.09}_{- 0.07} $ & $  1.07^{+ 0.04}_{- 0.05} $ & $  0.052^{+ 0.061}_{- 0.048} $ & $ \darkor{ 866^{+ 0}_{- 814}} $ & $  101^{+ 434}_{- 72} $ &  84500.11 \\
45 & NGC~6397 & Bulk $\mu$ & $  0.002 $ & \multicolumn{1}{c}{\darkg{\bf 0 }} & \multicolumn{1}{c}{\darkg{\bf 0 }} & $  3.119^{+ 0.538}_{- 0.069} $ & $  3.31^{+ 0.61}_{- 0.03} $ & $  1.07^{+ 0.08}_{- 0.02} $ & \multicolumn{1}{c}{--} & \multicolumn{1}{c}{--} & \multicolumn{1}{c}{--} &  144061.37 \\
46 & NGC~6397 & Bulk $\mu$ & $  0.001 $ & \multicolumn{1}{c}{\darkg{\bf 0 }} & \multicolumn{1}{c}{\darkg{\bf 0 }} & $  3.448^{+ 0.140}_{- 0.167} $ & $  3.28^{+ 0.04}_{- 0.07} $ & $  1.12^{+ 0.03}_{- 0.03} $ & \multicolumn{1}{c}{--} & \multicolumn{1}{c}{--} & $  522^{+ 186}_{- 124} $ &  144038.69 \\
47 & NGC~6397 & Bulk $\mu$ & $  0.008 $ & \multicolumn{1}{c}{\darkg{\bf 0 }} & \multicolumn{1}{c}{\darkg{\bf 0 }} & $  3.526^{+ 0.108}_{- 0.220} $ & $  3.27^{+ 0.04}_{- 0.06} $ & $  1.12^{+ 0.02}_{- 0.04} $ & $  0.037^{+ 0.010}_{- 0.034} $ & $  781^{+ 133}_{- 317} $ & \multicolumn{1}{c}{--} &  144036.01 \\
48 & NGC~6397 & Bulk $\mu$ & $  0.006 $ & \multicolumn{1}{c}{\darkg{\bf 0 }} & \multicolumn{1}{c}{\darkg{\bf 0 }} & $  3.502^{+ 0.137}_{- 0.187} $ & $  3.27^{+ 0.05}_{- 0.05} $ & $  1.13^{+ 0.02}_{- 0.04} $ & $  0.030^{+ 0.074}_{- 0.026} $ & $  665^{+ 53}_{- 620} $ & $ \darkor{ 14^{+ 530}_{- 0}} $ &  144039.97 \\
\hline
\hline
\end{tabular}
\end{table*}



\newpage

\section{Extra plots}

Plots available as online material.

\begin{figure*}
\centering
\includegraphics[width=0.45\hsize]{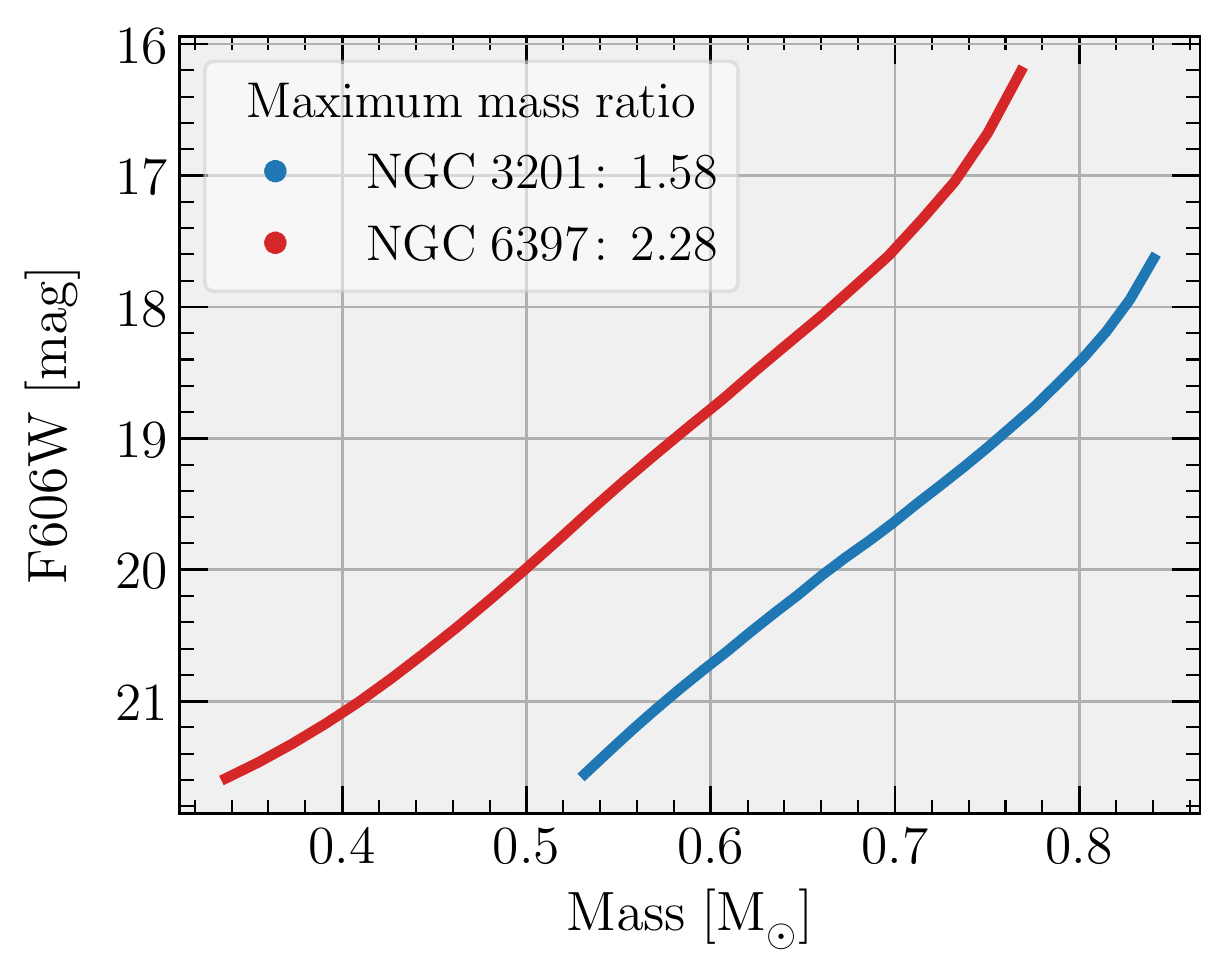}
\caption{\textit{Mass-magnitude:} Mass-magnitude (F606W) relation from \parsec\ isochrones, for NGC~3201 in \textit{blue} and for NGC~6397 in \textit{red}. The limits of the curves represent the respective limits of our cleaned data. The box indicates the ratio $m_{\rm max} / m_{\rm min}$ for the cleaned data.}
\label{fig: mass-mag-relation}
\end{figure*}

\begin{figure*}
\centering
\includegraphics[width=0.49\hsize]{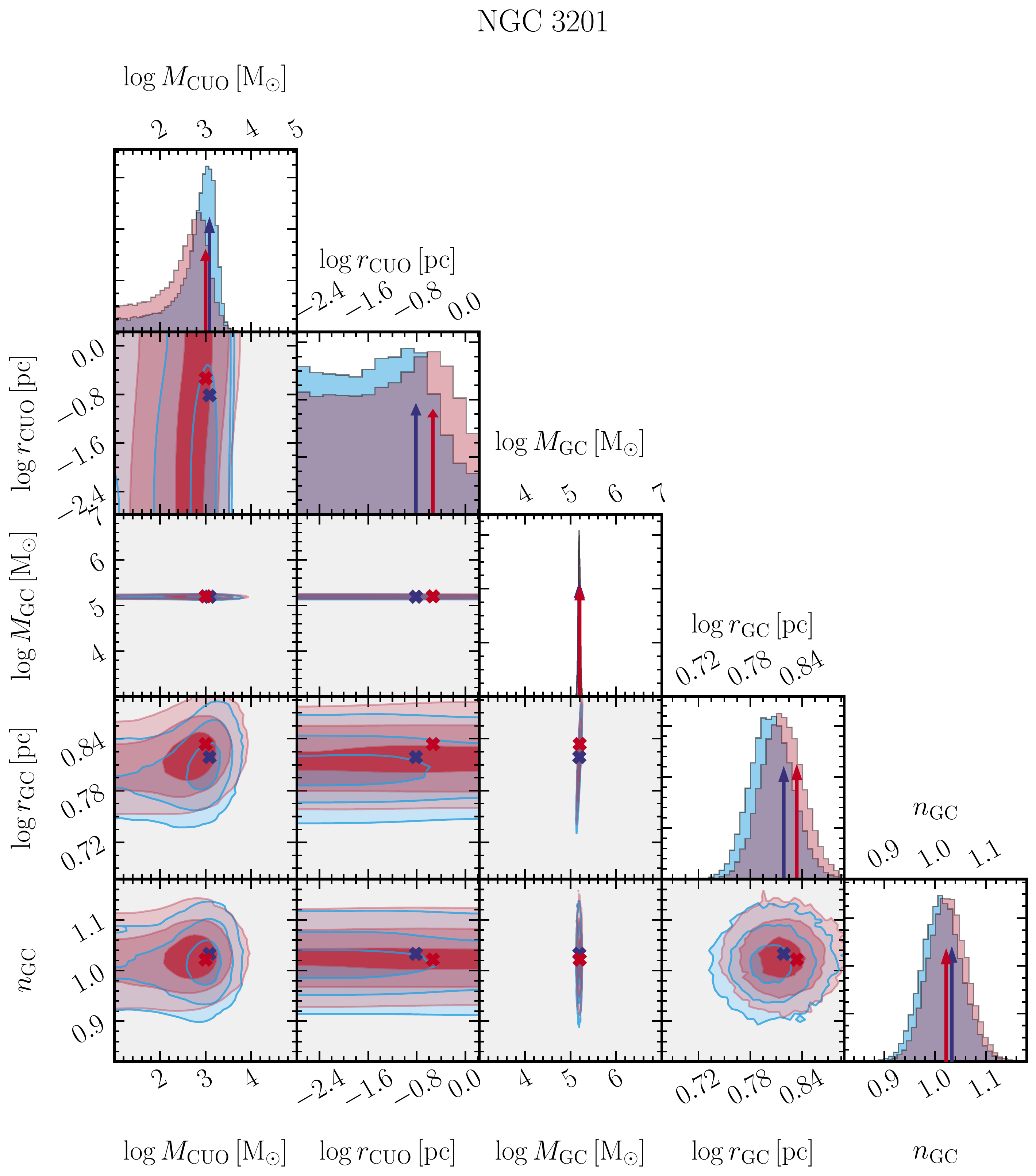}
\includegraphics[width=0.49\hsize]{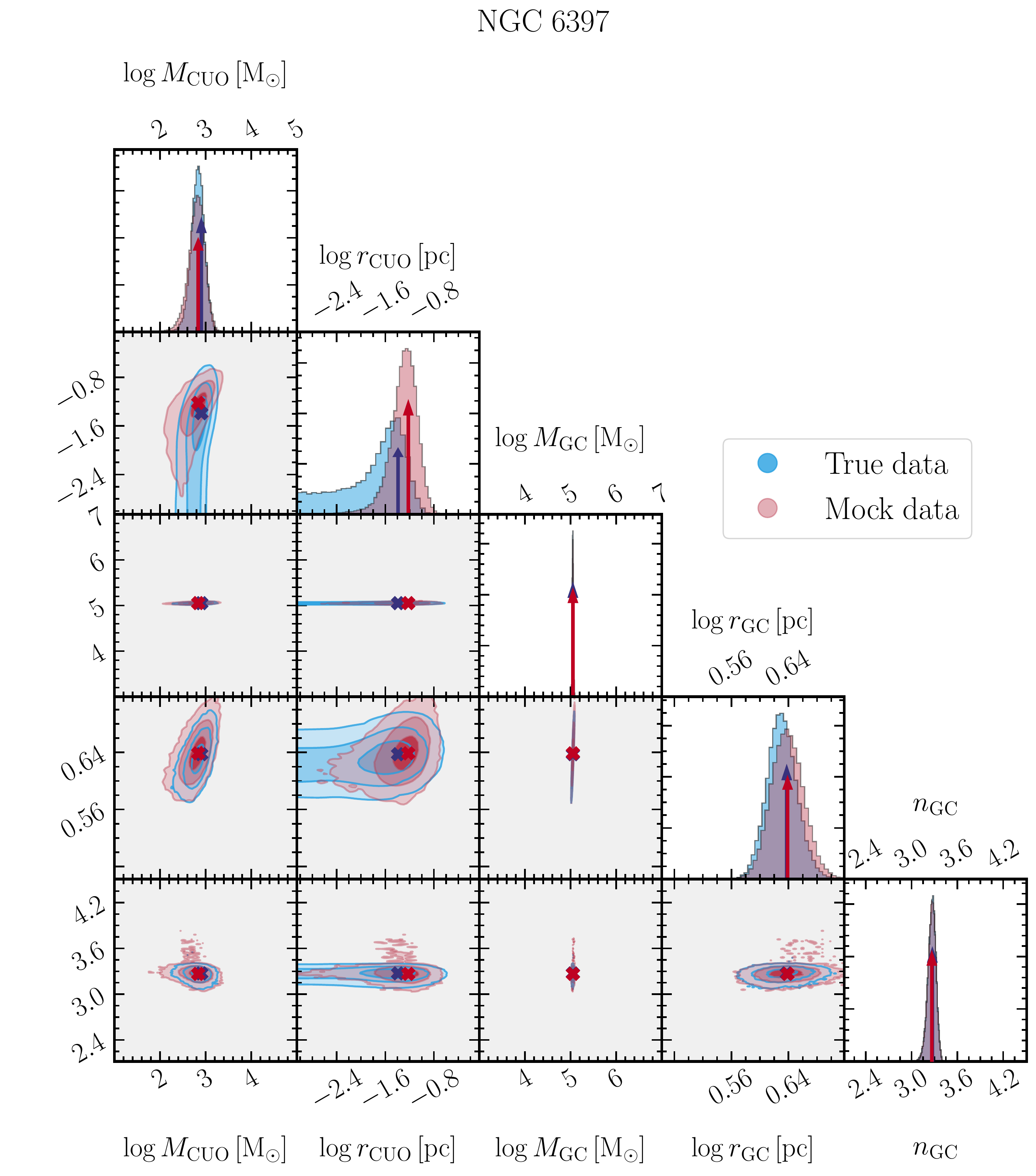}
\caption{{\it Corner plots:} \mpo\ corner plots of the mass modelling fit of NGC~3201 (\textbf{left}) and NGC~6397 (\textbf{right}). The free parameters are, from \textbf{top} to \textbf{bottom}:  mass 
of the sub-cluster of unresolved objects (CUO); Plummer 2D half-mass radius of the CUO; total stellar mass of  the cluster; half-projected mass radius $R_{\rm e}$ (in pc) of the GC stellar mass profile and the S\'ersic index $n$ of the GC stellar mass profile. 
Again, the priors are flat for $M_{\rm CUO}$ and $M_{\rm GC}$ within the plotted range and zero outside, while they are Gaussian for the two radii and S\'ersic index, centred on the middles of the panels and extending to $\pm3\,\sigma$ at the edges of the panels, and zero beyond.
In \textit{blue}, we indicate the outcome from the fits of the true data (i.e., \hst\ and \gaia\ EDR3), while the fits from mock data with a CUO prescription, constructed with \agama\ (see Section~\ref{ssec: mock-build}), are in \textit{red}.}
\label{fig: corner-comp-full}
\end{figure*}

\begin{figure*}
\centering
\includegraphics[width=0.12\hsize]{Figures/name_NGC3201.pdf} \\
\includegraphics[width=0.247\hsize]{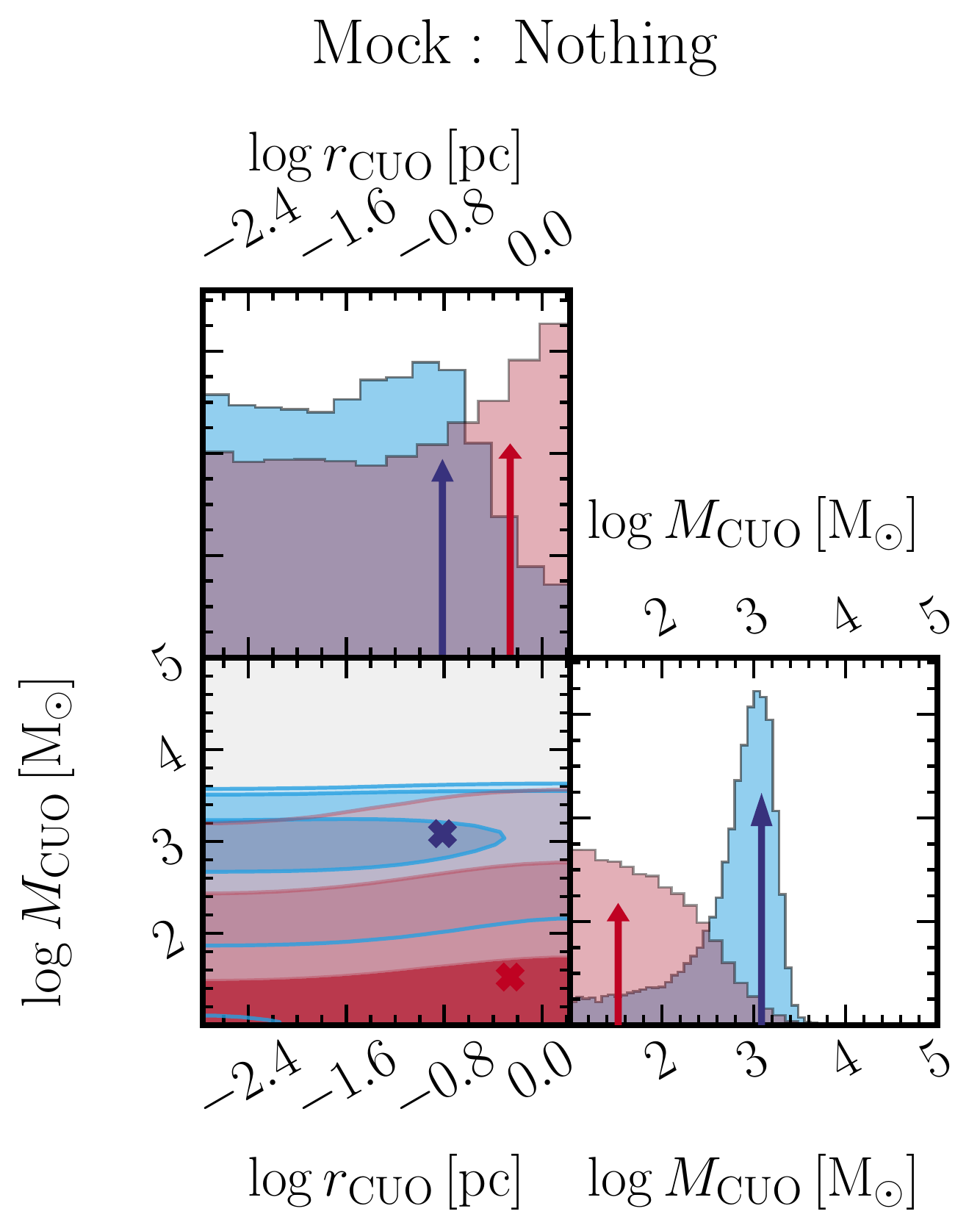}
\includegraphics[width=0.247\hsize]{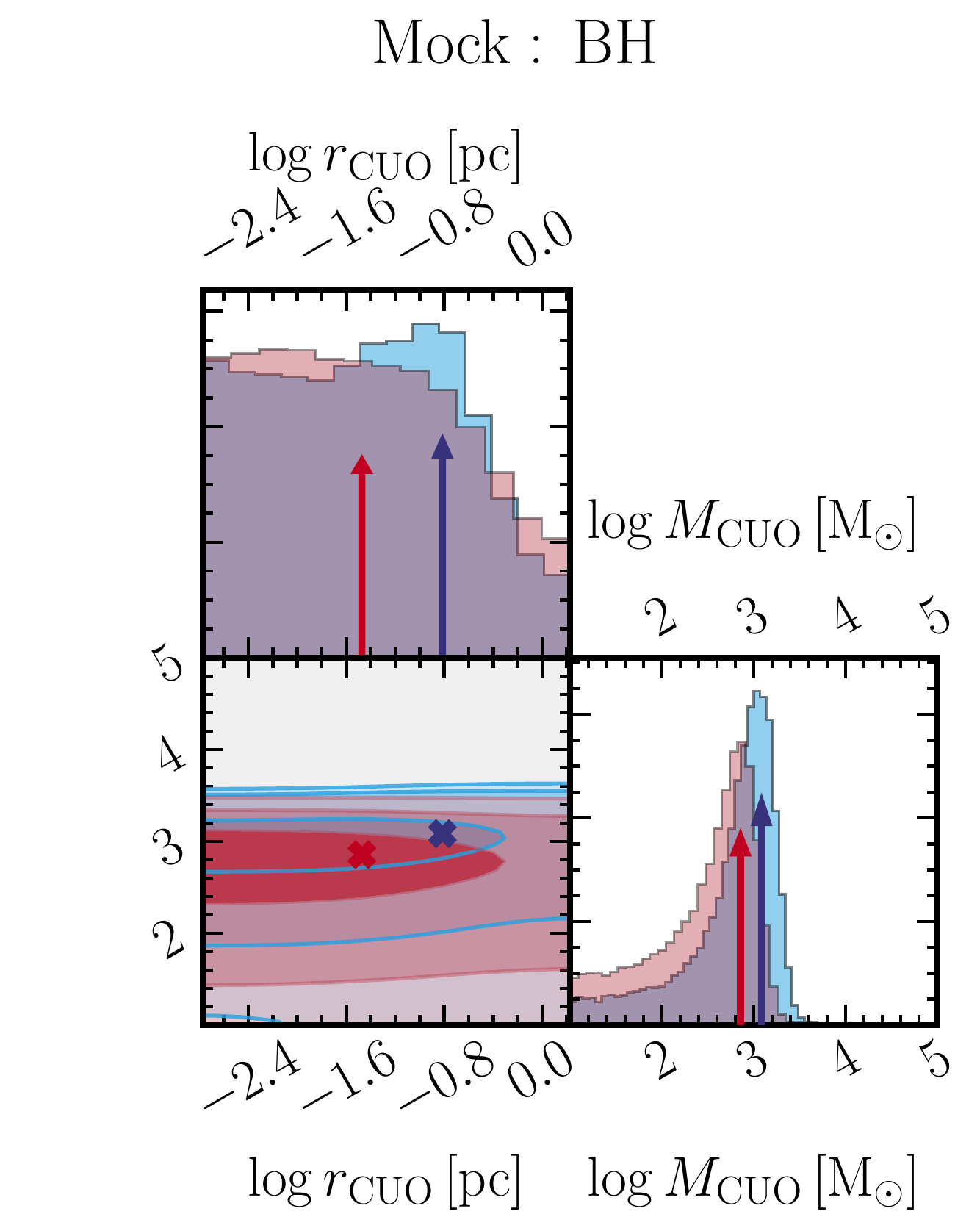}
\includegraphics[width=0.247\hsize]{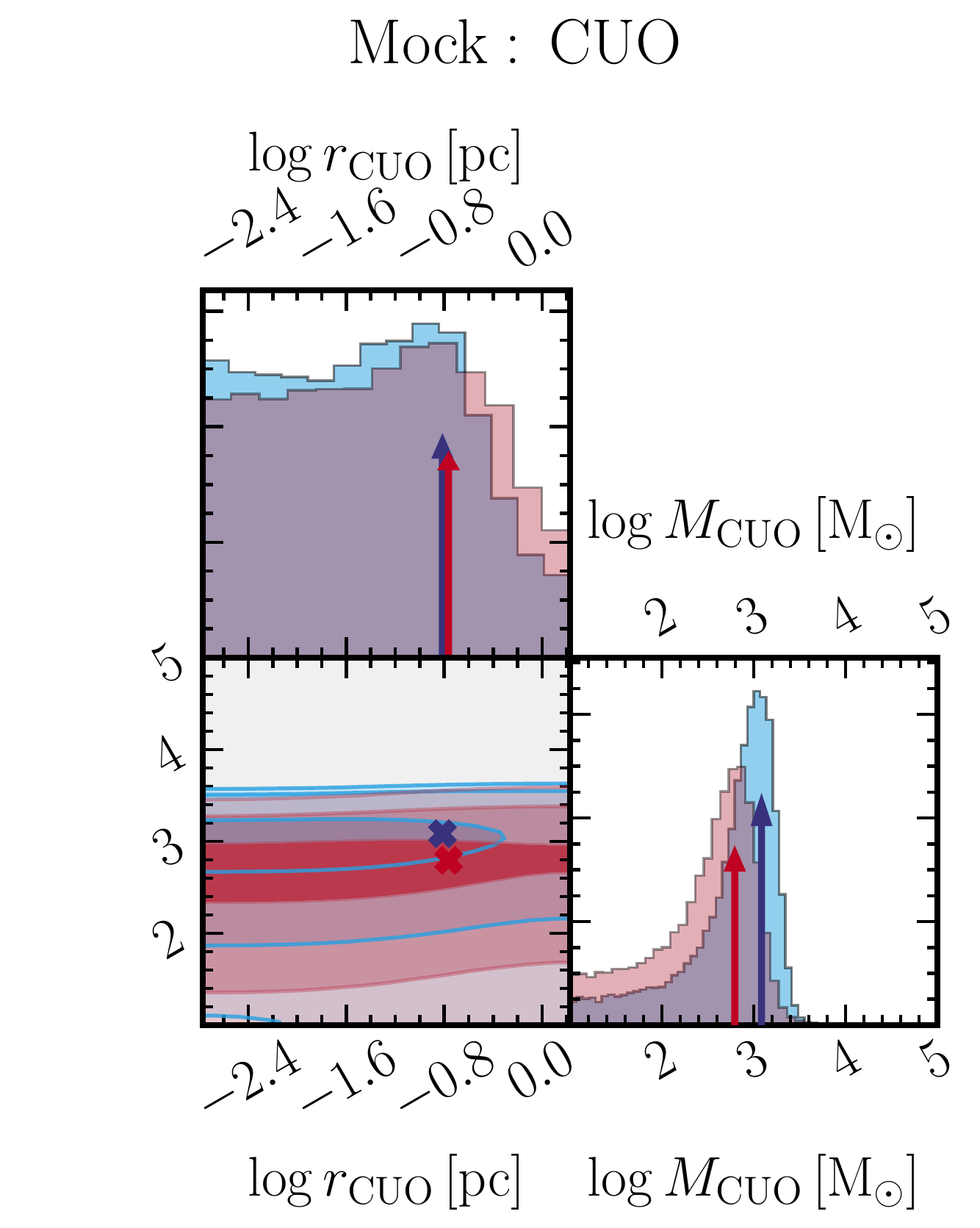}
\includegraphics[width=0.247\hsize]{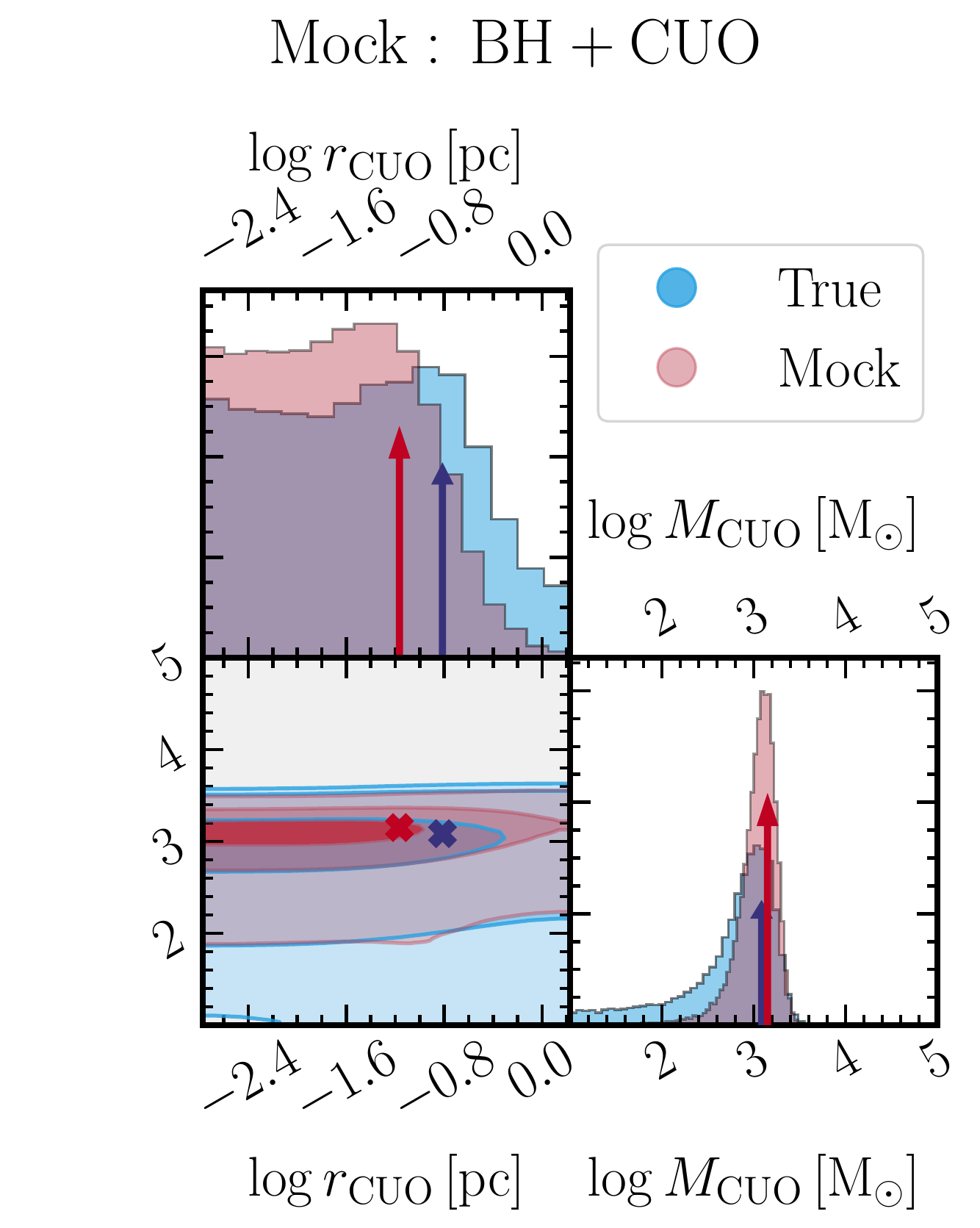}
\caption{{\it Mock data with underestimated errors:} Similarly to Figure~\ref{fig: corner-comp-partial}, we present the corner plots of the logarithm of the cluster of unresolved objects (CUO) mass (in M$_{\odot}$) and 2D Plummer half mass radius (in pc) for the true data (\hst\ and \gaia\ EDR3) in \textit{blue} and the mock data (constructed with \agama) in \textit{red}, for NGC~3201, which had different results when performing data cleaning with different error thresholds. 
The priors are flat for $M_{\rm CUO}$ within the plotted range and zero outside, while they are Gaussian for the scale radii, centred on the middles of the panels and extending to $\pm3\,\sigma$ at the edges of the panels, and zero beyond.
The mock data prescription is, from \textbf{left} to \textbf{right}: No central dark component (Nothing); a central black hole alone (BH); a central CUO (CUO) and both a central black hole and CUO (BH$+$CUO). The mocks were constructed with the best values of each respective isotropic mass model from Table~\ref{tab: results-full} (online version), but the errors provided to \mpo\ in the inner regions (i.e., up to $2 \, r_{\rm CUO}$) were underestimated by 10\%, in order to test the effect of underestimated errors in the true data. We notice that no strong mass overestimation, such as the mass peak for models with a central mass, is due to the underestimated errors.}
\label{fig: underestimated-err}
\end{figure*}

\begin{figure*}
\centering
\includegraphics[width=0.4\hsize]{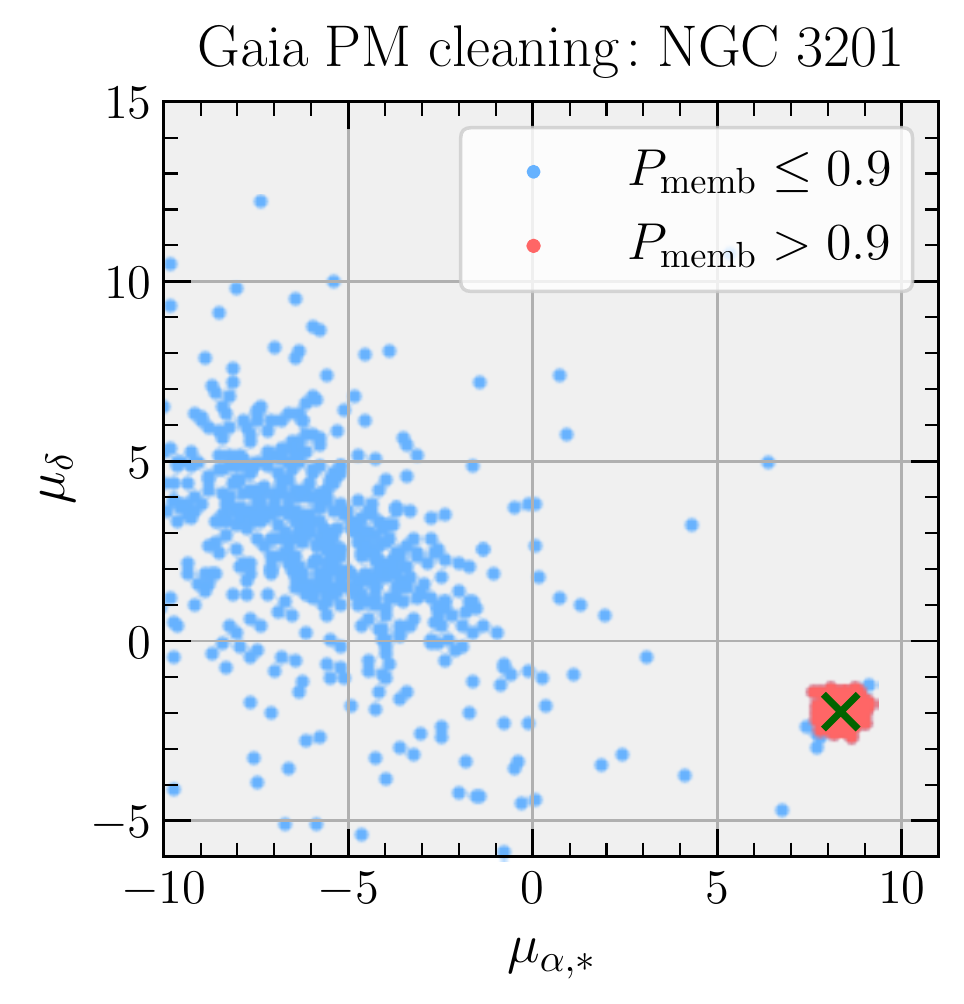}
\includegraphics[width=0.4\hsize]{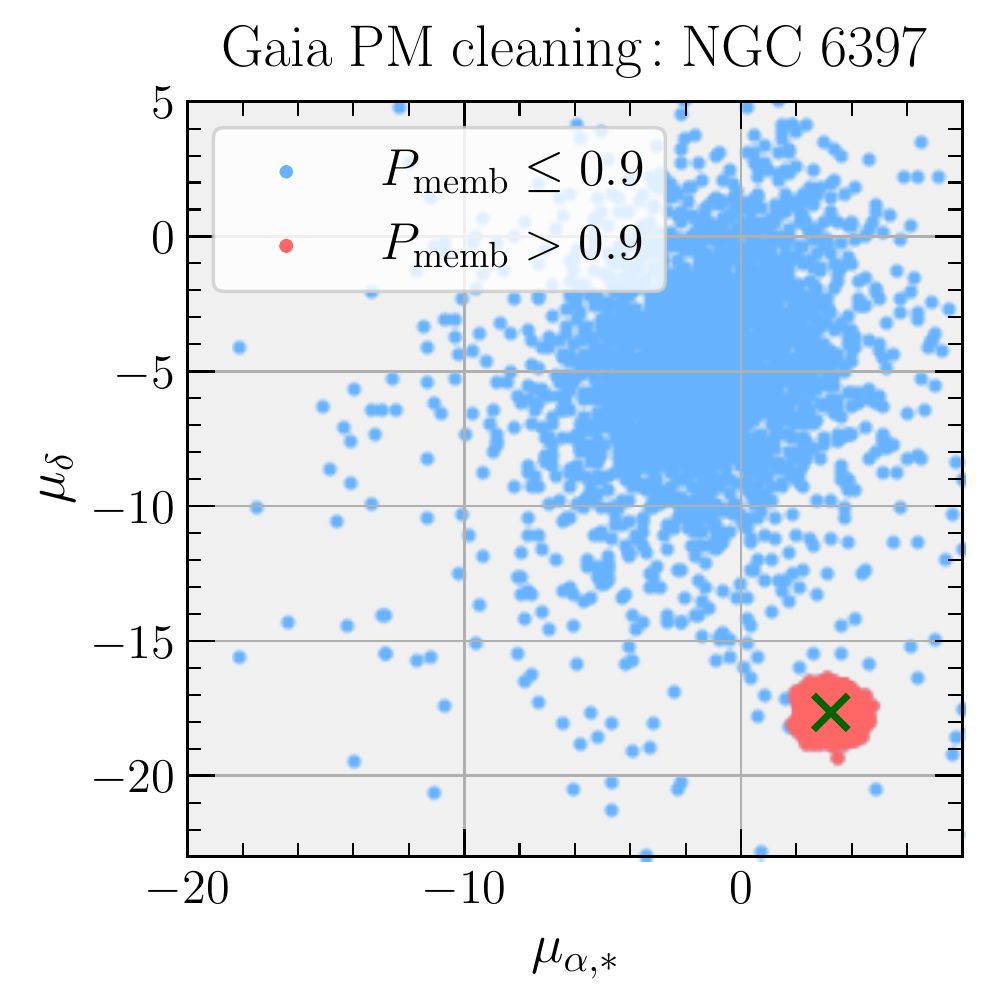}
\includegraphics[width=0.4\hsize]{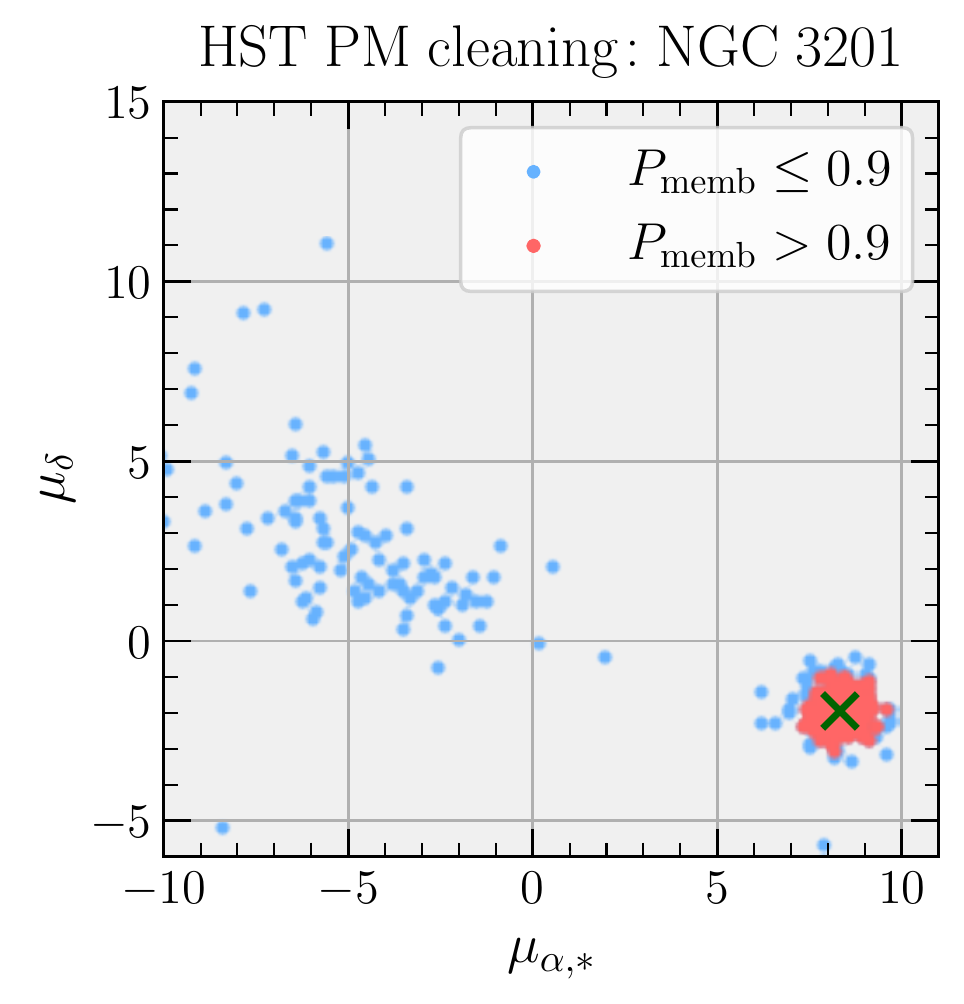}
\includegraphics[width=0.4\hsize]{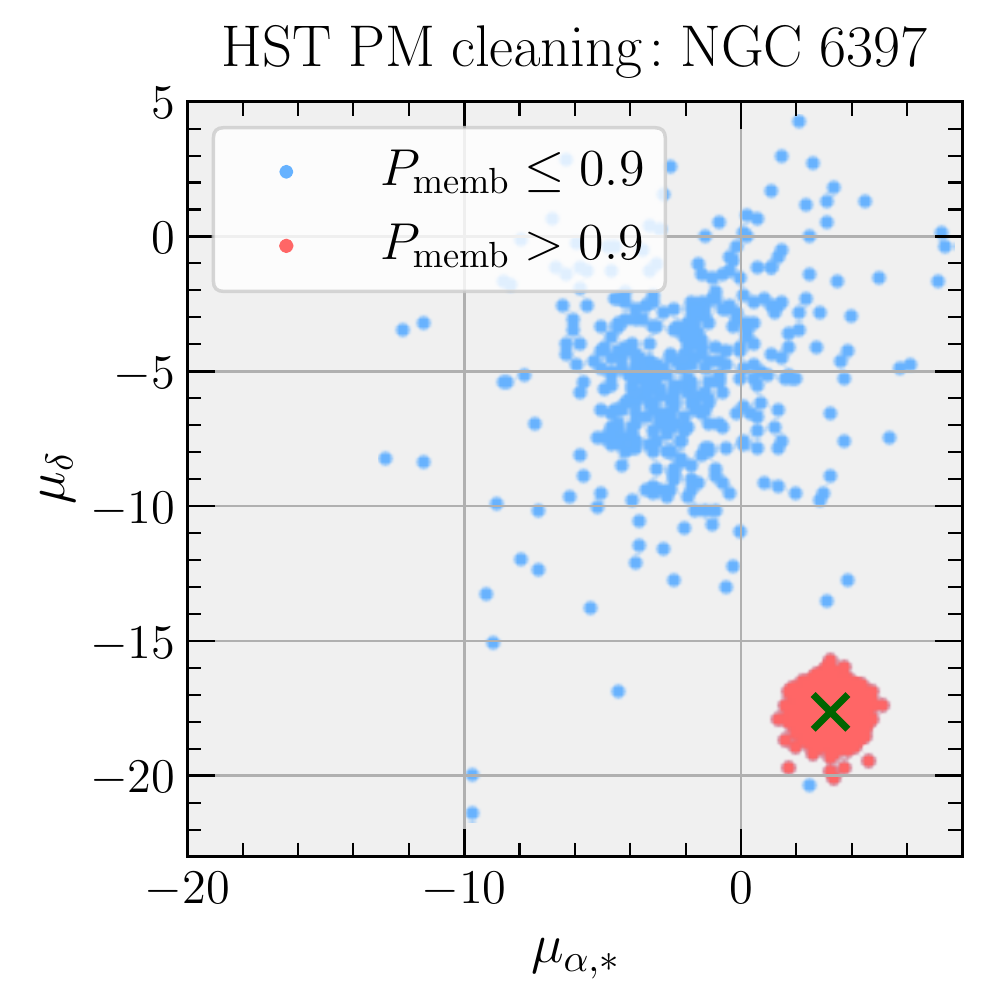}
\caption{\textit{Proper motion interloper filter:} Comparison between the proper motion subset before (\textbf{blue}) and after (\textbf{red}) the interloper proper motion cleaning described in Section~\ref{ssec: pm-ilop-cut}. \textit{Upper plots} display the \gaia\ EDR3 proper motion space, while \textit{lower plots} depict the \hst\ proper motion space. The stars from NGC~3201 are on the \textit{left} column, while the \textit{right} column shows the NGC~6397 stars. The \textbf{green crosses} represent the bulk proper motion of the clusters derived by Vitral (2021) with \gaia\ EDR3.}
\label{fig: pm-cut}
\end{figure*}

\begin{figure*}
\centering
\includegraphics[width=0.4\hsize]{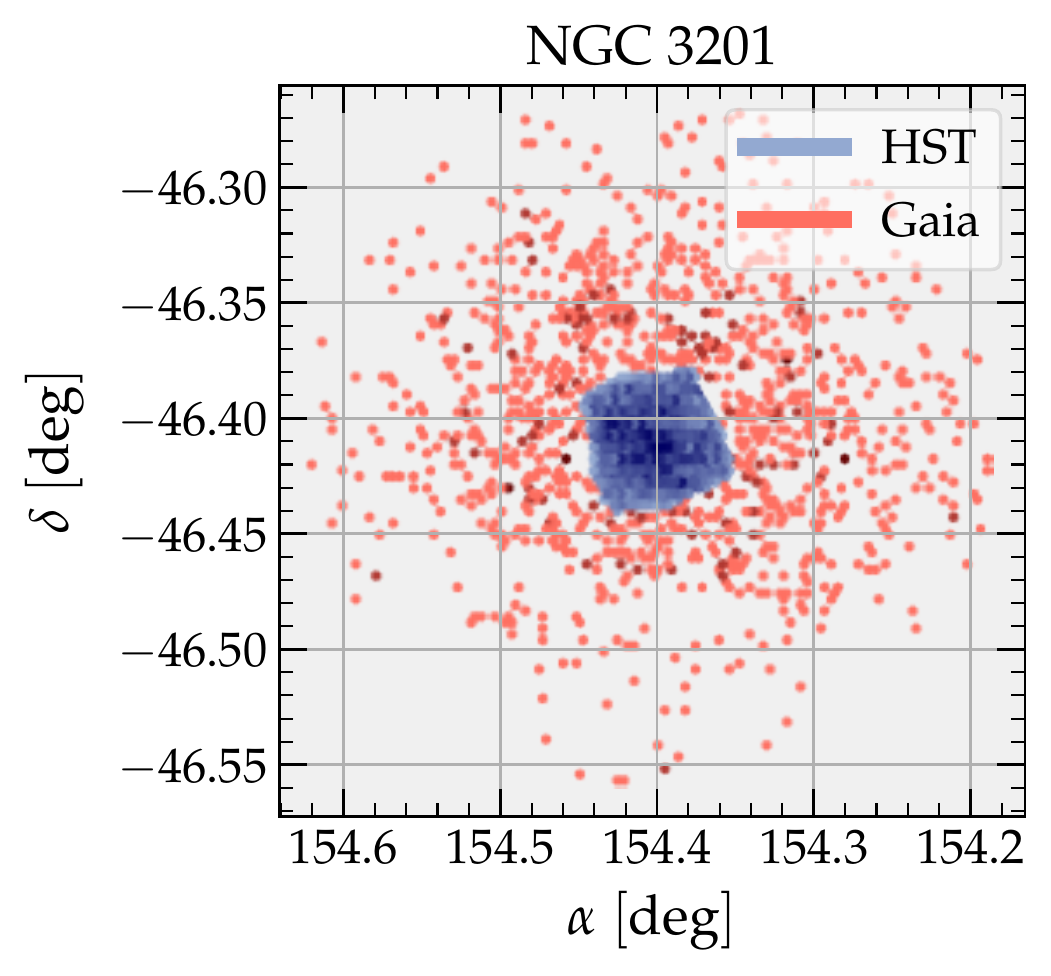}
\includegraphics[width=0.4\hsize]{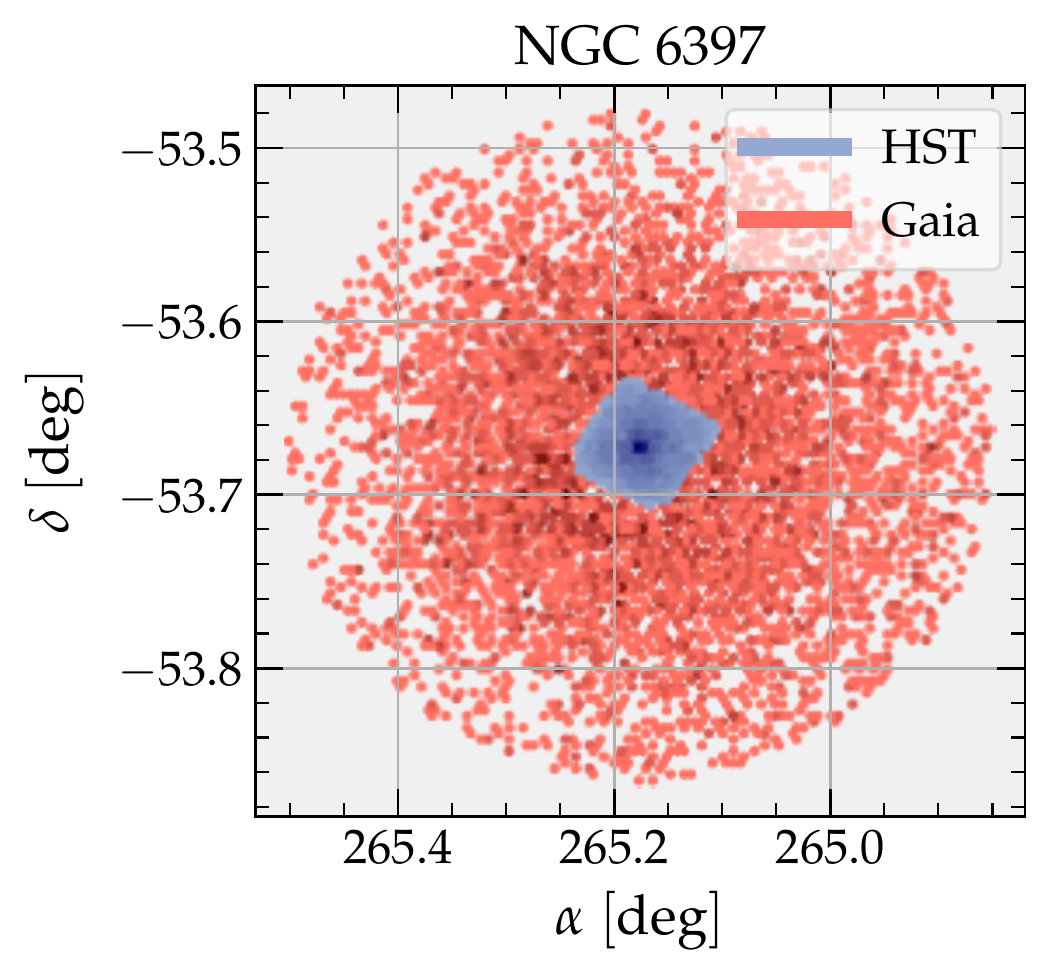}
\caption{\textit{Sky view:} Position in the sky of the data used in our modelling. Stars associated with \hst\ are in {\it blue}, while the ones associated with \gaia\ EDR3 are in {\it red}.}
\label{fig: sky-view}
\end{figure*}

\begin{figure*}
\centering
\includegraphics[width=0.45\hsize]{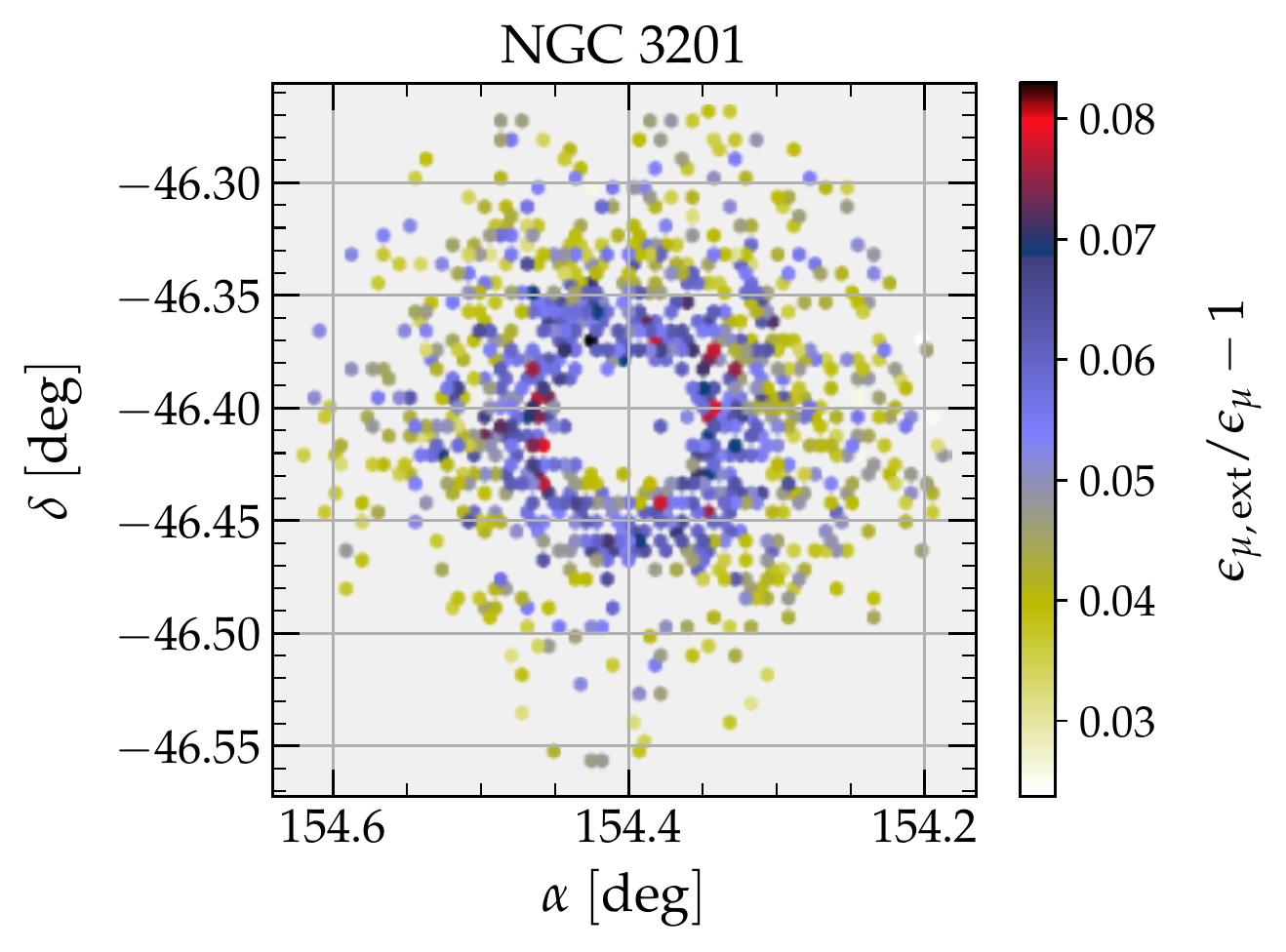}
\includegraphics[width=0.45\hsize]{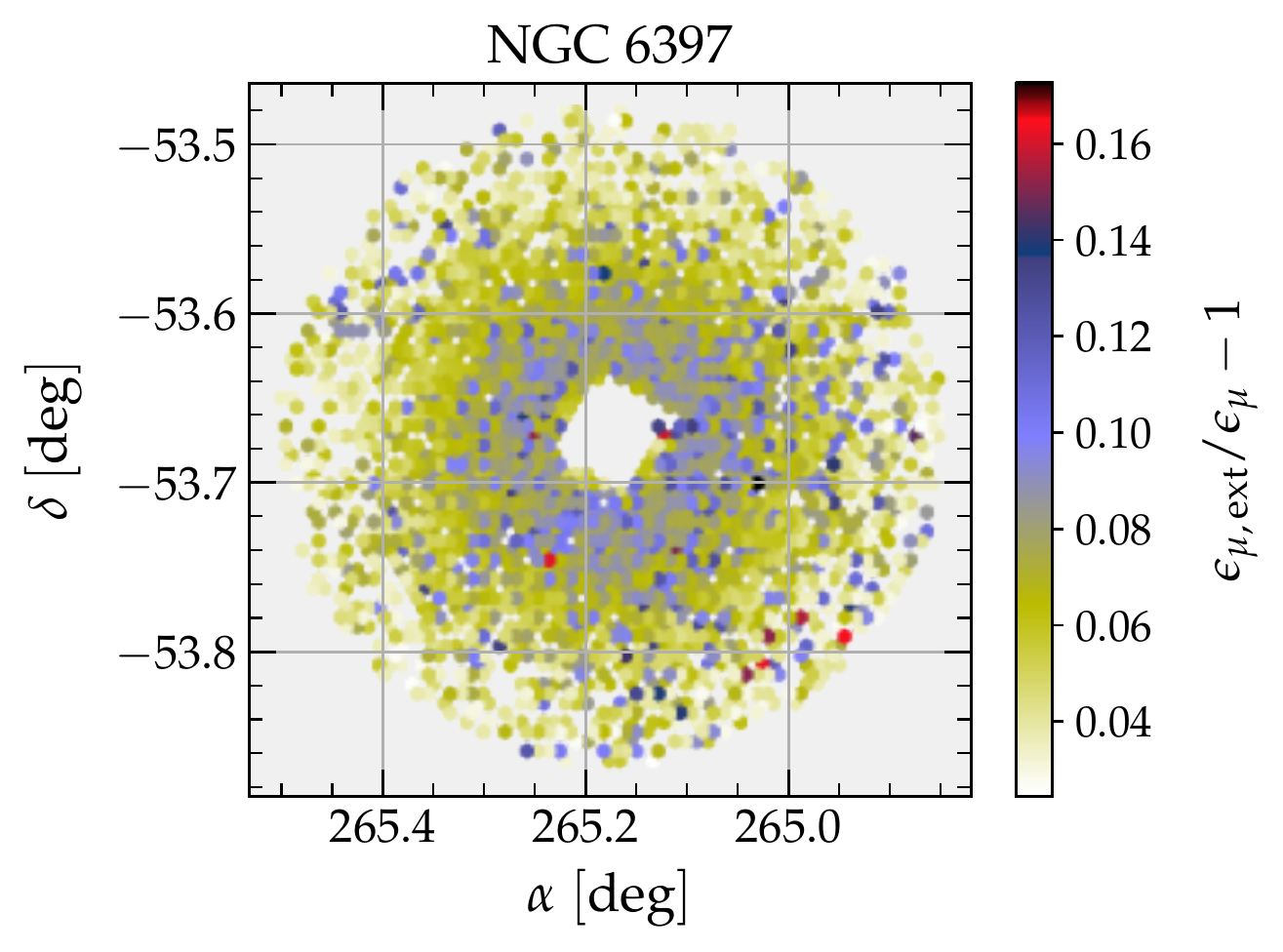} \\
\vspace{0.5cm}
\includegraphics[width=0.35\hsize]{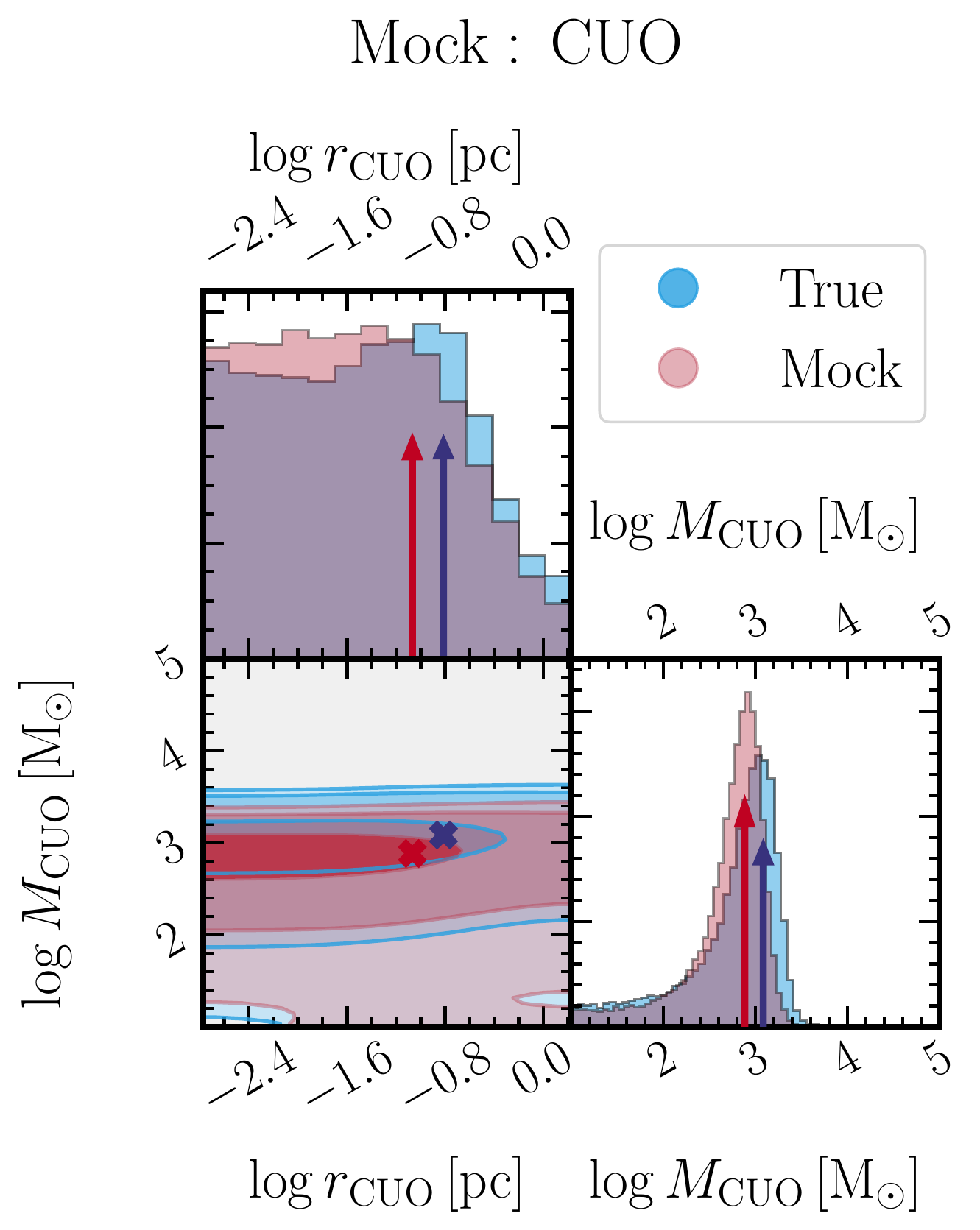}
\hspace{1.0cm}
\includegraphics[width=0.35\hsize]{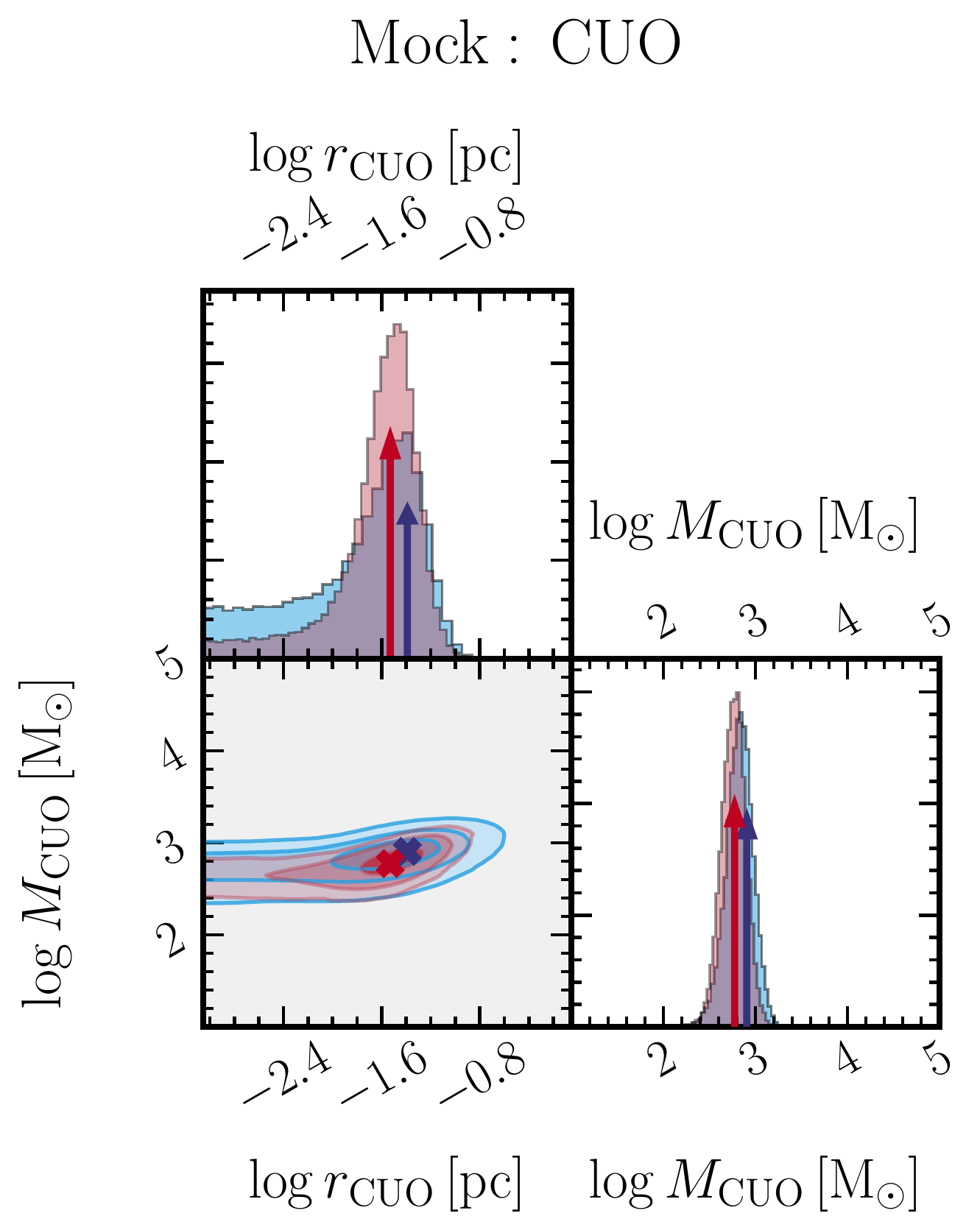}
\caption{\textit{Gaia systematics:} The \textbf{upper} plots evaluate the \textit{Gaia} systematics in our cleaned \textit{Gaia} data, computed according to eq.~3 from \protect\cite{Vasiliev&Baumgardt&Baumgardt21}, using the values from their table~1 (first line, with $\epsilon_{\mu, \rm sys} = 0.026 \ \masyr$). The bulk of the stars, quantified by the 84th percentile, have underestimated proper motion errors (i.e., $\epsilon_{\mu, \rm ext} / \epsilon_{\mu} - 1$) by only $< 6\%$ for NGC~3201 and $< 10\%$ for NGC~6397. The respective medians (50th percentile) are of 5\% and 7\% for NGC~3201 and NGC~6397. The \textbf{lower} plots display the comparison of marginal distributions of our standard fits (\textit{blue}) on real data (neglecting the \textit{Gaia} systematics) and the fits of mock data (\textit{red}) with a sub-cluster of unseen objects, constructed according to Section~\ref{ssec: mock-build}, but with \textit{Gaia} underestimated errors following the same pattern than evaluated above. One sees that the small magnitude of these \textit{Gaia} systematics do not significantly affect our fits of both mass and extent of the inner dark population. This gives us a reasonable marge to neglect these systematics in our modelling. Furthermore, their individual, separate effects on $\epsilon_{\mu \alpha}$, $\epsilon_{\mu \delta}$ and $\rho_{\mu \alpha \delta}$, which we use in our Jeans modelling, are not yet well quantified, and assuming the same factor for all these components could insert new systematics, which in turn are beyond the scope of our modelling.}
\label{fig: scaling}
\end{figure*}

\begin{figure*}
\centering
\includegraphics[width=0.95\hsize]{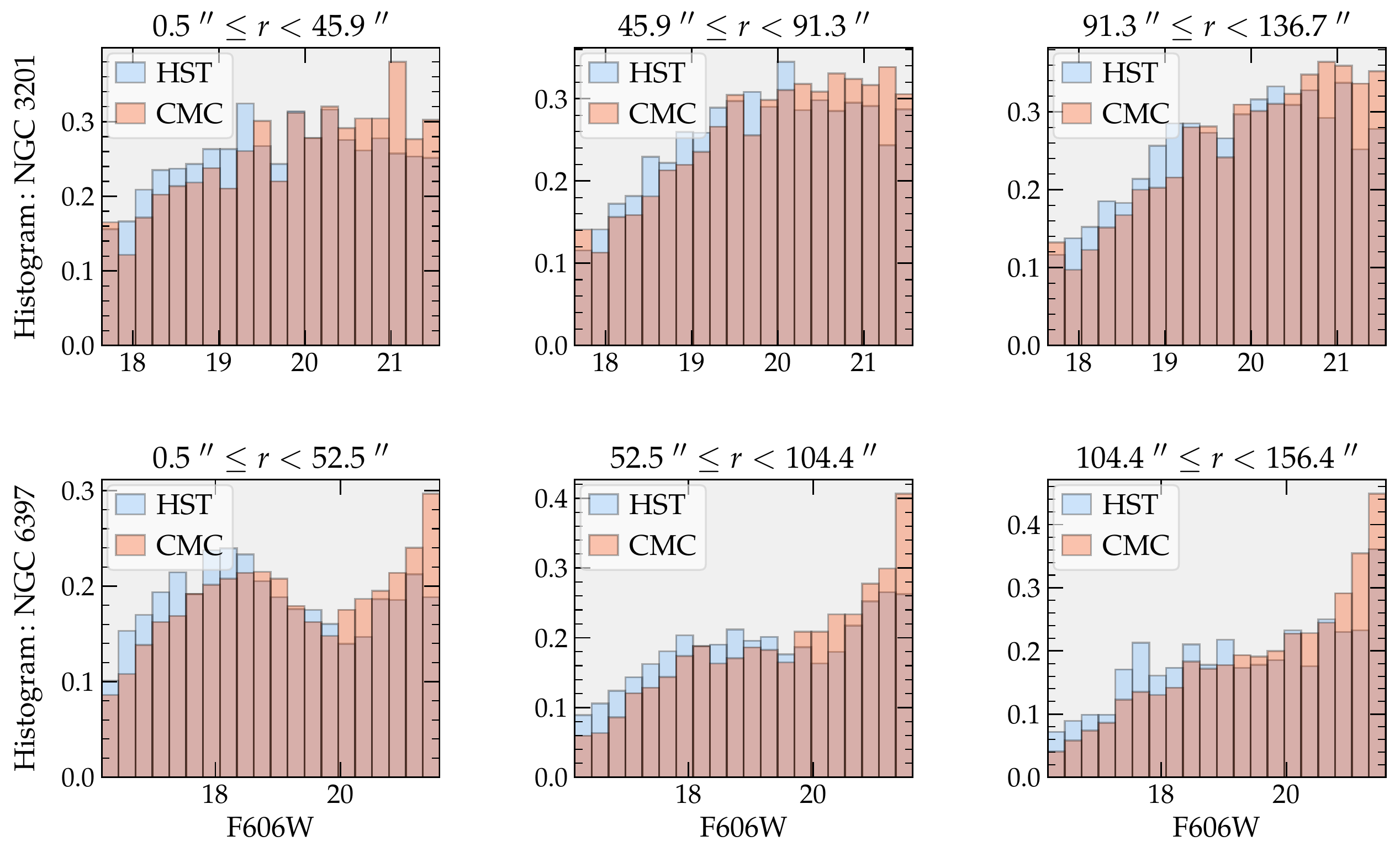}
\caption{\textit{Magnitude distribution:} \textbf{Normalised} histogram of \texttt{CMC} F606W magnitudes (orange, converted from masses with the \textsc{Parsec} isochrones) and \textit{HST} F606W magnitudes (blue), for three annuli surrounding the cluster’s centre. The upper row indicates the results for NGC~3201, while the lower row depicts the behaviour for NGC~6397. For a fairer comparison, we used the non-cleaned \textit{HST} subset, but still in the same mass/magnitude ranges of the clean data we actually use. We also forced spatial incompleteness, by assuming the following: (1) At the centre-most 500 pixels (20 arcsec), at F606W$=17$ one is 100\% complete, at F606W$=19$ one has a completeness of 90\% and at F606W$=21$, one has 70\% . (2) Between 500 and 1000 pixels (20 and 40 arcsec), at F606W$=17$ one is 100\% complete, at F606W$=19$ one has a completeness of 100\% and at F606W$=21$ one has 85\%. (3) At larger radii, one is always complete. This figure adds a strong reliability to the \texttt{CMC} models we used to compare our fits.}
\label{fig: mass-seg}
\end{figure*}

\begin{figure*}
\centering
\includegraphics[width=0.49\hsize]{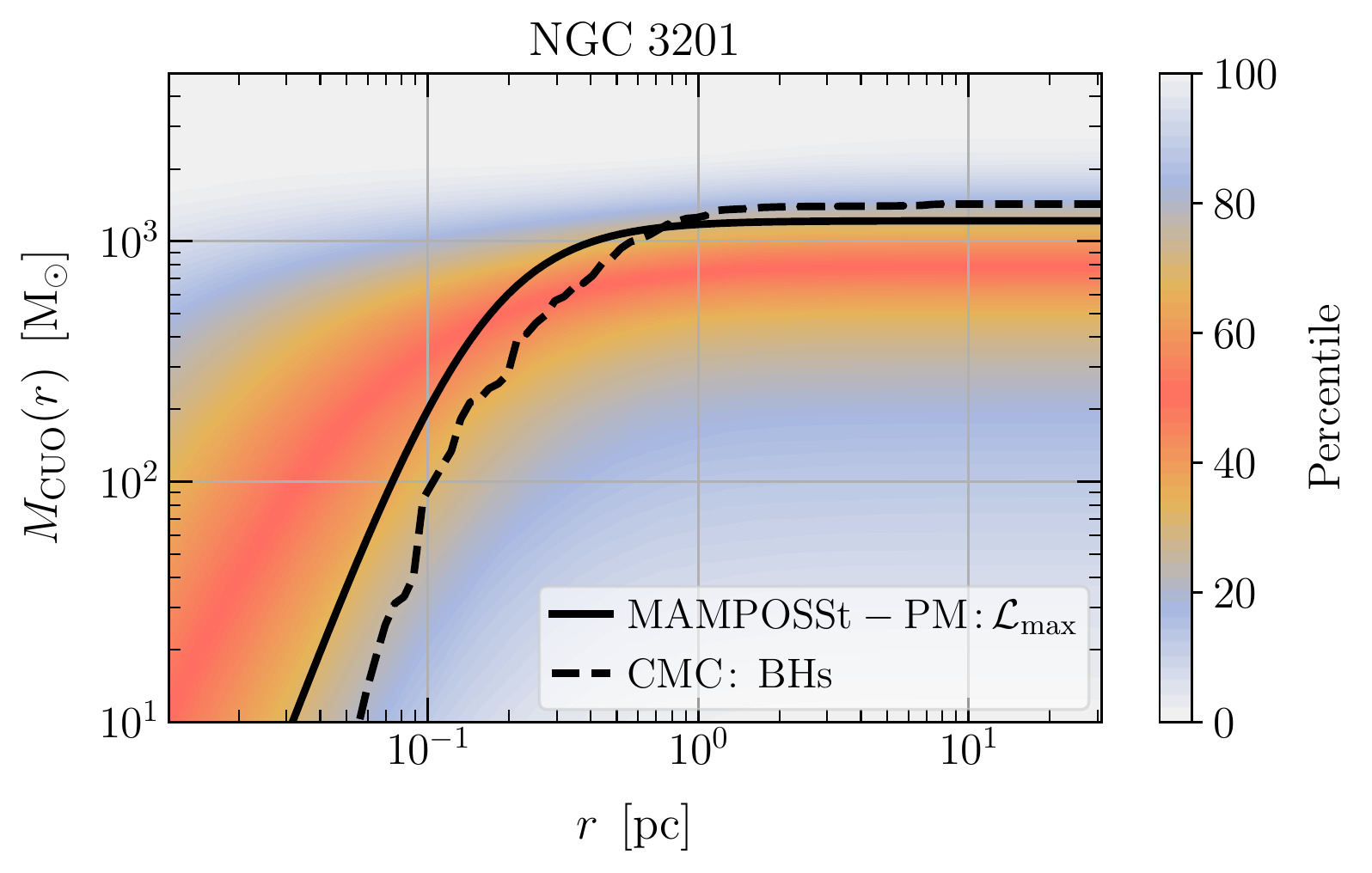}
\includegraphics[width=0.49\hsize]{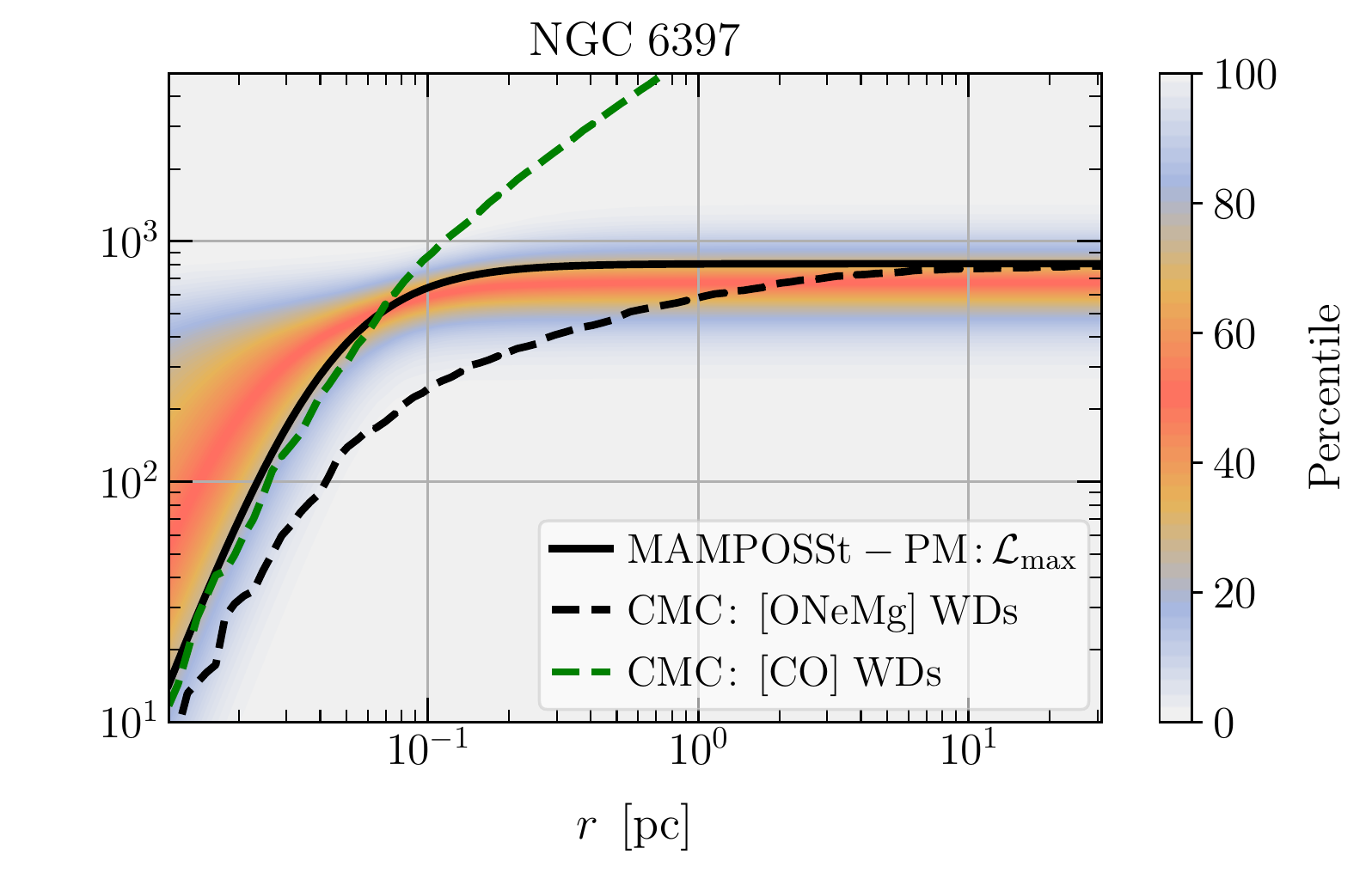}
\caption{\textit{Mass profiles:} Comparison of the cumulative mass profiles of the sub-clustered population of stellar remnants in NGC~3201 (\textit{left}) and NGC~6397 (\textit{right}), estimated by \mpo\ (best likelihood value in \textbf{thick black}) and by \cmc\ (\textbf{dashed black}, formed by black holes in NGC~3201 and mainly by [ONeMg] white dwarfs in NGC~6397). The colour bar indicates the percentile of the \mpo\ MCMC chain post burn-in phase (The skewed marginal \mpo\ mass distribution of the CUO in NGC~3201 leads to a higher mode than the median, much in line with the maximum likelihood; on the other hand, the marginal \mpo\ mass distribution of the NGC~6397 CUO is symmetric, and the maximum likelihood value is higher than the mode). For NGC~6397, we also display the population of [CO] white dwarfs in \textbf{dashed green}, which are actually segregated deeper in the cluster's gravitational potential (as our \mpo\ fits suggest), but ends up mixing with the luminous stellar population (hence, not forming a sub-cluster).}
\label{fig: mass-prof-cuo}
\end{figure*}


\bsp	
\label{lastpage}

\end{document}